\documentclass[aps,prab,twocolumn,groupedaddress,floatfix]{revtex4-2}
\usepackage{hyperref}
\usepackage{graphicx}
\usepackage{dcolumn}% Align table columns on decimal point
\usepackage{bm}% bold math
\usepackage{amsmath} % for matrices
\graphicspath{{figures/}}
\begin{document}

%\preprint{}

\title{The Continuous Electron Beam Accelerator Facility at 12 GeV}
\thanks{This material is based upon work supported by the U.~S.~ Department of Energy, Office of Science, Office of Nuclear Physics under contract DE-AC05-06OR23177.}



\author{P.~A.~Adderley$^{1}$, S.~Ahmed$^{1,l}$, T.~Allison$^{1,d}$, R.~Bachimanchi$^{1}$, K.~Baggett$^{1}$, M.~BastaniNejad$^{1}$, B.~Bevins$^{1}$, M.~Bevins$^{1}$,
 M.~Bickley$^{1}$, R.~M.~Bodenstein$^{1}$
% ...
}
\email{krafft@jlab.org}

%\thanks{}
\altaffiliation{\\$^a$Present Affiliation: Fermi National Accelerator Laboratory, Batavia, IL 60510\\
$^b$Present Affiliation: SLAC National Accelerator Laboratory, Menlo Park, CA 94025\\
$^c$Present Affiliation: Oak Ridge National Laboratory, Oak Ridge, TN 37830\\
$^d$Present Affiliation: European Spallation Source, Lund, Sweden\\
$^e$Present Affiliation: Phase Space Tech, Lund, Sweden\\
$^f$Present Affiliation: Northern Illinois University, DeKalb, IL 60115\\
$^g$Present Affiliation: Lawrence Berkeley National Laboratory, Berkeley, CA 94720\\
$^h$Present Affiliation: United States Department of Energy, Germantown, MD 20874\\
$^i$Present Affiliation: Facility for Rare Isotope Beams, Michigan State University, East Lansing, MI 48824\\
$^j$Present Affiliation: Daresbury Laboratory, Warrington, WA4 4AD, United Kingdom\\
$^k$Present Affiliation: Institute of Basic Science, Daejeon, 34047, Republic of Korea\\
$^l$Present Affiliation: ANSYS\\
$^m$Present Affiliation: xLight Inc., Palo Alto, CA 94306\\
$^n$Present Affiliation: Belgian Nuclear Research Centre - SCK CEN, 2400 Mol, Belgium}
% no Jay Benesch?
\affiliation{
$^1$Thomas Jefferson National Accelerator Facility, Newport News, VA 23606\\
$^2$Department of Physics, Center for Accelerator Science, Old Dominion University, 
Norfolk, VA 23529\\
$^3$Department of Physics, University of Connecticut, 
Storrs, CT 06269}


\date{\today}

\begin{abstract}
This review paper describes the energy-upgraded CEBAF
accelerator. This superconducting linac has achieved 12 GeV beam
energy by adding 11 new high-performance cryomodules containing eighty-eight superconducting
cavities that have operated CW at an average accelerating gradient of
20 MV/m. After reviewing the attributes and performance of the
previous 6 GeV CEBAF accelerator, we discuss the upgraded CEBAF accelerator
system in detail with particular attention paid to the new beam acceleration systems.
In addition to doubling the acceleration in each linac, the upgrade included improving
the beam recirculation magnets, adding more helium cooling capacity to allow the newly installed
modules to run cold, adding a new experimental hall, and improving numerous other
accelerator components.
We review several of the techniques deployed to operate and analyze the accelerator
performance, and document system operating experience and performance.
In the final portion of the document, we present much of the current
planning regarding projects to improve accelerator performance and enhance operating margins,
 and our plans for ensuring CEBAF operates reliably into the future. For the benefit of potential users of CEBAF, the performance and quality measures for beam delivered to each of the experimental halls is summarized in the appendix.
\end{abstract}

%\keywords{}

\maketitle


\tableofcontents
 
\section{Introduction}
During the years 1986-1995 the Continuous Electron Beam Accelerator Facility
(CEBAF) was built in Newport News, Virginia, USA, supporting
research in nuclear physics.
The main unique features of this accelerator are the combination of 100\% duty factor with high average beam current but low bunch charge, very high quality electron beam, and high energy, permitting coincident electron scattering and photon induced reactions probing both nuclear and nucleon structure.
CEBAF was the first large-scale deployment
of superconducting RF beam acceleration and the first large-scale application of
multipass beam recirculation \cite{ANRNPS}. Although originally designed to achieve 4~GeV,
by 2009 CEBAF produced beam energies of 6~GeV, allowing world-class
electron scattering experiments to be performed in three experimental halls.

In the three decades since the original design parameters of CEBAF were defined, the
understanding of the behavior of strongly interacting matter has evolved significantly
and important new classes of experimental questions have been identified that can be
addressed optimally by a CEBAF-type accelerator at higher energy. The original design of the facility,
coupled with developments in superconducting RF technology, made it feasible to double
CEBAF's beam energy to 12~GeV in a cost-effective manner, providing
a new research tool capable of addressing the science. The science motivating the 12~GeV Upgrade
included breakthrough programs launched in four main areas
(they are described in detail in \cite{LH2}): (1) probing potential new physics through high precision
tests of the Standard Model using precision, parity-violating electron scattering experiments;
(2) discovering the quark structure of nuclei and indentifying hidden flavors;
(3) understanding the spin and flavor dependence of valence parton distributions;
and (4) by measuring generalized parton distributions with high precision,
discover the three dimensional structure of nuclei.

In addition, this project was exciting as it allows one to
experimentally study the physical origins of quark confinement.
A theoretical explanation, supported by lattice QCD calculations,
is that quark confinement stems from the formation of a stringlike ``flux tube''
between quarks. This idea, and the proposed mechanisms for flux tube formation,
can be tested by determining the spectrum of the gluonic excitations of mesons,
sometimes referred to as ``hybrid'' mesons. In order to provide the
requisite excitation energy, the most basic requirement of the new
project was to achieve 12~GeV electron beam energy after 5 1/2 passes through
the recirculated linacs. The beam generated photons in a new
experimental hall, allowing here-to-fore impossible experiments in precision
QCD spectroscopy to be performed.  The major construction of the
accelerator upgrade to 12~GeV was completed in a six-month shut-down at the end
of 2011 and a long shut-down throughout 2013.
Since 2016, nuclear physics experiments at the higher beam energy have been performed.
For reference, tables giving the beam performance requirements for the delivery
of beam to each of the
experimental halls is found in the Appendix.

This paper begins with a review of the main technical features of the accelerator
as it was configured and operated in the 6~GeV era. The main body of the paper
consists of technical descriptions of the upgrades to the accelerator allowing
CEBAF to operate at 12~GeV. In the following sections we present information about the
accelerator performance in the new configuration, as well as a review of
some of the significant technical systems allowing the recirculated linacs to
operate properly and with high efficiency. We conclude with a forward-looking
discussion on CEBAF's future.

\begin{figure*}
\includegraphics[width=6.8in]{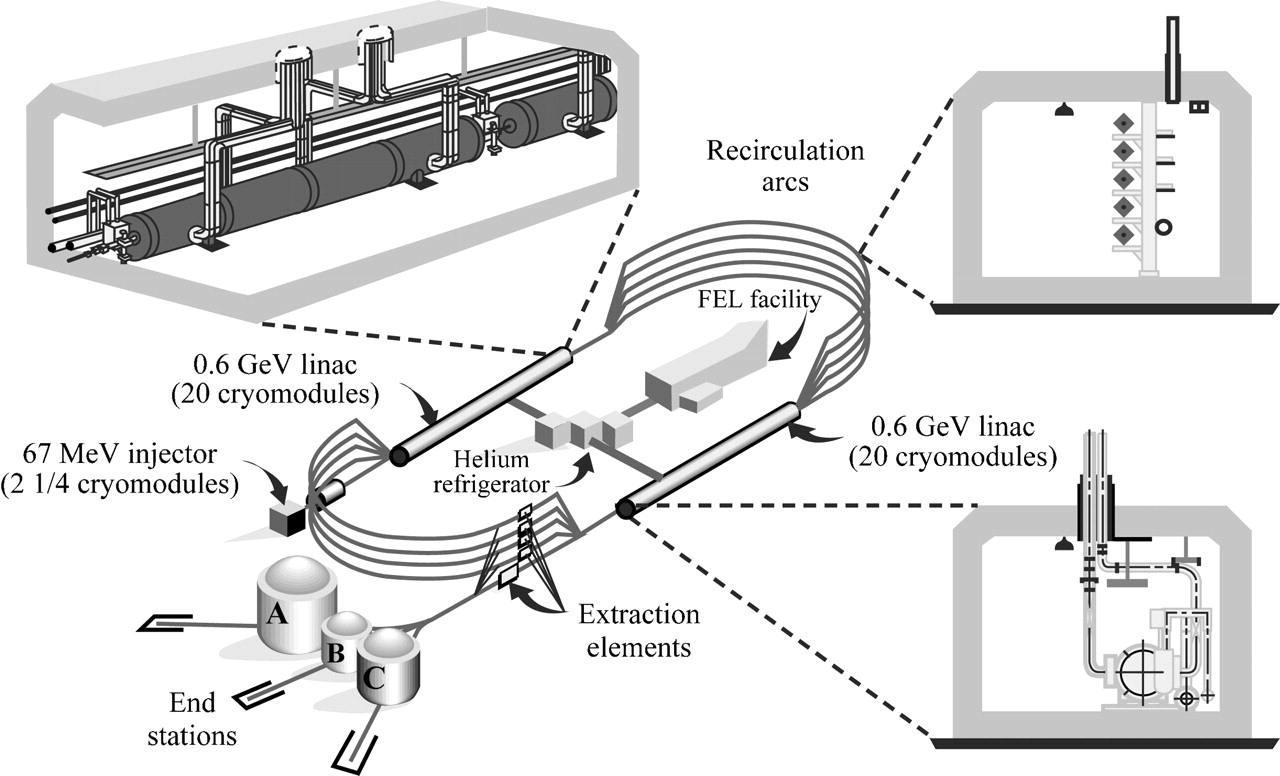}% Here is how to import EPS art
\caption{\label{fig:scm6GeV} Schematic of the 6~GeV CEBAF. (From \cite{ANRNPS}).}
\end{figure*}

\section{6 GeV CEBAF}

Before the upgrade activity, CEBAF was a 5-pass, recirculating CW electron linac operating
at up to 6~GeV. The layout appears in Figure~\ref{fig:scm6GeV} and
Table~\ref{tab:tablesum} summarized the principal accelerator parameters.

Beams of high average current up to 200 \textmu A
with 90\% polarization, of low geometric emittance less than 10$^{-9}$ m rad, and
of low relative
momentum spread less than $3\times10^{-5}$ were produced. By combining five-pass recirculation, a three laser
photocathode source, and subharmonic-rf-separator-based extraction,
three beams at different energies could simultaneously be delivered to three
end stations (Halls A, B, and C). The operating hall-to-hall
current ratios could approach 10$^6$ and the beam delivered
with a specified orientation of the beam
polarization.
The linacs were built up from cryomodules, each of which contained eight
CEBAF/Cornell \cite{PKN1} 5-cell superconducting RF (SRF) cavities.
Originally designed as a 4-pass 4~GeV recirculating linac with 50 cryomodules \cite{HG1},
5-pass recirculation with 40 cryomodules
was adopted as a cost control and optimization measure in 1988. Because the CEBAF tunnel
layout and construction had already started, during the
6~GeV era 5 cryomodule slots in each linac were left vacant.
These slots were filled with high-performance cryomodules as part of the upgrade.

\begin{table}[b]%The best place to locate the table environment is directly after its first reference in text
\caption{\label{tab:tablesum}
Principal Parameters for CEBAF in the 6 GeV era. (From \cite{ANRNPS}).
}
\begin{ruledtabular}
\begin{tabular}{ll}
Energy&6 GeV\\
Average Current (Halls A and C)&1-150 \textmu A\\
Average Current (Hall B)&1-100 nA\\
Bunch Charge&$<$ 0.3 pC\\
Repetition Rate&499 MHz at hall\\
Beam Polarization& 90\% \\
Beam size ($rms$ transverse) & $\sim$80 \textmu m \\
Bunch length ($rms$)& 300 fs, 90 \textmu m \\
Energy Spread & 2.5 $\times$ 10$^{-5}$ \\
Beam Power & $<$ 1 MW \\
Beam Loss & $<$ 1 \textmu A \\
Number of Passes & 5 \\
Number of Accelerating Cavities & 338 \\
Fundamental Mode Frequency & 1497 MHz \\
Accelerating Cavity Effective Length&0.5 m \\
Cells/cavity & 5 \\
Average $Q_0$ & 4.0 $\times$ 10$^9$ \\
Implemented $Q_{\rm ext}$  & 5.6 $\times$ 10$^6$\\
Cavity Impedance ($R/Q$) & 480 $\Omega$ \\
Average Cavity Accelerating Gradient & 7.5 MV/m \\
RF power & $<$ 3.5 kW/cavity \\
Amplitude Control & 1 $\times$ 10$^{-4}$ \\
Phase Control & 0.1$^{\circ}$ \it{rms} \\
Cavity Operating Temperature & 2.1 K \\
Liquifier 2 K Cooling Power & 5 kW \\
Liquifier Operating Power & 5 MW \\
\end{tabular}
\end{ruledtabular}
\end{table} 

\subsection{Design Summary}

Many considerations went into the design of CEBAF. For example, increased
siting flexibility of the more compact design and cost drove the decision to deploy
two antiparallel linacs instead of one long linac.
Many of CEBAF's features 
derived from the high cost of superconducting beam acceleration.
In order to take maximum advantage of the accelerating gradient
possible from each cavity, CEBAF was run as a linac, with the electron
bunches close to the peak of the accelerating voltage. This possibility exists
because phase focusing was not needed for the highly relativistic beam,
but this choice implies that the recirculation arcs be designed to be
isochronous ($M_{56} < 10$ cm). The pass-to-pass requirement for phase
control dictated that the recirculation pathlength be within 100 \textmu m
of an integer number of RF wavelengths. In practice, this requirement
was accomplished by measuring the pathlength (see Section~\ref{PLControl}) and
varying the path through individual ``pathlength'' chicanes
placed in each recirculation arc.
 
Vertical stacking of the various energy beam lines
was chosen largely for practical reasons. Vertical
dispersion was introduced, and the choice must be made between constructing individually achromatic
vertical bends or correcting vertical dispersion only at the end of the complete
arc. At CEBAF the vertical dispersion was corrected locally \cite{RYDRD1}.
This choice made operational, real-time
analysis of beam behavior through the arcs as transparent as possible, and
avoids vertical phase space growth driven by synchrotron radiation.
For the same reasons, a philosophy of functional modularity in the optics design
was adopted, resulting
in the following breakdown of transport sections from linac to linac \cite{RYDRD1,DRD1}: achromatic
vertical bend to separate different energies, matching section, 180$^\circ$ horizontal
achromatic bend based on a regular lattice operated with matched beam-envelope
functions, matching section, achromatic vertical bend back to linac level, with the
whole system globally isochronous. The two matching sections downstream of the
linacs were long, one containing pathlength-adjusting doglegs, the other containing
doglegs and beam extraction elements, while the matching regions immediately
upstream of the linac sections were short and have no additional functions.

A decision was made to keep the recirculation arc radii
large enough to allow later upgrades in energy by avoiding excessive degradation
of beam emittance and energy spread from synchrotron radiation. Magnets were
designed as low-field, low-current-density devices to minimize power consumption.
As a consequence, the
6~GeV beam transport system could be upgraded to 12~GeV by merely
replacing power supplies, increasing the saturated field strength in the recirculation arc dipoles, and exchanging a small number of other magnets. With a
completely new lattice and magnets, the arc tunnel radius is large enough to allow
a future upgrade to about 25~GeV  \cite{ANRNPS,DRDRef2}.

Operating the three end stations simultaneously was a desideratum,
and with the use of multipass beam recirculation, additional degrees
of freedom became available to achieve this goal. Three beam
operation was implemented by creating three interlaced 499 MHz beams at
the source. The bunches going to each of the separate Halls
were spaced apart by 120$^\circ$ of rf phase at 499 MHz. Together they form a 1497 MHz
beam in which each bunch has properties, particularly charge, that may differ
from its immediately preceding and trailing pair of neighbors, but which repeated
every third bunch. Such a current profile was achieved by using three
independent rf-gain-switched lasers \cite{MP1,CHMP1} directed at a single photocathode,
each laser with a third subharmonic 499 MHz bunch repetition frequency.

Extracting the beams to each of three end stations
was achieved by using properly phased rf deflecting cavities (``rf separators'') operating
at 499 MHz. For example, rf separators were installed in the various recirculation paths downstream of each
full pass making it possible
to serve different halls simultaneously with beams of different but correlated
energies. In addition, distributing three full-energy beams at the same time was possible using a single separator located after the fifth pass through the accelerator. In contrast to the cylindrically symmetrical rf deflector designs available at the time, the Jefferson Lab separators were ahead of their time in being fully
three dimensional \cite{Yao1}.

\subsection{CEBAF Injector and Its Upgrade}

The CEBAF photoinjector provided independent beam delivery of spin-polarized electron beams to
each experiment hall simultaneously over a wide range of requested current: from 100 pA to 180 \textmu A.
The design of the CEBAF injector was based on an injector for a microtron accelerator \cite{AHRef1P,AHRef1},
and CEBAF's injector design and layout \cite{AHRef1,AHRef2} have not changed materially since the injector
was initially installed for 4~GeV CEBAF. The injector started with a DC electron gun producing electrons that
are transported to a chopper system.  The chopper system constrained the longitudinal beam extent (bunch length)
to ensure proper bunching is initiated in a downstream single cell re-entrant RF cavity (buncher).  Next a five-cell
side-coupled graded-beta RF cavity (capture section) accelerated the bunched beam from the gun energy to 500 keV
\cite{AHRef1,AHRef2}. See Figure 1 in \cite{AHRef28} for a photo and simulation model.  A booster (quarter cryomodule) with
two five-cell SRF cavities accelerated the captured beam to 5--6 MeV.  Along a 6 m optics transport line,
the near relativistic beam continued to drift and bunch while being transported to two full cryomodules,
each with eight 5-cell cavities. The full modules accelerated the beam to the required injection energy
for the target machine energy. An injection chicane merged the injector beam into the main accelerator
\cite{AHRef4}.
The injector included several spectrometer dump lines for cavity phasing and energy measurements.
The initial design proved to be robust and flexible as it has been easily adapted to the
increasing demands of the 6~GeV and 12~GeV eras.

During the 6~GeV era, the injector changed in four ways \cite{AHRef13,AHRef20,MPRef9}.
The first was installing the 499 MHz three-beam chopper system, which supported operating three
experimental halls concurrently.  The second was installing improved full cryomodules capable
of accelerating the beam to 67.5~MeV, needed to operate at 6~GeV. The third was transitioning from
100 kV thermionic gun operations \cite{AHRef1} to 100 kV polarized source operations.
The polarized source was a DC photocathode gun capable of producing three interleaved polarized
electron beams at 499 MHz and is thoroughly discussed in the next section
\cite{Card1,MPAdded}.  The fourth was installing
a two Wien and solenoid system to set the spin delivered to the experimental halls.
The Wien system was installed between the gun and the chopper system, so the polarized source is
further away from the chopper system than in the original injector design.
To compensate for bunch lengthening of high current bunches and ensure the polarized electron bunches
match the longitudinal acceptance of the chopper system, an additional buncher cavity (prebuncher) was
installed between the gun and the chopper system.  Also, solenoids between the Wien system and the quarter
cryomodule were changed from single wound to counter wound solenoids to preserve the spin of the electrons
set by the Wien system.

The CEBAF injector was very capable, and considerable effort in the 6~GeV era was dedicated to standardizing
the injector set up process to produce the small bunch length, small energy spread, and suitable transverse
phase space required for both routine and challenging beam operations to support nuclear physics operations.
A summary of the main beam parameters from the injector at the close of the 6~GeV era
appears in Table~\ref{tab:tableinjsum}. Conditions for a CEBAF energy of 6~GeV are assumed.

\begin{table}[t]%The best place to locate the table environment is directly after its first reference in text
\caption{\label{tab:tableinjsum}
Principal beam parameters for the CEBAF injector when operating at 6~GeV \cite{ANRNPS}.
}
\begin{ruledtabular}
\begin{tabular}{lc}
Energy&67.5 MeV\\
Average Current (Halls A and C)&1-150 \textmu A\\
Average Current (Hall B)&0.1-100 nA\\
Bunch Charge&$<$ 0.3 pC\\
Repetition Rate&499 MHz at hall\\
Beam Polarization& 90\% \\
Transverse Beam Size ($rms$) & $\sim$500 \textmu m \\
Beam Normalized $rms$ Emittance& $\sim$0.5 mm mrad\\
Bunch Length ($rms$)& 300 fs, 90 \textmu m \\
Relative Energy Spread & 1 $\times$ 10$^{-3}$ \\
Beam Longitudinal $rms$ Emittance& $\sim$20 keV-ps\\
Number of SRF Accelerating Cavities & 18 \\
\end{tabular}
\end{ruledtabular}
\end{table}

\begin{figure}[b]
\includegraphics[width=3.4in]{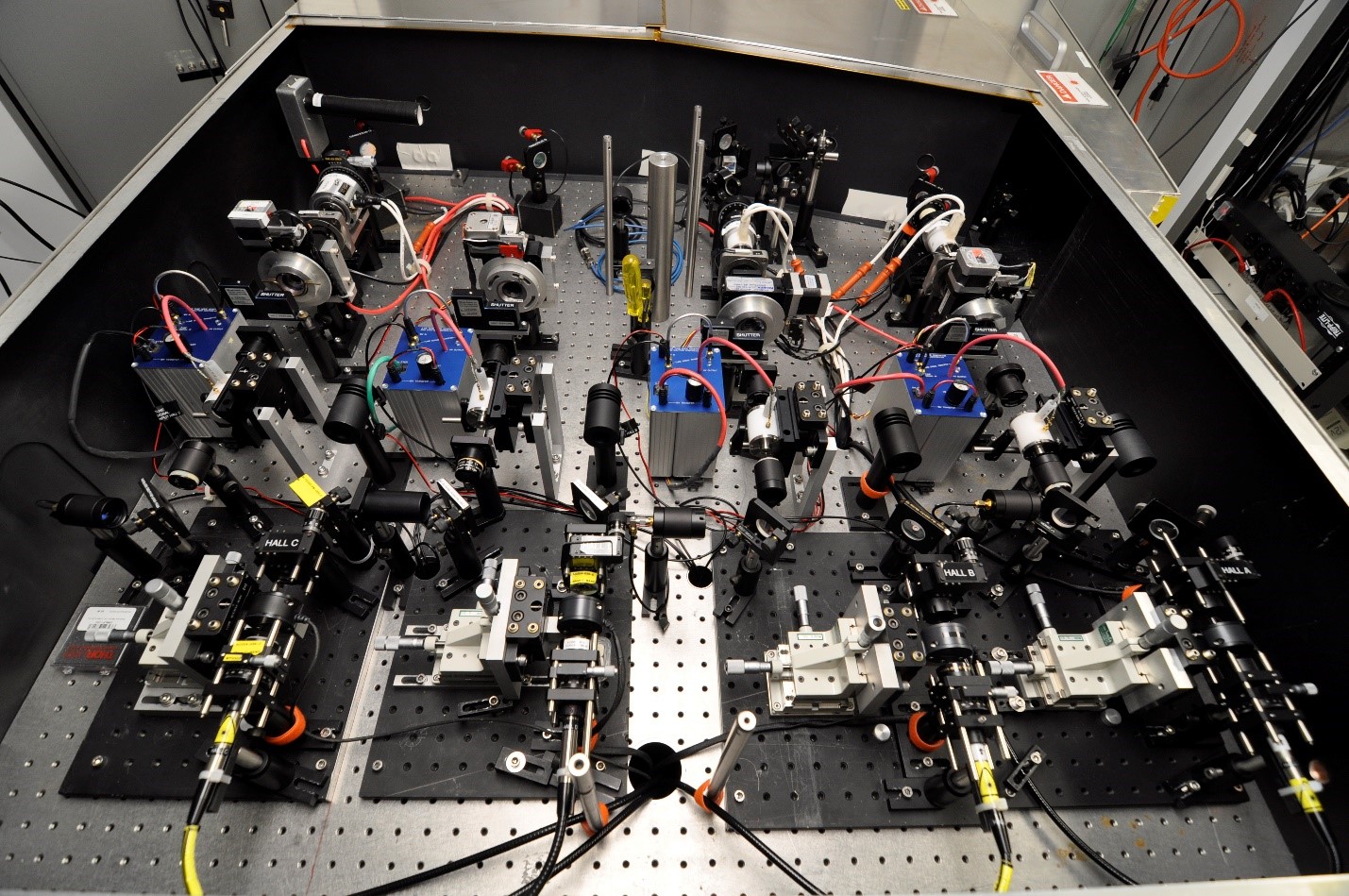}% Here is how to import EPS art
\caption{\label{fig:mpfig5} 
Photograph of the 4-laser system used to create interleaved electron beams, one laser for each
experiment hall.  Laser beams are combined using partially reflective mirrors and polarization
sensitive optical elements.}
\end{figure}

\subsubsection{Injector Improvements for 12 GeV CEBAF}

Most components of the 12~GeV CEBAF photoinjector are located as in the 6~GeV CEBAF photoinjector,
but the layouts of the injection chicane and full energy injector spectrometer are
adjusted to accommodate the new beam line that is part of the upgrade.
The main improvements to the injector
are: (1) to increase the gun voltage to 130 kV DC, (2) to increase the overall injector
energy capability to 123~MeV \cite{AHRK123MeV}, and (3) to add a fourth drive laser and required RF equipment to allow
four beams to be produced simultaneously, for
delivery to four experiments. Here, we briefly discuss the changes. More detail will be
found in the referenced sections in the main body of this paper.

The first change reduces space charge effects and
results in more consistent beam setups in cases
when the bunch charge to the different halls is
very different (e.g., when the Hall B charge is 0.01 fC provided
simultaneously with 0.2 pC to Hall A).
It is the first step in a longer term
project to achieve 200 kV electron kinetic energy from the gun, described
in greater detail in Section~\ref{MPrefsec}.

The second change follows directly from the desire to have the beam on
the same orbits at 12~GeV as it had operating at 6~GeV. For this desire to be achieved, throughout CEBAF the beam energies in the
various passes and arcs must be at nearly the same energy ratios.
In the 12~GeV CEBAF design, when Hall D is operating at 12~GeV,
the maximum beam energy to the original Halls A-C is 10.9~GeV.
Therefore, the beam energy at injection must be scaled up by a
factor of 1.82 (10.9/6) going from 6~GeV in Halls (A-C) to 12~GeV in Hall D, or from 67.5~MeV to 123~MeV. 
Such an energy increase is comfortably
accomplished by replacing the final cryomodule in the injector
with a cryomodule of the new upgrade design. As discussed
in Section~\ref{R100ref}, the specific new cryomodule in the injector, called
``R100'' \cite{r100mar}, is built with capabilities largely equivalent to the cryomodules
placed in the linacs for the upgrade. Because there is
no recirculation in the injector, the RF power required to accelerate the
beam load in the injector is much less than in the main linacs,
meaning the RF systems driving the R100 can be largely reused
after the upgrade. 

Thirdly, a major difference between the 6 GeV and 12 GeV CEBAF photoinjectors is that now four halls
can receive beam
simultaneously instead of just three.  This upgrade
is accomplished by adding a fourth drive laser (Fig.~\ref{fig:mpfig5}) and
by modifying the extraction/separator system \cite{MPRef12}.  Whereas during 6 GeV CEBAF
operation, interleaved laser pulse trains at 499 MHz were used, now lasers can operate at 249.5 MHz,
the 6th subharmonic of 1497 MHz.
Interleaved pulse trains at 249.5 MHz permit simultaneous 4-hall operation, albeit with ``empty buckets''
that pose no problems for the nuclear physics program. The  
required modifications to the RF extraction/separator system are described further in Section~\ref{fourhall}.

\begin{figure}[b]
\includegraphics[width=3.4in]{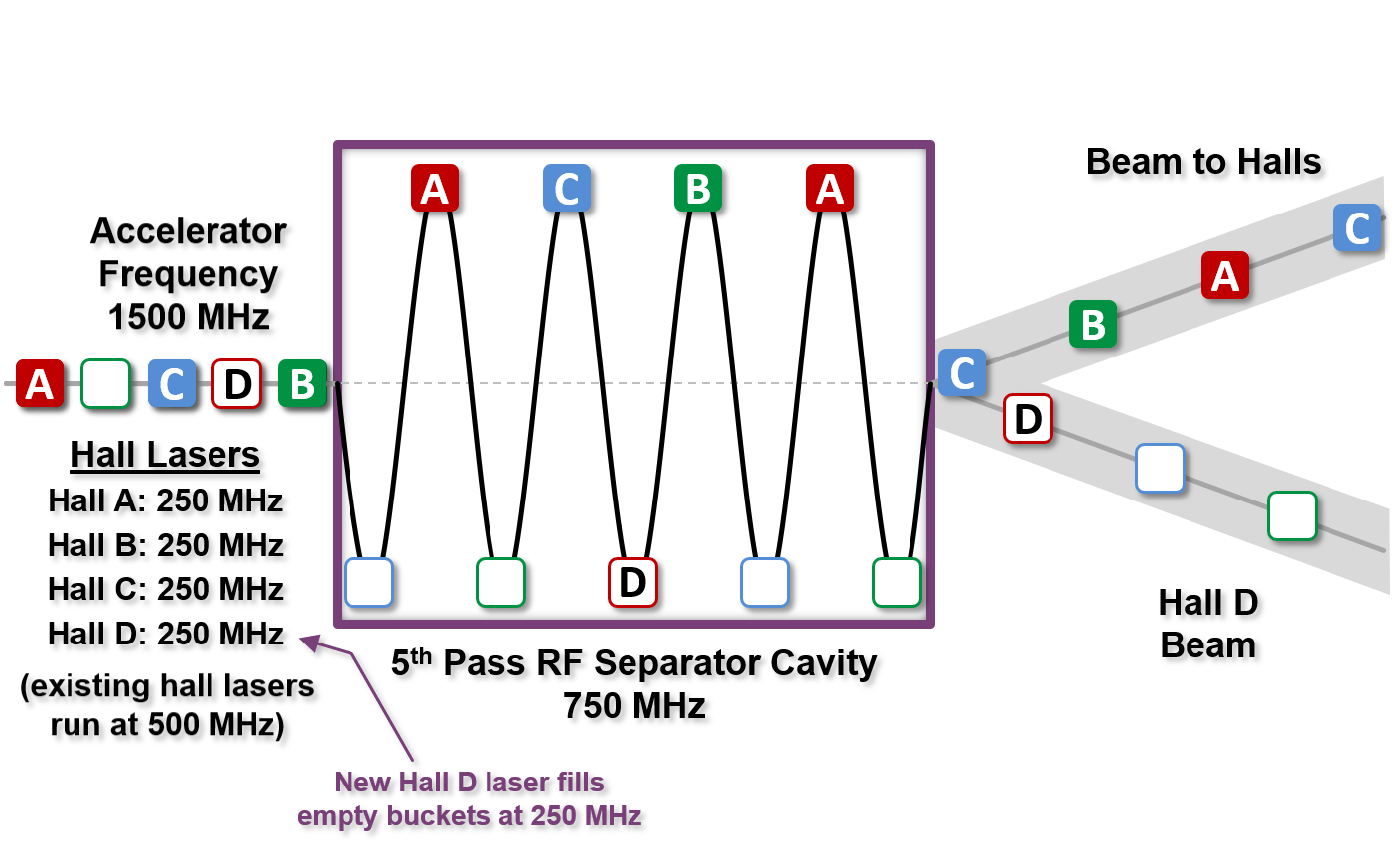}% Here is how to import EPS art
\caption{\label{fig:mpfig6} 
Schematic showing how beams are interleaved, separated and delivered to each experiment hall. (From \cite{SpataRef3})}
\end{figure}

\subsection{Polarized Source and Polarization to Halls}
All polarized beams originate from a single photocathode inside a DC high voltage photogun
biased at 130 kV \cite{MPRefa}.  Successful uninterrupted production of polarized electron
beams requires expertise with GaAs-based photocathodes, high voltage, ultrahigh vacuum, and drive laser technology.
Over more than two decades, a wide variety of technologies and improvements have been implemented to improve beam
quality and decrease downtime.

\subsubsection{Polarized Electron Source}
\begin{figure}[b]
\includegraphics[width=3.4in]{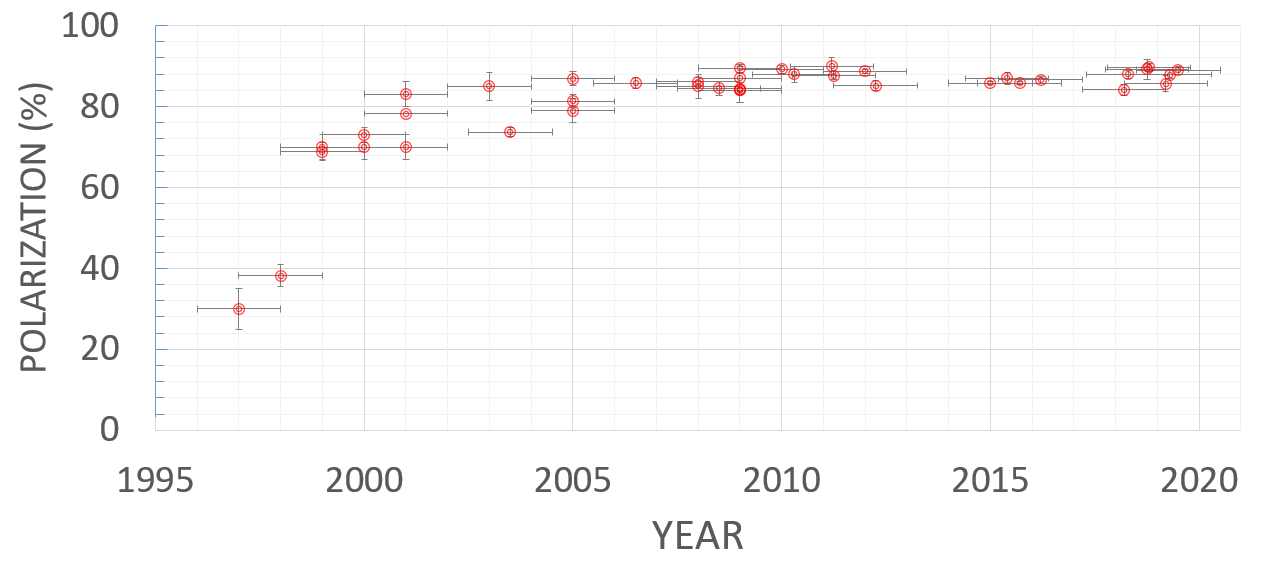}% Here is how to import EPS art
\caption{\label{fig:mpfig1} 
Evolution of beam polarization provided to experiment halls at CEBAF using three different types of photocathodes: bulk GaAs,
strained GaAs/GaAsP and strained superlattice GaAs/GaAsP}
\end{figure}

Bulk GaAs is very inexpensive and provides very high quantum efficiency (QE), but unfortunately polarization
is just 35\% \cite{MPRefb} due to degenerate energy levels in the valence band.  The nuclear
physics program benefits from significantly higher polarization obtained by introducing an axial strain within
the GaAs crystal lattice, accomplished by growing GaAs on a substrate with different lattice constant \cite{MPRefc}.  The evolution of beam polarization at CEBAF is shown in Figure~\ref{fig:mpfig1},
where over the span of 23 years, beam polarization has increased from 35\% to 90\%, with beams produced today using a
strained-layer GaAs/GaAsP superlattice photocathode \cite{MPRefd}.  New photocathodes including
the strained-layer GaAs/GaAsP superlattice photocathode grown atop a distributed Bragg reflector \cite{MPRefe}
and strained-layer GaAs/GaAsP superlattice photocathode manufactured using a nano-pillar array \cite{MPReff}\cite{MPRefg} promise high polarization but with significantly higher QE that is needed for proposed high
current applications, such as polarized positron generation \cite{MPRefh}.

One of the biggest obstacles to successful polarized beam production is field emission and high voltage breakdown
within the DC high voltage photogun.  Field emission at even picoAmpere levels \cite{MPRefi} can
degrade photogun operating lifetime.  A photogun with inverted-insulator geometry provides the electron beams
at 12 GeV CEBAF \cite{MPRefa}. With this design, there is no exposed high voltage because the ceramic insulator
extends into the vacuum chamber, and serves as the support structure for the cathode electrode in which the
photocathode is housed.  This design helps to minimize the amount of metal biased at high voltage, which in
turn helps to minimize field emission.  Another innovation employed at Jefferson Lab relates to electrode
polishing.  To prevent field emission and breakdown, electrodes must possess an extremely smooth surface
free of embedded contamination.  Electrodes are no longer polished by hand using diamond paste, which is a
very laborious and time-consuming process.  Now, electrodes are barrel polished, with a smooth surface achieved
in only hours \cite{MPRefj}.

The operating lifetime of the photocathode is limited by ion bombardment, the process whereby residual gas
becomes ionized by the extracted electron beam, with ions attracted to the negatively biased photocathode. 
Ions that bombard the photocathode can sputter away the thin layer of chemicals applied to the surface used
to create the required negative electron affinity condition, or they can become implanted within the
photocathode material reducing the electron diffusion length \cite{MPRefk}.
The best way to minimize
ion bombardment is to operate the photogun under the best vacuum conditions possible.  At 12 GeV CEBAF this
is accomplished using a photogun with load-lock design, where photocathode heating and activation steps are
performed outside the photogun high voltage chamber behind a closed valve.  In addition, photogun vacuum chamber
components are pre-baked at 400~$^\circ$C to reduce material outgassing \cite{MPRefl},
and some surfaces are coated with
non-evaporable getter material to provide distributed pumping
\cite{MPRefm}.  These steps (and others) result in
extremely good photogun vacuum, in the low $10^{-12}$ torr range, such that hundreds of coulombs of charge can be
delivered before the photocathode must be heated and reactivated.  
Typically once or twice a year the entire photocathode emission area is activated to support physics running, and always during planned accelerator down periods.
 
The drive lasers used to generate interleaved optical pulse trains are composed of 1560 nm gain-switched,
fiber-coupled telecom diode lasers followed by fiber amplifiers that produce 35 ps optical pulses at
249.5 or 499 MHz repetition rates. This light is then frequency-doubled to produce watts of power at
780 nm \cite{MPRefn}.  Gain-switching is a purely electrical pulse forming technique that
does not depend on the laser optical cavity length.  As a result, the optical pulse trains never lose
phase lock to the accelerator rf frequency. 
Although gain-switched lasers possess unique characteristics such as great simplicity, high stability, and easy tuning of frequency and pulse width, their relatively low pulse contrast tends to produce a low level of DC light which complicates beam delivery to experiment halls when operating at low current. 

\subsubsection{Polarization to Halls}

\begin{figure}[b]
\includegraphics[width=3.4in]{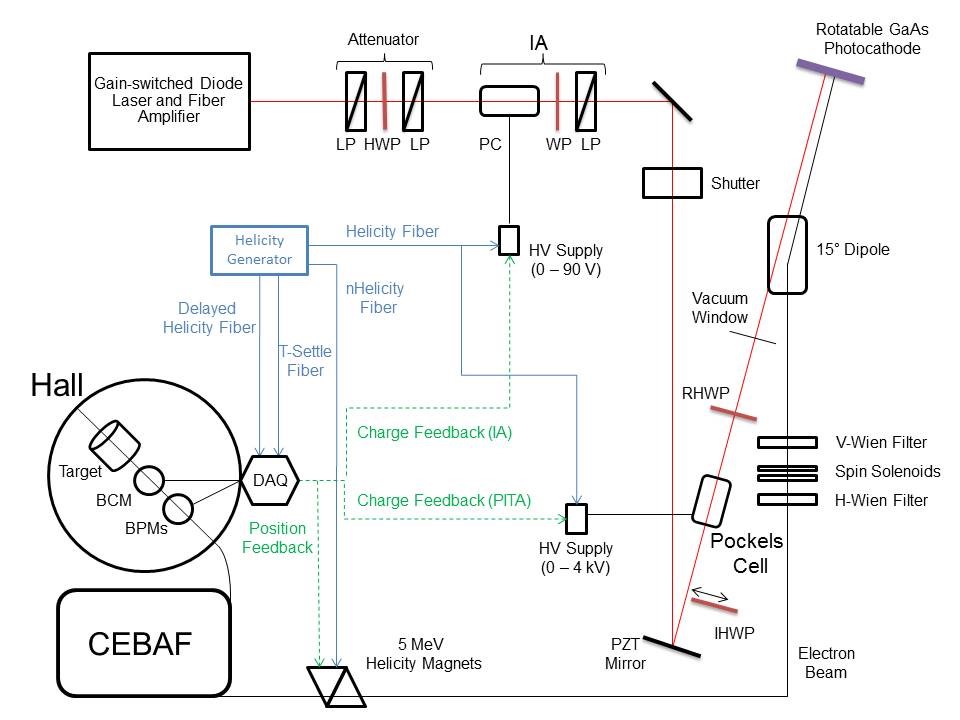}% Here is how to import EPS art
\caption{\label{fig:mpfig2}
Schematic diagram showing the devices used to measure helicity-correlated beam asymmetries and the
accelerator systems used to minimize these asymmetries with feedback algorithms (LP: linear polarizer, HWP:
half-wave plate, PC: Pockels cell, WP: waveplate, RHWP: rotatable halfwave plate, IHWP: insertable halfwave
plate, IA: charge asymmetry controller, V and H Wien: vertical and horizontal Wien filters, BCM: beam
current monitor, BPM: beam position monitor)}
\end{figure}

Parity-violating electron scattering experiments represent one class of physics experiments performed at
12~GeV CEBAF \cite{MPRefo} that place challenging demands on the accelerator.
These experiments study the parity violation phenomenon or they use the phenomenon to explore nuclear structure.
Since the measured scattering asymmetries of parity violation experiments are very small (ppm, ppb),
it is important that beam properties be identical in the two helicity spin states. Minimizing so-called helicity
correlated beam asymmetries \cite{MPRefp} was an important R\&D focus for 12 GeV CEBAF and necessary
for successful completion of new, proposed parity-violating electron scattering experiments \cite{MPRef3}. 
All helicity correlated beam asymmetries originate from the Pockels cell, the electro-optical element used to
create circularly polarized laser light which is required to produce polarized electron beams
from GaAs-based photocathodes. Charge asymmetry, beam position asymmetry, and beam size asymmetry are
the most frequently cited metrics. If the laser light polarization could be made perfectly circular, helicity
correlated beam asymmetries would vanish, but small imperfections in the optical devices on the laser table,
and birefringence of the vacuum window through which the laser light passes en route to the photocathode, result
in some small amount of linear polarization within the laser light.  This residual linearly polarized light combined
with the QE anisotropy of the photocathode \cite{MPRefd},
produces non-zero helicity
correlated beam asymmetries that must be minimized using precise alignment techniques performed at the
photoinjector and feedback algorithms that rely on laser table components depicted in Figure~\ref{fig:mpfig2}. 
In addition, there are methods to flip the polarization of the electron beam to provide a systematic check
on the physics measurement \cite{MPRef9}

For electrons leaving the photocathode, the electron spin direction is parallel (anti-parallel) to the beam
trajectory, but the spin direction rotates in the magnetic field of the arc magnets en route to the experiment
halls.  Since most polarized-beam experiments require longitudinal polarization at the target, a means to
counter this spin precession is required.  At 6 GeV CEBAF, a ``Z''-style spin manipulator \cite{MPRefs} was
first employed - it provided full 4$\pi$ spin rotation capability, but it was composed of many short-focal length
elements and was difficult to use in practice.  The 12 GeV CEBAF photoinjector employs a ``Two Wien'' spin
manipulator \cite{MPRef9}.  It provides full 4$\pi$ manipulation of the spin direction
but is compact and
much easier to use.   For parity violation experiments, it provides a comparatively simple means of introducing
a 180$^\circ$ spin flip that is required to reduce multiple systematic effects that
cannot be directly measured.

\begin{figure}[t]
\includegraphics[width=3.4in]{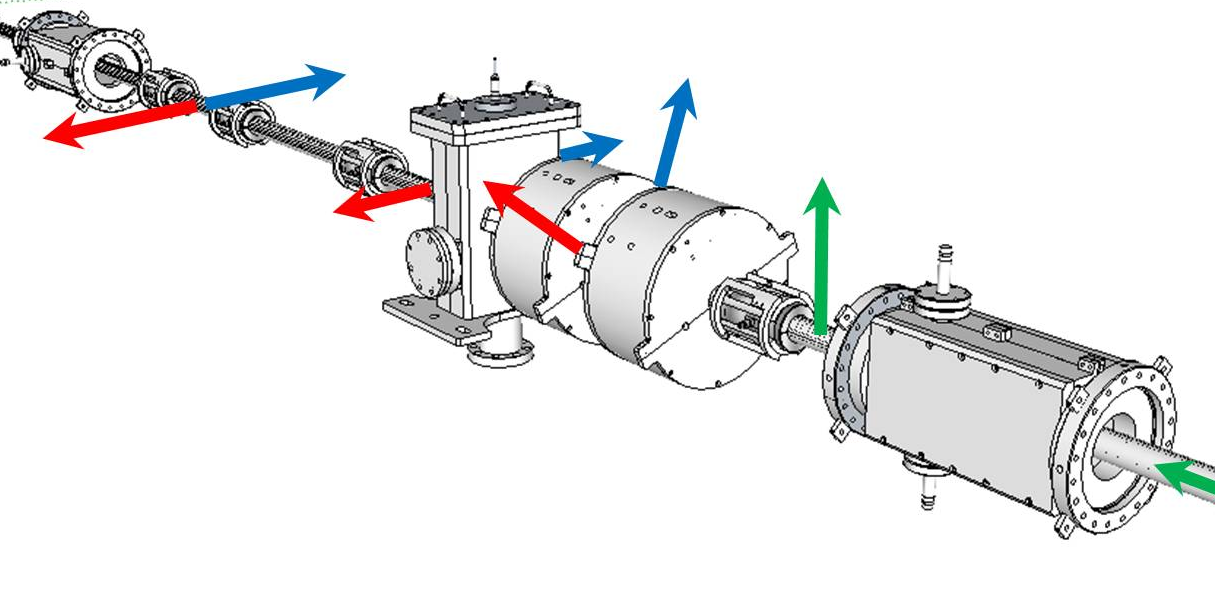}% Here is how to import EPS art
\caption{\label{fig:mpfig3} 
Schematic view of the 4$\pi$ spin manipulator used at 12 GeV CEBAF, composed of two Wien filters and intervening
spin-rotator solenoids (beam traveling left to right). The colored arrows denote the spin direction
after passage through each element.}
\end{figure} 
  
\begin{figure}[t]
\includegraphics[width=3.4in]{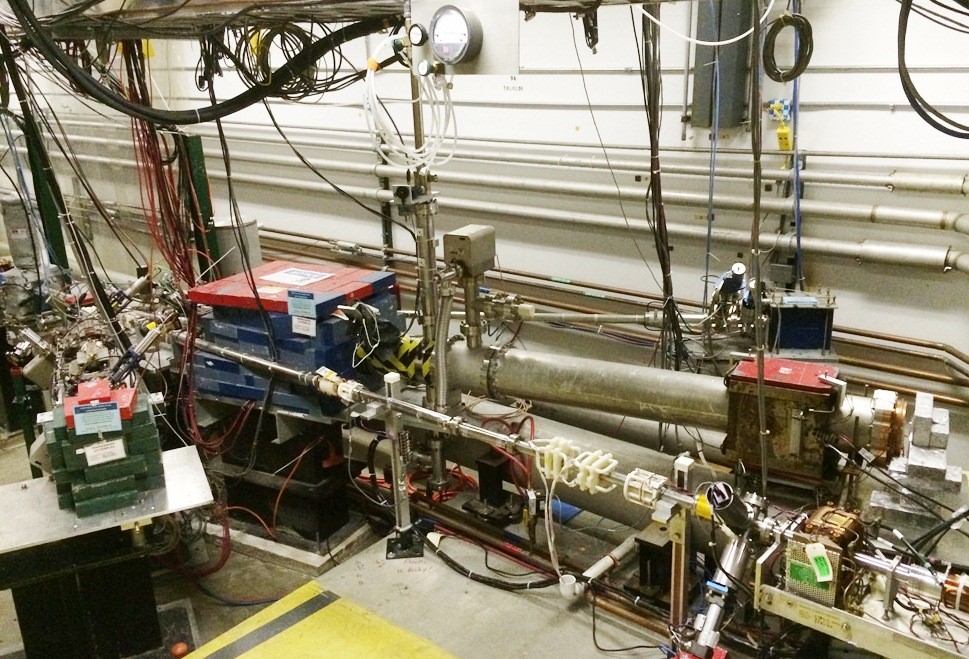}% Here is how to import EPS art
\caption{\label{fig:mpfig4} 
The 5 MeV Mott-scattering electron polarimeter, positioned between the ``straight ahead'' beamline leading to
the injector linac, and a spectrometer beamline.}
\end{figure}

A Mott-scattering polarimeter located in the 5 MeV region of the 12 GeV CEBAF photoinjector is used to measure
beam polarization (Fig.~\ref{fig:mpfig4}) and provides a valuable cross-check of polarization measurements
made at the experiment halls \cite{MPReft}. The polarimeter was recently assigned a level of
precision/accuracy by performing the so-called foil thickness extrapolation, with Mott scattering
asymmetries measured from multiple gold target foils of different foil thickness. Extrapolating
to ``zero thickness'' provides a measurement of single-scattering asymmetry which can be compared
to theoretical predictions. The statistical precision of the polarimeter is less than 0.25\%, with
the measured asymmetry unaffected by $\pm$1 mm shifts in the beam position on the target foil, and by
beam current changes and deadtime effects over a wide range of beam currents.  The overall uncertainty
of a beam polarization measurement at the injector is 0.61\% and is dominated by the uncertainty in the
foil thickness extrapolation as determined from fits to the measured asymmetries versus foil thicknesses;
the estimated systematic effects; and the dominant uncertainty from the calculation of the theoretical
Sherman function \cite{MPRefu}.

What distinguishes CEBAF from almost all other accelerators in terms of beam transport quality is the exacting demand imposed by parity-violation (PV) experiments discussed in the Introduction.  These experiments aim to discern tiny asymmetries in scattering cross sections between opposite spin directions, or helicities, of the electron beam, and are extremely prone to contamination by false signals from other helicity-dependent inputs, such as beam coordinates (offset and angle) entering the detector.  As CEBAF parity-violating experiments probe such asymmetries down to a few ppm’s, false signals must be controlled to well below this level.  Nominally due to phase space conservation, beam coordinate dependence on helicity, originating from the Injector at 335 keV/c (momentum), is damped down by about a factor of 100 when it reaches the detector at 3 GeV/c for PV experiments.  This effect, known as adiabatic damping, is however realized only if the beam is transported exactly as designed over almost four orders of magnitude in momentum.  Small deviation from design transport, such as minor coupling or local near singularity tolerable to other applications, can lead to helicity-dependent contamination overwhelming PV signatures.  At CEBAF, an initiative was launched to ensure global adherence to design transport for PV experiments across the entire momentum range \cite{MPRef2}.  Efforts were focused on three fronts: (1) Ensuring close adherence to design transport using existing diagnostic and control provisions from the Injector exit to the experimental halls, (2) Installing additional diagnostic and control elements as needed to correct for deviation from the 4-dimensional transverse design transport in the Injector complex, with special attention to off-diagonal coupling and on-diagonal near singular transport, both of which can translate into gross magnification of helicity-dependent coordinates not correctable at higher energy, and (3) Using a global signal activated with 30 Hz piezo kickers (PZT) from the Injector to the experimental halls as an end-to-end tuning guide for real time global transport correction.  It is important to ensure the absence of near singular transport at any point along the entire transport path, as it increases beam sensitivity, complicates correction efforts, and can magnify otherwise benign projected emittance growth beyond repair.  Successful execution of this program has eventually resulted in adiabatic damping of helicity-dependent beam coordinates at 3 GeV as expected, and unprecedented precision of CEBAF based PV experiments, with asymmetry determined to better than 100 ppb in some cases.

\subsection{6 GeV CEBAF Performance Summary}
At the end of operations at 6 GeV, the CEBAF accelerator performed as designed. In this section of the paper two specific aspects of the performance of technical systems that were instrumental in achieving desired performance are summarized.

\subsubsection{Performance and Control of SRF Cavities}

When CEBAF was initially designed in 1985 and 1986,
to obtain 4 GeV from four passes required 1 GeV from $50\times8=400$
cavities, or 2.5 MV per cavity. An SRF cavity modified from one
run with high current beam at Cornell University's CESR collider \cite{sundelin} \cite{PKN1}
was adopted very early in the CEBAF project \cite{CRReview}. Operating at 1497~MHz and with
five 10 cm long elliptical cells, the initial accelerating gradient
requirement for the CEBAF accelerating cavities was 5 MV/m, on average.
Similarly, in order to fall comfortably within the cooling capacity of a
5~kW (at 2 K) Helium cooling plant, the $Q_0$ requirement for the
cavities was $2.4\times10^9$. At 5 MV/m each cavity would dissipate
5.4 W of dynamic heat. As an eight cavity cryomodule was expected to produce at
least 20 MV, the cryomodules in the originally installed complement are
now known generically as ``C20s''.

\begin{figure}[b]
\includegraphics[width=3.4in]{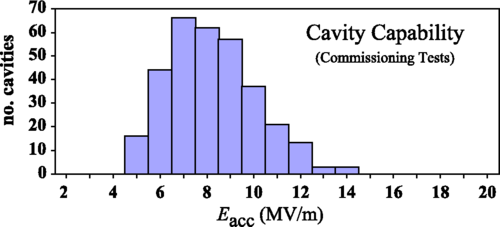}% Here is how to import EPS art
\caption{\label{fig:crfig5} Six GeV CEBAF cavity usable gradient after initial complement of cavities installed. (From \cite{CRRef38})}
\end{figure}  

After the transition of CEBAF to 5-pass recirculation the number of linac
cavities was reduced from 400 to 320. However to still
achieve 4 GeV beam energy, these basic
performance
requirements for individual cavities remained unchanged.
During the course of the CEBAF project
the installed cavities significantly exceeded the basic requirements,
and cavities installed later in the project performed better than
those installed earlier. Fig.~\ref{fig:crfig5} shows a histogram of the installed
gradient capability of the initial complement of CEBAF cavities \cite{CRRef38}. The average
possible gradient was above 7.5 MV/m, exceeding the project goals by
50\%. The average $Q_0$ of the cavities was also much
better than the requirement \cite{CRRef38}. Nominally, operating CEBAF above 6 GeV should have been possible. During the years 2001 to 2003
CEBAF was usually run at between
5.5 and 5.8 GeV maximum beam energy, limited by RF trips of
the cavities. Presently, twenty-seven of
the forty C20 cryomodules originally installed remain in CEBAF.

The maximum beam current to be delivered to the Halls at any one
time was 200 \textmu A, and so the maximum beam load in the
linacs was $200\times4$ passes $=800$ \textmu A in the first project
specifications \cite{Leesum}. To minimize the RF power required
at this load, the cavity $Q_L$ was specified to be 6.6$\times 10^6$,
yielding a cavity 3 db bandwidth of 220~Hz. As the maximum anticipated
beam load was 2~kW, and margins were required to control against
the fluctuations of the cavity resonance frequency away from
the RF source frequency (the so-called microphonics), a saturated
power of 5~kW for the klystrons was specified. The klystron adopted
for CEBAF used a permanent magnet and had a modulating anode.
It had four cells so the gain was a modest 40 dB.
Cathode voltage and current were typically 11 kV and 1A. 

During this early period the RF controls for CEBAF used analog feedback mated with an
x86 Intel processor \cite{sim91}. As usual at the time, the Low Level RF (LLRF) controls were
all analog with primitive remote controls. The controls allowed operators to change system gains,
observe signals (cavity field, forward and reflected power, etc.) and had built in health checks.
The design used separate phase and amplitude controls, typical for controlling normal conducting (NC) cavities.  

\begin{figure}[t]
\includegraphics[width=3.4in]{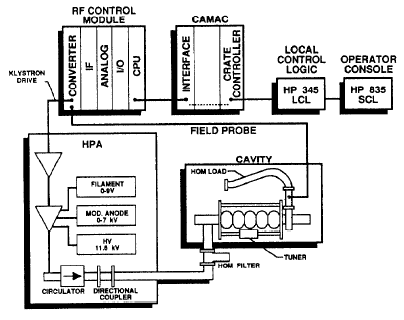}% Here is how to import EPS art
\caption{\label{fig:chfig1} 6 GeV CEBAF RF system showing cavity, klystron, and controls. (From \cite{sim91})}
\end{figure}

The SRF cavities posed new challenges for LLRF control. Unlike NC cavities
where the $Q_L$ is much lower, SRF cavities have greatly enhanced susceptibility to any
detuning of the cavity frequency, e.g. from microphonics or helium cooling
pressure fluctuations. To meet the cavity field control specifications required for the CEBAF LINAC,
high feedback gain LLRF controls were needed to suppress the 
gradient and phase fluctuations produced by detuning. 
Figure~\ref{fig:chfig1} shows a diagram of the CEBAF RF system.  

One of the modernizations at the time for the RF control system was making it a modular system.
The division was made through a crate that supported the different electronic boards.
The crate had the following cards, RF, IF (intermediate frequency), analog, digital and processor board.
The RF board converted the cavity frequency from 1497 MHz down to an intermediate frequency (IF) of 70 MHz.
The IF board performed the signal processing needed to control phase and amplitude of the cavity.
The analog provided separate feedback gain channels for phase and amplitude control, each with
the ability to vary the gain. The digital board had analog to digital converters (ADC), digital
to analog converters (DAC) and TTL digital I/O that was used for component on the RF, IF and analog boards.
The digital board talked directly to the processor board which had a 8186 Intel processor. The RF control system
communicated through the CAMAC crate with the EPICS controls.  Five different control cards made up a C20 LLRF Control Module \cite{sim91}.

Another novel concept at the time was to calibrate the LLRF system in an automated test stand.
All of the RF channels, both receiving and transmitting, were calibrated against a standard reference
or power meter. In addition, each LLRF control module was placed inside an environmental chamber
and cycled to characterize and correct temperature drifts on its RF channels. RF components
were susceptible to both phase and amplitude drifts and the measurements
and resulting calculated corrections
allowed the LLRF system to operate with minimized drifts \cite{HCalProc}.  
The original LLRF control systems have been very reliable with few issues during their operational life. 

\begin{figure}[t]
\includegraphics[width=3.4in]{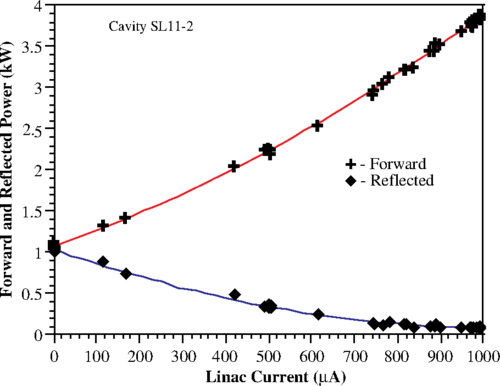}% Here is how to import EPS art
\caption{\label{fig:CRMatched} Forward and reflected power observed in high
power beam test in CEBAF. (From \cite{CRHighPow})}
\end{figure}

In 1997 CEBAF was run up to the 200 \textmu A maximum CW beam current and demonstrated full 5-pass beam power of 800~kW \cite{CRReview,CRHighPow}.
Figure~\ref{fig:CRMatched} shows the forward and reflected power measured as a function
of beam current in one of the cavities in the South Linac
during the test.
The cavity was operated at 6.5 MV/m field gradient accounting for a total 5-pass beam load of 3.5~kW. The measurements demonstrated convincingly several
significant features
of the SRF design: full reflection of the incident RF power
at low beam load, near matched beam load at the highest
operating current and thus appropriate cavity $Q_L$,
and near unity overall RF to beam conversion efficiency
when operating at the matched load.

\subsubsection{C50 Linac Improvement Program}

Starting in 2006, Jefferson Lab initiated a maintenance and upgrade program
for cryomodules. The ten worst performing cryomodules of the original
complement were removed and the cavities were reprocessed. The refurbished cryomodules
had many improvements: dogleg input waveguide couplers and a revised
vacuum window were added which largely eliminated previous
gradient limitations due to window charging and arc discharges \cite{CRRef36},
the $Q_L$ of the input coupler was raised 20\% to $8\times10^6$
supporting operating
at higher accelerating gradient, the mechanical tuners were improved,
and any damaged or worn components in the cryomodule were replaced.
After the cavities were refurbished, an operating goal of 12.5~MV/m
at an increased $Q_0$ of $6.8 \times10^9$ was
established, so an overall energy gain of 50 MV per cryomodule was
indicated. Naturally, the cryomodules came to be called ``C50s''. 
After installing the eighth C50 in the spring of 2009 CEBAF could
be operated at 6 GeV \cite{CRRef37}. The two additional modules served to increase
reliability and reduce RF trip rates during 6 GeV operations.
Following the end of the official 10-cryomodule C50 program,
an eleventh C50 module was installed in CEBAF in
2013 and a twelfth C50 module was installed in 2019.
Only minor changes to the RF drive hardware were made in this
program, so at the highest beam loads the voltage delivered
by the C50 cavities had to be lowered.

\subsubsection{Beam Performance}

In order to achieve the best beam performance in a CEBAF-like
machine, several conditions had to be achieved. As described
in the previous section, a beam
of appropriate transverse and longitudinal phase space had to
be created and injected into the recirculated linacs. Once the
bunches were in the linacs, in order to minimize the energy spread
of the accelerated bunches, the phase of RF had to be chosen so that
the bunches are on the crest of the accelerating wave
on the first pass and all subsequent passes. In practice these conditions
were achieved in a two-step process where firstly the phases of
the RF cavities are chosen to maximize the energy after the
linacs through a spectrometer measurement, and secondly the pathlengths of the individual beam passes are adjusted so that the
higher pass bunches have the same average RF phase as the first-pass
bunch. Spectrometer measurements allow the phase to be set
to crest phase to less than a degree (1.9 ps).

The total time for one beam recirculation is 4.2 \textmu s. As described in Section~\ref{PLControl}, by establishing a 4 
\textmu s beam macropulse, the phase of the individual higher beam passes were measured using a longitudinal pickup cavity \cite{PathLength1} and the pathlength adjusted so that the pass to pass phase difference was under 0.1$^\circ$ (200 fs). By energy modulating the beam, the same type of measurement was used to verify that the individual arcs were isochronous~\cite{arcteam}.

CEBAF had quadrupoles installed in the warm regions between each cryomodule. Usually, CEBAF was run as a FODO system so that the phase advance through the linacs is constant on the first beam pass through the linac. Therefore, on higher beam passes the linac beam optics tended to be dominated by free drift optics modulated by the periodicity generated by the linac quadrupoles. Ideally, the beam optical dispersion and its derivative vanished when the beam traverses each beam pass through the linac and at the entrance and exit of each recirculation arc. As discussed in Section~\ref{Optcontrol}, standard beam optics sets were downloaded into CEBAF, with focusing quadrupoles set proportionally to the beam energy delivered to the experimental Hall receiving the highest beam energy.

In CEBAF the beam optics was verified through a series of toggling measurements \cite{Lebnim}. Low average power pulsed beam was established where the beam macropulses are 60 Hz power line-synced. For transverse beam optics verifications, aircore magnets were excited with 30~Hz, line-synced
current pulses, transversely kicking the beam in both the horizontal and vertical directions at two locations for each direction. The two output locations were measured at all BPMs downstream of the kickers, and the difference orbit so obtained compared to the machine beam optics model. Deviation from the machine model were found when the measured Courant-Snyder invariant for the difference orbit fails to agree with the expectations from the machine model.

Likewise, the dispersion (and implicitly its derivative) were determined simply by modulating the beam energy via the gradient sets in several superconducting cavities,
also toggled in a line-synced manner. In this case, the dispersion was simply proportional to the measured difference orbit. Specific beam optics correction procedures were applied to ensure the dispersion patterns are correct throughout the accelerator.

Early post-commissioning experience \cite{Lebnim} revealed serious discrepancies between predictions of the optics
model and actual beam displacements in the machine. Major sources of irreproducible behavior in the beam optics were linked to
focusing effects of bending dipoles and several quads exhibiting few percent focusing errors. The effect was
particularly large for vertical dipoles of the spreaders and the recombiners, which were not measured with sufficient
accuracy before installation. Simultaneous fitting of six independent difference orbits by varying the focusing
terms of each dipole in the spreaders and recombiners yielded a unique set of body gradients for all dipole magnets.
A similar process was performed for the horizontal dipoles of nine arcs.
Once understood and corrected, excellent agreement between models and measurements were present throughout 6 GeV running.

\begin{figure}[b]
\includegraphics[width=3.4in]{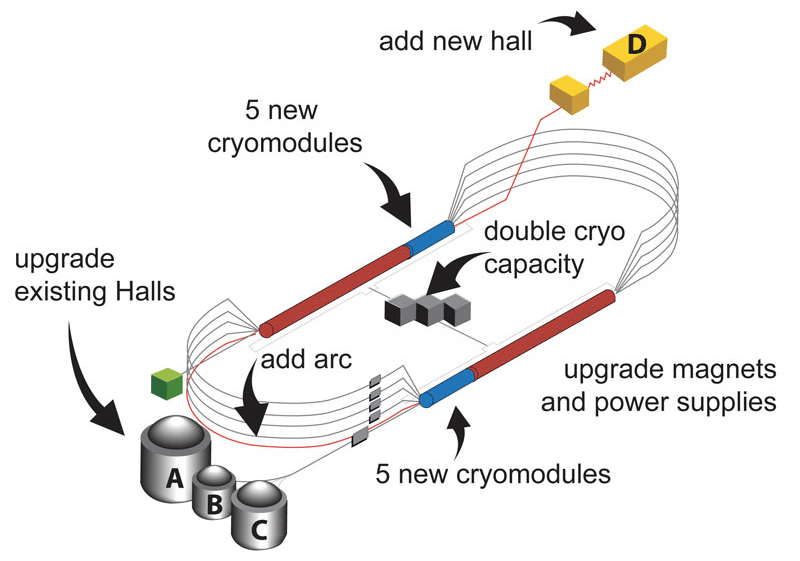}% Here is how to import EPS art
\caption{\label{fig:scm12GeV} 12 GeV Upgrade Project Technical Scope.}
\end{figure} 

\section{Elements of the 12 GeV Upgrade}

The 12 GeV Upgrade project, a major project sponsored by the DOE Office of Nuclear Physics,
substantially expands the research capabilities of CEBAF by doubling the maximum energy and adding
a major new experimental apparatus \cite{LH1}.  The technical scope of
the upgrade project is illustrated in Figure~\ref{fig:scm12GeV}
and includes: doubling
the accelerating voltages of the linacs by adding 10 new high performance cryomodules and
the rf power systems to support these cryomodules;
expanding the 2K cryogenics plant by a factor of two as required;
upgrading the beam transport system from 6 to 12 GeV through extensive re-use and/or modification of existing hardware;
adding one recirculation arc, a new experimental area (Hall D), and the beamline to it; constructing the
major new experimental equipment for this area; and finally, upgrading the experimental equipment in the pre-existing
Halls A-C. This section provides high level descriptions of the changes made;
many details about the accelerator upgrade are found in the numerous references.
The principal parameters for 12 GeV CEBAF are given in Table~\ref{tab:12GeVsum}.

\begin{table}[b]%The best place to locate the table environment is directly after its first reference in text
\caption{\label{tab:12GeVsum}
Principal Parameters for CEBAF in the 12 GeV era \cite{LHFUND1}}
\begin{ruledtabular}
\begin{tabular}{ll}
Energy (Hall D)&12 GeV\\
Energy (Halls A, B, and C)&11 GeV\\
Average Current (Halls A and C)&1-90 \textmu A\\
Average Current (Hall B)&1-100 nA\\
Average Current (Hall D)&0.1-5 \textmu A\\
Bunch Charge&$<$ 0.5 pC\\
Repetition Rate&249.5 MHz/hall\\
Beam Polarization& 90\% \\
Beam size (\it{rms} transverse) & $\sim$150 \textmu m \\
Bunch length (\it{rms})& 300 fs, 90 \textmu m \\
Energy Spread & 2 $\times$ 10$^{-4}$ \\
Beam Power & $<$ 1 MW \\
Beam Loss & $<$ 1 \textmu A \\
Number of Passes & 5.5 \\
Number of Accelerating Cavities & 418 \\
Fundamental Mode Frequency & 1497 MHz \\
Amplitude Control & 1 $\times$ 10$^{-4}$ \\
Phase Control & 0.1$^{\circ}$ \it{rms} \\
Cavity Operating Temperature & 2.1 K \\
Liquifier 2 K Cooling Power & 10 kW \\
Liquifier Operating Power & 10 MW \\
\end{tabular}
\end{ruledtabular}
\end{table}

\begin{figure}[t]
\includegraphics[width=3.4in]{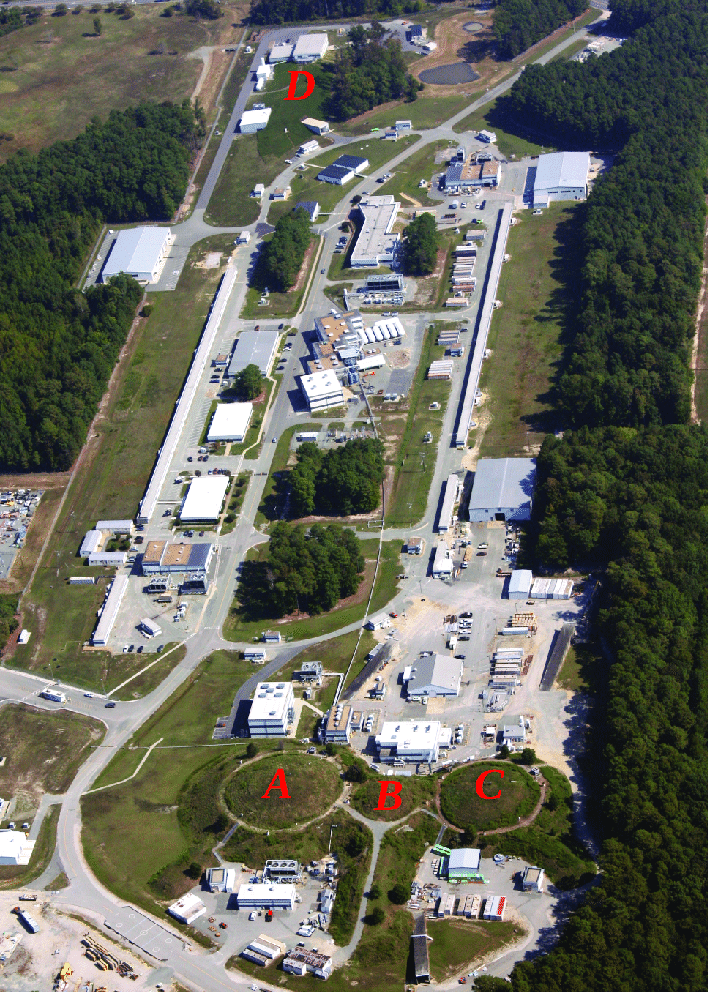}% Here is how to import EPS art
\caption{\label{fig:air12GeV} Aerial view of CEBAF with Experiment Halls Labelled in Red Text.}
\end{figure}

A recent aerial view of CEBAF is given in Figure~\ref{fig:air12GeV};
the camera is held facing in a roughly easterly direction.
The service buildings for the two side-by-side linacs are
readily visible, as well as the shielding mounds for the
original experimental Halls A-C in the
lower portion of the photograph. The newer
Hall D is located in the upper center
part of the photo. The linacs lie roughly perpendicular
to the earth longitude line at the CEBAF site. Therefore, the
linac left-located in the photo is known as the North Linac and
the right-located linac is called the South Linac. The recirculation
arcs for the accelerator lie under the half-circle roads aligned with
the linacs. Analogously, the arcs located on the far side of the
photo are called
East Arcs and arcs located on the near side are called
West Arcs. As in the 6 GeV era, all splitting of the beams for delivery to all the Halls occurs
at the downstream end of the South Linac. The new splitting procedure
allowing all four Halls to be fed beam simultaneously is described in
Section~\ref{fourhall}.

\subsection{New SRF}

\begin{table}[b]%The best place to locate the table environment is directly after its first reference in text
\caption{\label{tab:table1}%
Requirements for the eight cavity cryomodules in the CEBAF 12 GeV
Upgrade Project. Design Requirements for the individual SRF cavities in a
C100 cryomodule.
}
\begin{ruledtabular}
\begin{tabular}{lcc}
Quantity&Value&Units\\
\colrule
Cryomodule Requirements&&\\
Number of Cryomodules&11&\\
Cryomodule Length&10.4&m\\
Number of SRF Cavities & 8 & \\
Average Cavity Gradient& 19.3 & MV/m \\
Energy Gain & 108 & MeV \\
Static Heat (@ 2 K)& 30 & W\\
Dynamic Heat (@ 2 K) & 250 & W \\
\colrule
Cavity Requirements&&\\
Frequency&1497&MHz\\
Cells/cavity&7&\\
Length & 0.7 & m \\
Energy Gain & 13.5 & MeV \\
Average Accelerating Gradient & 19.3& MV/m \\
Average $Q_0$& 9$\times$10$^9$ & \\
$Q_L$ & 3$\times$10$^7$ & \\
3 dB Bandwidth&50&Hz\\
Cavity Impedance ($R/Q$) & 670 &$\Omega$ \\
Geometric Factor&281&$\Omega$ \\
Matched Current & 465 & \textmu A \\
RF Power at Matched Load&6.4&kW\\
RF Power & $<$ 10& kW/cavity \\
Lorentz Detuning& 2 & Hz/(MV/m)$^2$ \\
\end{tabular}
\end{ruledtabular}
\end{table}

Reference~\cite{CRReview} and the references therein
thoroughly document the development of SRF accelerator systems supporting the operation of CEBAF,
and changes to the cavities, cryomodules, and supporting superstructure that have been made.
In this section focus is applied to those developments in the accelerating cavities and cryomodules
that were needed to prepare for and build the 12 GeV upgrade project. This section is an edited and condensed
version of Section XVI of Ref.~\cite{CRReview}, and the Figures in this Section appeared in that publication
and the given references.
In particular, much additional information and additional references on cavity processing that are omitted here, may be found in
Ref.~\cite{CRReview}.

\subsubsection{Developing the Upgrade (C100) Cryomodule}
In the earliest thinking about the 12 GeV Upgrade, the fundamental cryomodule component
was to be new 70 MV cryomodules. Seventeen were to be built: ten deployed in the vacant zones,
six to replace the weakest cryomodules in the 6 GeV CEBAF, and one to upgrade the injector
so it could achieve the required injection energy. The cavities were to be driven by nominal 5~kW klystrons,
but operated at higher voltage and a power of 8~kW. Two prototype cryomodules were built and tested during 2001 and 2002 \cite{CRRef11,CRRef12}.
The second prototype was installed in Jefferson Lab's Free Electron Laser and run at 82 MV total energy gain.
However, lack of complete confidence in this approach, plus an overall cost optimization led to a different solution:
build eleven new higher performance cryomodules to fill the vacant zones and upgrade the injector. 
The eleven new modules were designed to provide over 100 MeV energy gain
and were therefore, in analogy to the previous naming conventions,
dubbed ``C100''. As they were to fill spaces
left vacant in the 6 GeV CEBAF tunnel, the modules
were naturally designed
to be the same length as in the old linac; the installed
warm transitions between the cryomodules could then be identical
to those in the rest of the linacs. A goal of 108 MV per
cryomodule was adopted, but achieving this performance required higher power 13~kW klystrons \cite{CRRef1,CRRef2,CRRef3,CRRef4}. The highest level requirements for the new cryomodules
and SRF cavities for
the 12 GeV Upgrade project are summarized in Table~\ref{tab:table1}.

A new cavity was designed to achieve high total voltage within the pre-existing cryomodule length.
Adopting seven-cell cavities to replace CEBAF/Cornell five-cell cavities maximizes the active length within
the footprint \cite{CRRef5}. The total volume of the helium vessels was significantly reduced by enclosing each
seven cell cavity closely with a vessel. A new tuner was developed.\cite{CRRef6,CRRef7,CRRef8}.
The higher order mode (HOM)
damping scheme was modified to coaxial out-coupling of the HOM power. Wave guide coupling of the
incident power was retained from the 6 GeV era, but the coupler was modified. The modifications
greatly reduced cavity sensitivity to fabrication errors and eliminated a field asymmetry
in the original design, thus
  reducing transverse beam kicks \cite{CRRef9,CRRef10}. Finally, the nominal $Q_{\rm ext}$
for the input coupler was adjusted up from 6.6 $\times 10^6$ to 3 $\times 10^7$
to better match the maximum beam load expected during 12 GeV operations.

\begin{figure}[b]
\includegraphics[width=3.4in]{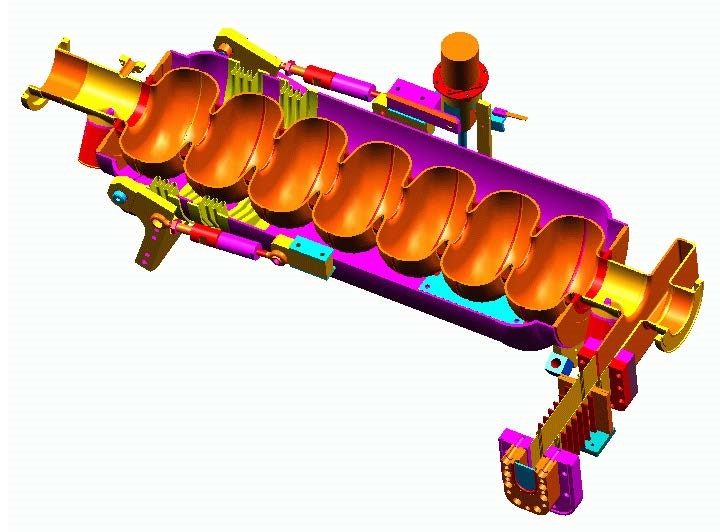}% Here is how to import EPS art
\caption{\label{fig:crfig1} Cross-section of {\it Renascence} cavity, couplers, helium vessel, and tuner. (From \cite{CRReview})}
\end{figure} 

\begin{figure}[t]
\includegraphics[width=3.4in]{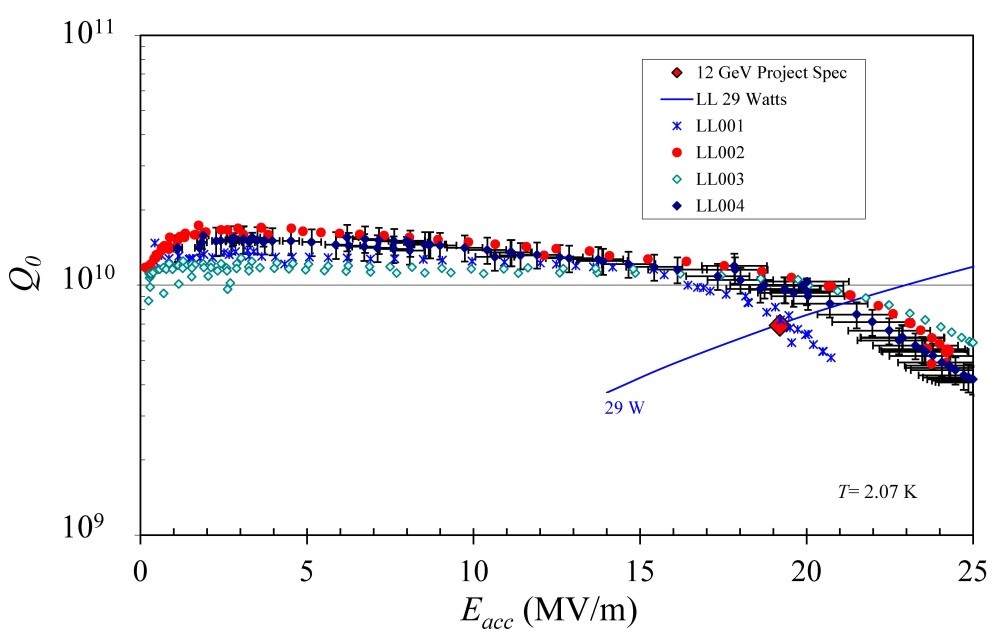}% Here is how to import EPS art
\caption{\label{fig:crfig2} Qualification Tests for the Four Low Loss Cavities in
{\it Renascence}. (From \cite{CRRef31})}
\end{figure} 

\subsubsection{{\it Renascence} Prototype C100 Cryomodule}

A cryomodule prototype project called {\it Renascence} was designed, built, and installed in CEBAF
to prepare for the 12 GeV Upgrade project. The basic requirements for {\it Renascence} were to achieve 108 MV
acceleration with dynamic heat load less than 250 W at 2.1 K, the specifications adopted for the upgrade project \cite{CRRef13,CRRef14}.
A cross-section diagram of a single 7-cell cavity and its attachments is shown in Figure~\ref{fig:crfig1}. In parallel,
two specific cavity shapes were tested in {\it Renascence}. The ``high gradient'' (HG)
Cell shape was optimized to minimize peak surface electric field. The ``low loss'' (LL) design was optimized to
attain the highest accelerating voltage per cooling power \cite{CRRef15}. The cavities achieved 19.2 MV/m accelerating
gradient with less than 29 W heat at 2 K in vertical tests \cite{CRRef16}. Cavity performance curves for the 4 LL
cavities in {\it Renascence} are given in Figure~\ref{fig:crfig2}.

Several technical improvements deployed in {\it Renascence} should be noted. A high thermal conductivity RF feedthrough was developed to be used with the DESY-type HOM couplers \cite{CRRef17,CRRef18,CRRef19}, and this achievement has been incorporated into
the LCLS-II cryomodules. New beamline flange clamps based on ``Radial-Wedge'' geometry
were developed and patented \cite{CRRef20}. A new clean ultra-high vacuum seal based on a ``serpentine gasket'' was developed
for mating the RF input waveguide to the cryomodules \cite{CRRef21}. 
 
After {\it Renascence} was installed in CEBAF, multipass beam-breakup instability was observed \cite{CRRef22,Chao2011}.
The beam current was limited to as low as 40 \textmu A due to a 2.156 GHz transverse deflecting mode
in cavity 5 (HG002) in this cryomodule. Because the HOM damping in {\it Renascence} was expected
to meet the 12 GeV project specifications, the observation was initially surprising. Subsequent
investigation found that, indeed such a HOM in cavity HG002 was not damped to the specified level
because of non-standard conditions as the cavity was fabricated \cite{CRRef23}. As a result, additional
quality assurance steps were added to the project cavity fabrication procedures,
including loaded $Q$ measurements for all relevant HOMs for cavities to be installed in the recirculated linacs.
A novel pole-fitting routine was utilized to quickly analyze HOM data \cite{marhauser3}.

Many other improvements to the C100 were made, both well before and as a result of {\it Renascence}.
For example, there were several significant changes to the cavity assembly procedure \cite{CRRef24}.
The project was built around the LL cavity structure, but to simplify tuning cavity stiffening rings were removed.
A HOM damping scheme with two couplers located more optimally led to a significant reduction in heat losses at sensitive pick-up probes. Stainless steel was substituted for titanium
for the helium vessels to save costs and improve reliability \cite{CRRef25}.

\subsubsection{Producing the C100 Cryomodules and Pre-installation Performance\label{R100ref}}

The 12 GeV Upgrade project formally started in early 2009. Eighty-six C100 cavities were built by
Research Instruments (RI), and delivered by March 2011 \cite{CRRef26}. High RRR ($>$ 250) fine grained niobium sheets
were provided by Tokyo Denkai and used to fabricate the cavities. These cavities were incorporated into the
cryomodules installed in the linacs. In parallel with this activity, and in order to get an early start on
the injector upgrade, eight LL cavities were fabricated in-house at Jefferson Lab to include in the injector
R100 cryomodule. The R100 cavities were fabricated to higher standards to establish high confidence
in the new HOM configuration \cite{r100mar,marhauser2}. By April 2011, the R100 cryomodule was finished \cite{marhauser1}.

\begin{figure}[b]
\includegraphics[width=3.4in]{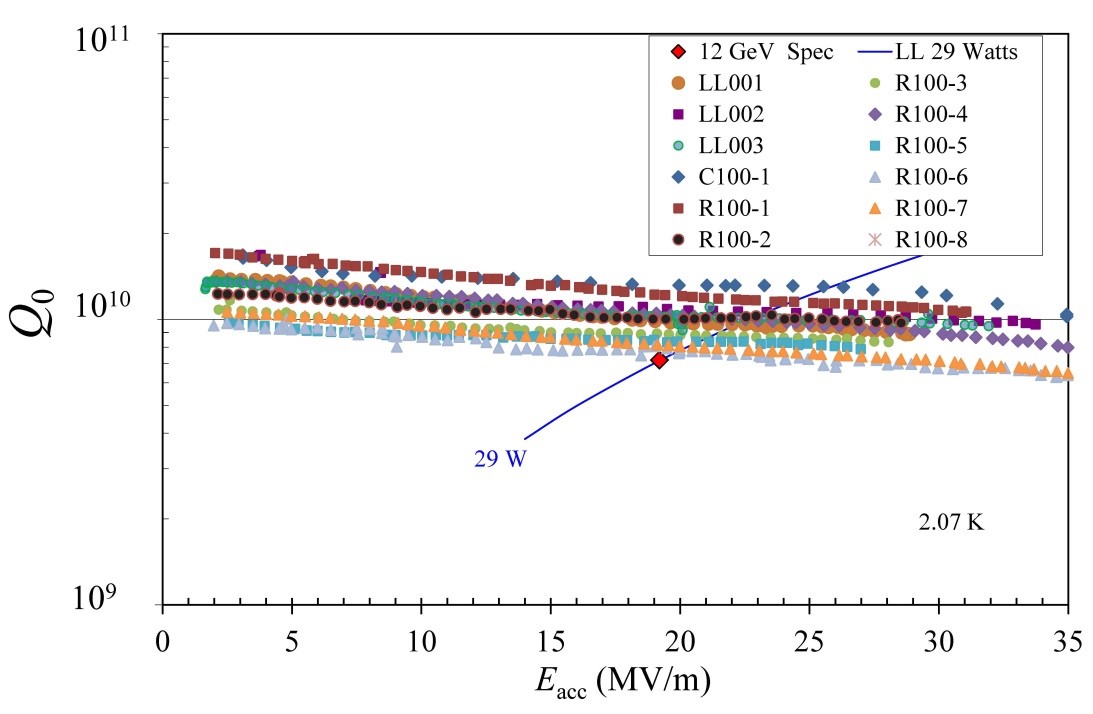}% Here is how to import EPS art
\caption{\label{fig:crfig3} Performance of Twelve CEBAF 7-cell Accelerating Cavities. (From \cite{CRRef31})}
\end{figure}

The new cryomodules benefited from contemporaneous results obtained from the International Linear Collider (ILC)
R\&D program \cite{CRRef28,CRRef29}. In tests using the early cavities, a final surface electropolish was incorporated into the
cavity fabrication procedure \cite{CRRef30,CRRef31}. Figure~\ref{fig:crfig3} summarizes the performance of the first twelve
7-cell, in-house built LL cavities that received a light electropolish as the finishing step, including all of
the R100 cavities. The electropolish helped to guarantee excellent cavity performance for all cavities and the
project adopted a final 30 \textmu m electropolish followed by 24 hour bake at 120 $^\circ$C just prior to cold testing
as a performance risk reduction measure. 

\begin{figure}[b]
\includegraphics[width=3.4in]{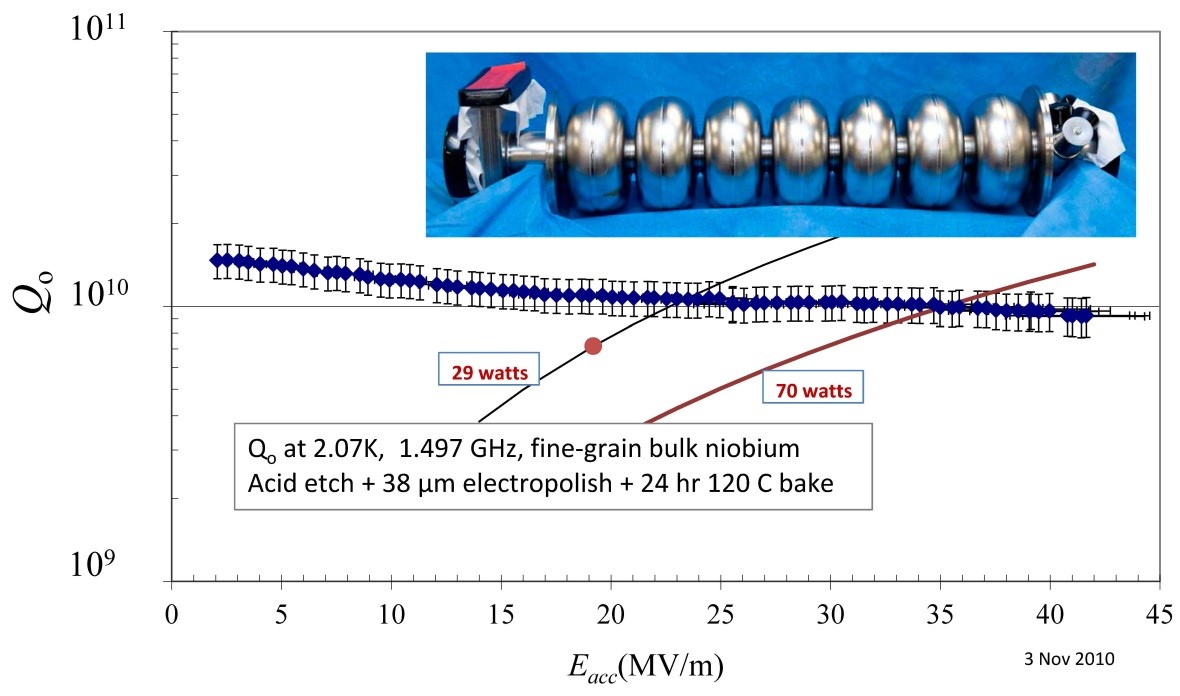}% Here is how to import EPS art
\caption{\label{fig:crfig4} Performance Test of Cavity C100-6 Without a Helium Vessel. (From \cite{CRReview})}
\end{figure}

Having had plenty of process development time prior to the arrival of the production stream of cavities and
the excellent performance of the cavity vendor, the 12 GeV cavity production line ran very smoothly
\cite{CRRef32,CRRef33,CRRef34,CRRef35}.
The cavity performance during VTA testing significantly exceeded requirements such that most of the cavities
were not actually tested to their limits, but were only tested to an administratively constrained 27 MV/m. The
electropolishing process and cavity performance was so stable and reliable that the decision was made for
efficiency to only test the cavities after the helium vessels were welded on. One early production cavity that was
tested to its limits was C100-6. Its excellent performance is illustrated in Figure~\ref{fig:crfig4}. Subsequently,
after the addition of the helium vessel around a cavity, the maximum cooling capacity at the 2.1 K test
temperature was $\sim$ 70~W. 

\subsection{New RF}
 
In order to power and control the cavity fields
in each C100 zone, new klystrons are needed
for this application and a completely
new approach for RF control is necessary.
To support the higher
gradients and higher $Q_L$ of the eighty newly installed cavities
in the ten linac C100 zones,
all have new klystrons, waveguides, RF control
systems, and other associated equipment \cite{CH1,AK1}. Next we
summarize the performance of the newly installed
systems.
 
\subsubsection{C100 RF System} 

\begin{figure}[b]
\includegraphics[width=3.4in]{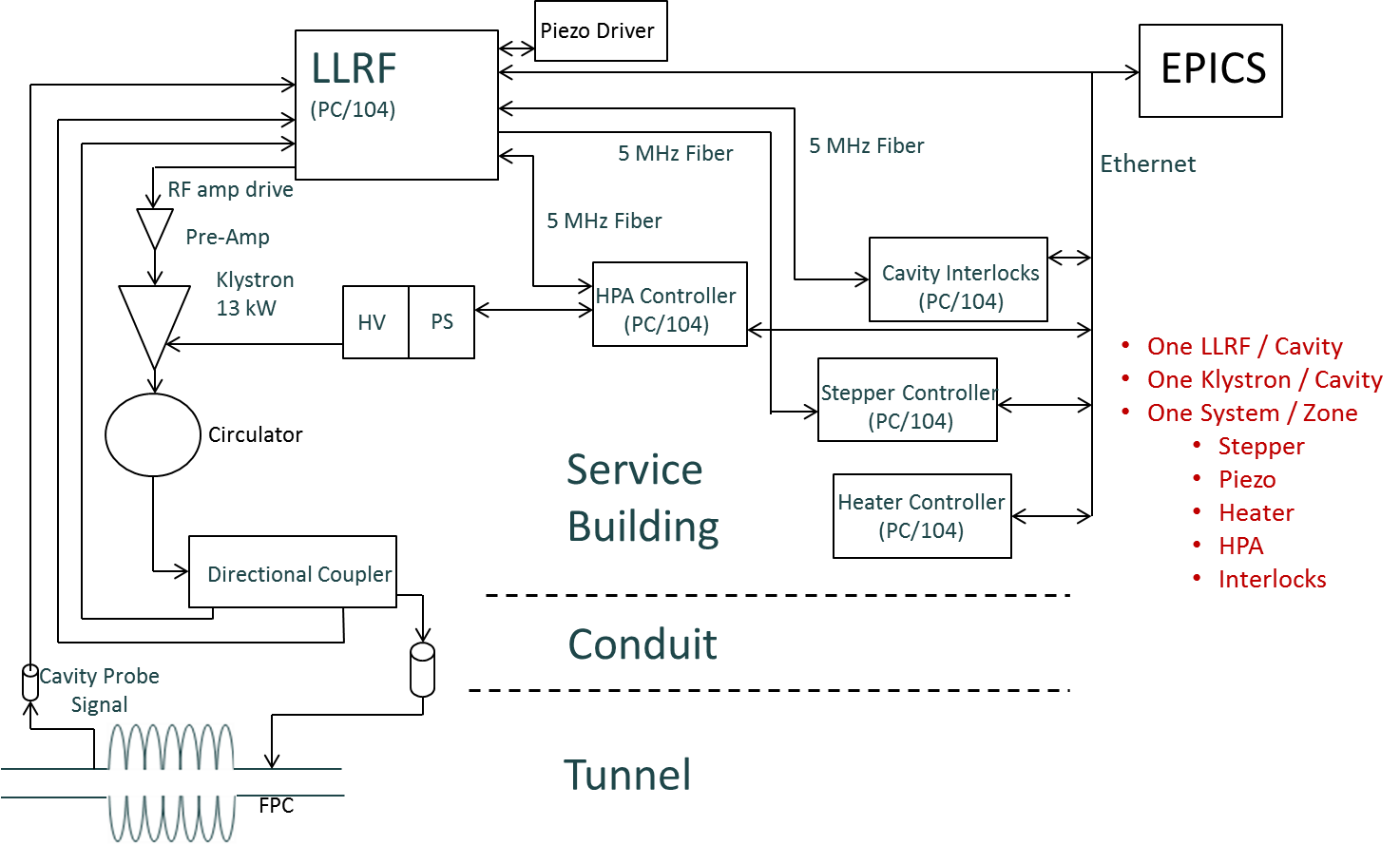}% Here is how to import EPS art
\caption{\label{fig:RFscheme} Schematic diagram of C100 cavity RF System \cite{RB1}.}
\end{figure}

\begin{figure*}
\includegraphics[width=6.8in]{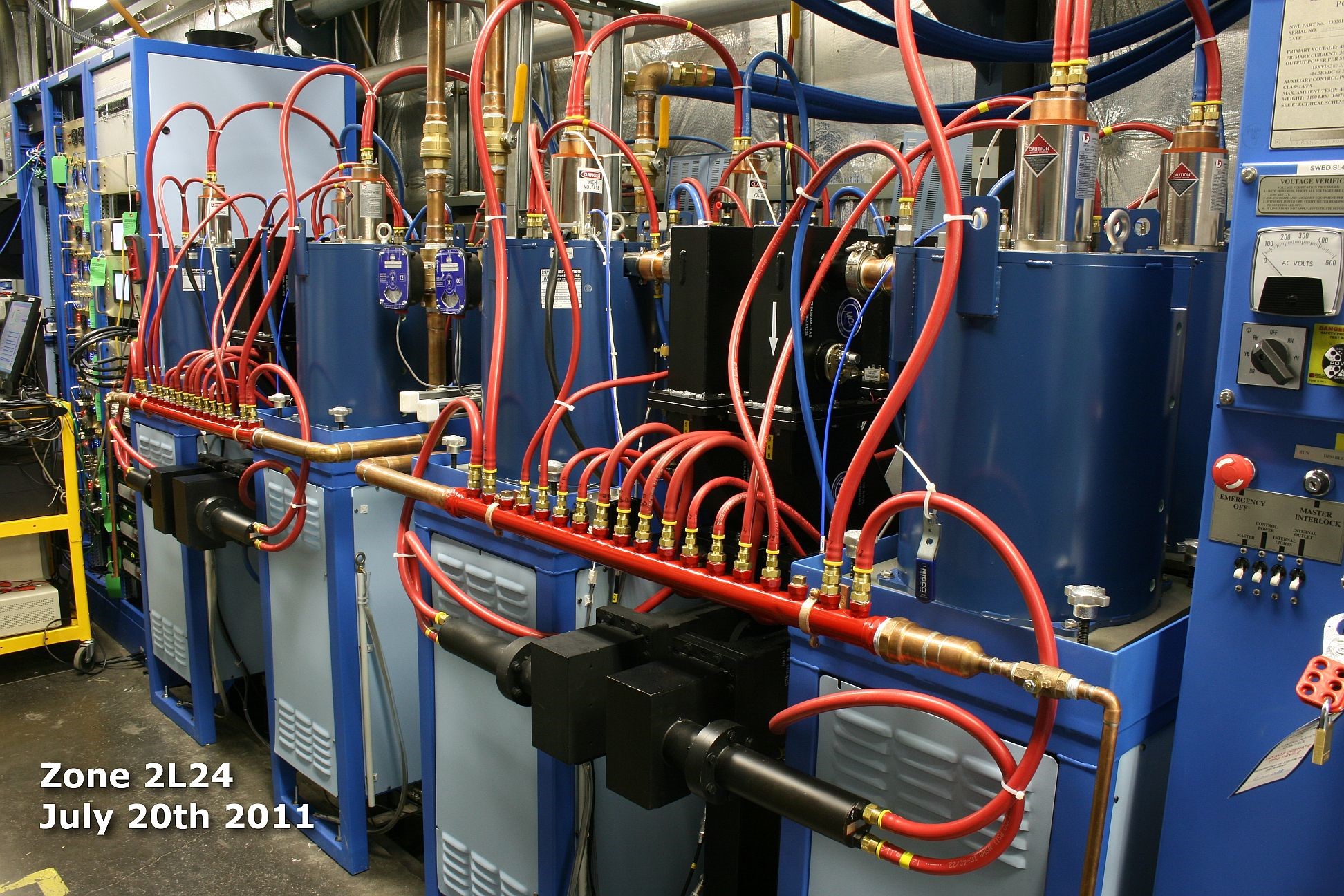}% Here is how to import EPS art
\caption{\label{fig:kly} Eight 13~kW klytrons installed in the South Linac.}
\end{figure*}

The C100 cavities are designed to operate CW at a maximum accelerating gradient of 19.3 MV/m. 
A single klystron powers an individual cavity and its accelerating gradient is controlled by a low level RF (LLRF) system,
as shown schematically in Figure~\ref{fig:RFscheme}.  The upgrade klystrons produce 12~kW of linear power
and up to 13~kW saturated. The water-cooled klystron is a five cavity tube with solenoid focusing,
made by L3 Communications.  The power requirement
includes the power needed to accelerate the beam at the maximum beam load and that needed to compensate for static and microphonic detuning.
An RF zone contains four high voltage power supplies, with each powering two klystrons.
The high power amplifier (HPA) system includes additional power supplies necessary for klystron operation
as well as multiple interlocks for protection of these devices. A photo of an installed zone of new klystrons
is shown in Figure~\ref{fig:kly}.
Klystrons and the RF control hardware reside in the linac service buildings about 7 m above the linacs.

\subsubsection{Controls/Low Level RF} 

The RF controls use a traditional heterodyne scheme and digital down conversion at an intermediate
frequency. The cavity field and resonance control PID algorithms are contained in one large field-programmable gate array (FPGA). The RF controls are unique
in that they incorporate a digital self-excited loop (SEL)
that has been implemented at 12 GeV CEBAF with great success
\cite{JDAllisonPAC,JDAllison}.
Analog SELs had first been used for heavy ion superconducting cavities \cite{JD1}.
Using the SEL allows the cavity to be turned on quickly
no matter how far detuned. Interfaces to the controls and interlocks provided by
both the HPA and LLRF controls are made via EPICS. 
The functions that required five cards
in the older system now reside in a single chassis, with additional capabilities. 

An operational issue, especially relevant for high-gradient low bandwidth
superconducting cavities is the radiation pressure detuning (Lorentz detuning)
that is observed at cavity turn on. The Lorentz detuning is proportional to the square of the cavity
gradient and is determined by cavity stiffness. Typically the detuning is measured during commissioning
and a Lorentz coefficient is assigned to the cavity. For the C100 cavities a Lorentz coefficient of 2 Hz/(MV/m)$^2$
is typical. For 20 MV/m operating gradient, detuning is 800 Hz from RF off to RF on. The
typical method for recovering a Lorentz detuned cavity is to use a piezo actuator and to compensate for
the detuning at turn on. At Jefferson Lab a different approach is taken. A digital SEL that tracks
the cavity up to the operational gradient is employed.
Once on frequency and at the requested gradient, a digital firmware
application then  locks the cavity to the reference. The cavity turn on sequence utilizes both firmware and
EPICS application software.

Cavity faults in the cryomodule present another operational challenge. Mechanical coupling between adjacent
cavities is roughly 10\%. For example if a cavity detunes 800 Hz due to the Lorentz effect when faulted, nearby cavities will
see 80 Hz of detuning, beyond the nominal 50~Hz bandwidth. The klystron does not have the
overhead at higher gradients to compensate for such a detuning. To keep the adjacent cavities at gradient when
a cavity trips off, they are immediately switched into SEL excitation. Once the faulted cavity is cleared and
brought to gradient, all the cavities are returned to external lock using an EPICS application.  

Cavity microphonics are measured continuously by determining the detuning angle from the cavity signal and the
forward power.  Both peak and rms  tune excursions are displayed for each cavity
in EPICS.
 \begin{figure}[b]
\includegraphics[width=3.7in]{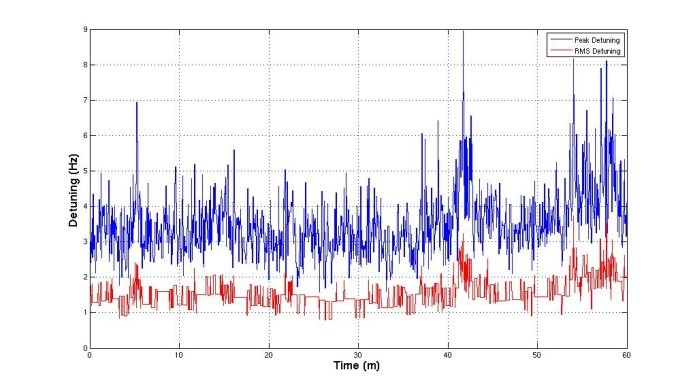}% Here is how to import EPS art
\caption{\label{fig:CHFigure2} Peak microphonic detuning (upper) and rms average detuning (lower) of a cavity.}
\end{figure} 
Figure~\ref{fig:CHFigure2} shows the
detuning in Hz for rms  and peak for a typical cavity.  In addition the cavity field regulation
(phase and amplitude)
is also measured continuously. Any excursion is noted on the EPICS RF screen so an operator can investigate.

A useful feature of digital LLRF control systems is the use of data buffers. The hardware allows
the operator to catch and post-analyze real time data from the cavity-control system. This is extremely
useful when diagnosing cavity faults or measuring microphonics.
 \begin{figure}[b]
\includegraphics[width=3.7in]{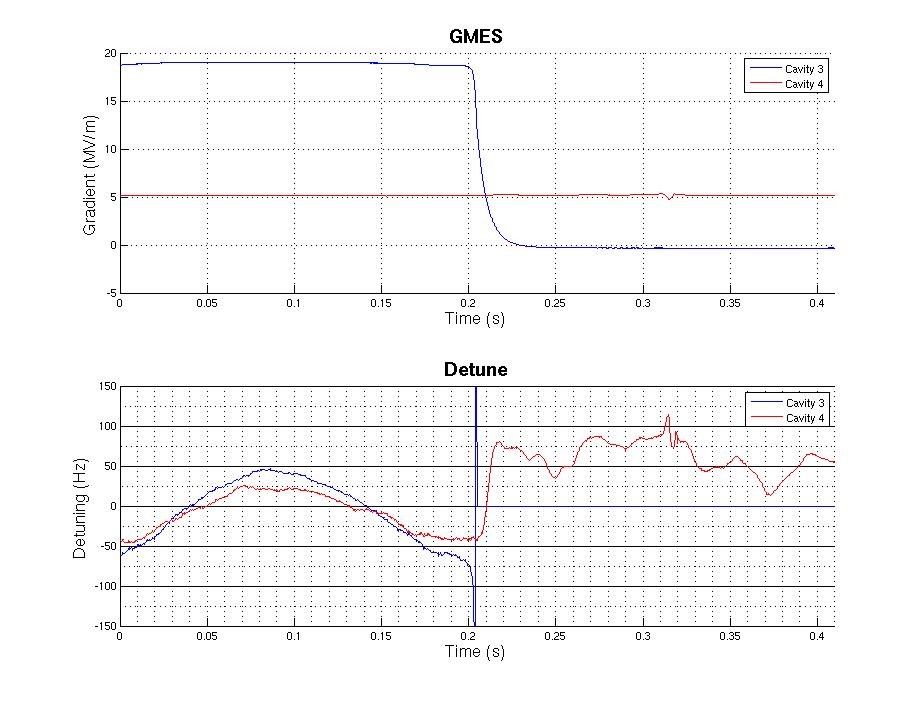}% Here is how to import EPS art
\caption{\label{fig:CHFigure3} Graph of gradient and detuning (Hz) as a cavity is faulting (blue). GMES is the measured cavity gradient as the fault progresses.}
\end{figure} 
Figure~\ref{fig:CHFigure3} is a plot of a cavity fault. The top graph displays the
cavity gradient of the faulted cavity and the adjacent cavity. The bottom graph shows each
cavity's detuning in Hz. The red curve on the bottom graph shows the sharp reaction
(Lorentz contraction from the faulted cavity) of the non-faulted cavity. The adjacent cavity was
operating at a fairly low gradient, 5 MV/m, so the klystron had more than enough overhead to absorb the 77 Hz detuning.   

Cavity frequency control is provided by a mechanical stepper motor. The stepper motor can tune the cavity to ±1 Hz of the reference, which meets the requirements for the RF system (control and power). The tuner is automated to keep the cavities close to the master reference. A piezo tuner was included in the design, and is available for cavity tuning studies \cite{MD1}. Activating them has not been needed for CEBAF operations.

The C100 LLRF systems have been in operation for 7 years,
and our cavity control methods have been adapted by
other newer SRF accelerators including LCLS-II \cite{CHLCLSII1,CHLCLSII2}. 

\subsection{Central Helium Liquifier Upgrade CHL II}

The original Central Helium Liquefier (CHL), now named CHL I, provides up to 4.8 kW refrigeration at 2.1~K for CEBAF’s SRF cavities, 12~kW at 35 K for cryomodule heat intercepts, and an additional 10 g/s liquefaction \cite{Rode1990}. A second refrigeration plant of equal capacity is required to meet the refrigeration requirements of the accelerator at 12 GeV \cite{Arenius2008}. The new refrigerator, CHL~II, has nearly identical capacity as CHL I, except for an increased liquefaction rate to 20 g/s. Each plant is capable of supporting one of the two linacs during 12~GeV operations, or both linacs simultaneously during 6~GeV operations. 

\begin{figure*}
    \centering
    \includegraphics[width=2.3in]{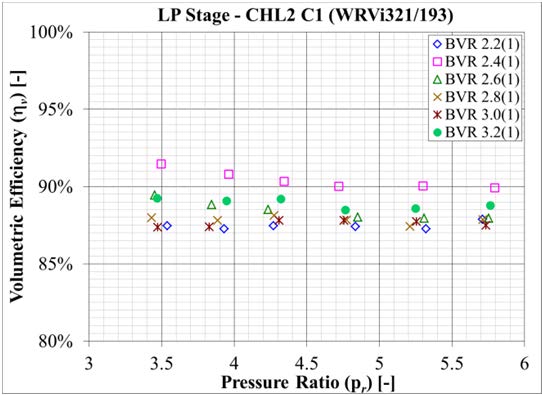}\hfill
    \includegraphics[width=2.3in]{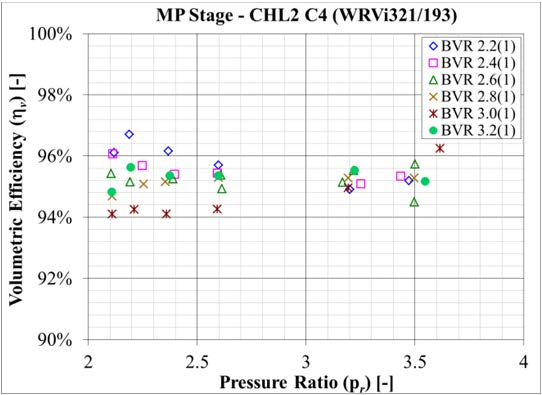}\hfill
    \includegraphics[width=2.3in]{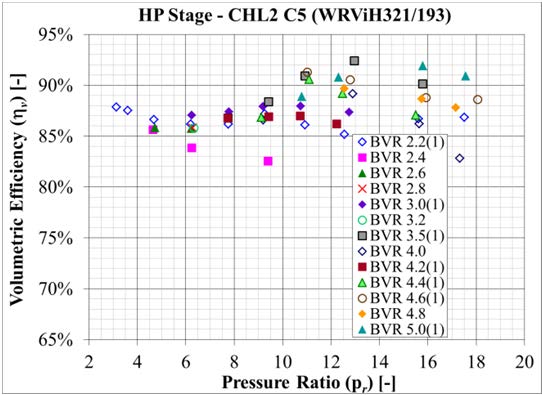}
    \includegraphics[width=2.3in]{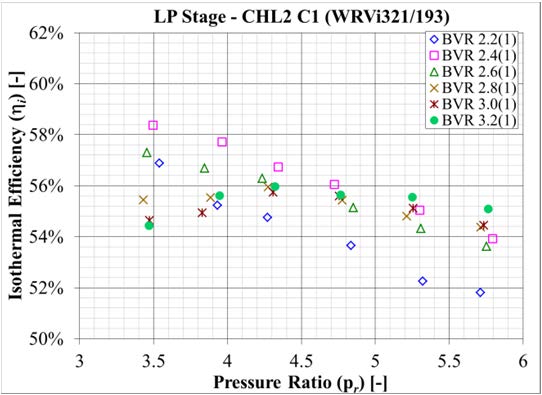}\hfill
    \includegraphics[width=2.3in]{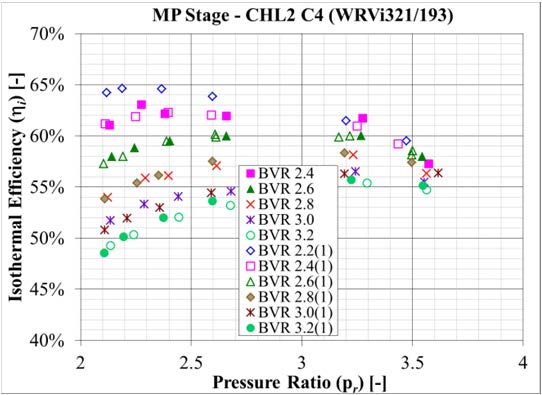}\hfill
    \includegraphics[width=2.3in]{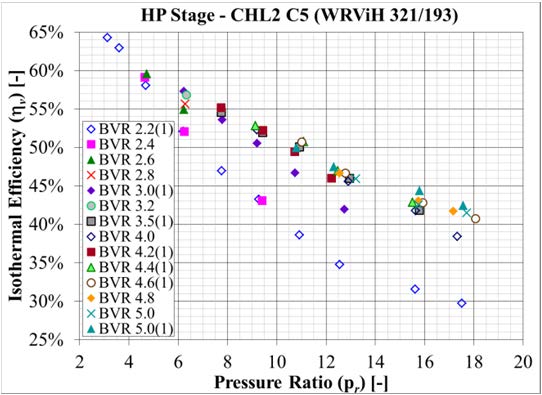}
    \caption{\label{fig:cmpeff} Measured volumetric (top row) and isothermal (bottom row) efficiency as a function of pressure ratio and built-in volume ratio for the CHL II low (left column), medium (center column), and high (right column) stage compressors, demonstrating good efficiency across a wide operating envelope. (From \cite{Knudsen2015})}
\end{figure*}

CHL II fully utilizes Jefferson Lab’s patented Floating Pressure – Ganni Cycle process, a constant pressure ratio process wherein the helium pressure in the refrigeration system naturally varies to compensate for changes in the load while the overall thermodynamic efficiency remains constant \cite{Ganni2010}. The 12 GeV Upgrade scope includes a warm helium compression system and a 4.5 K refrigeration system, which is comprised of two separate cold boxes. A third cold box contains a five-stage cryogenic centrifugal compressor system and 2.1 K subcooler heat exchanger, and produces the subatmospheric conditions required to maintain 2.1 K in the Linac. It was originally constructed as a redundant subatmospheric cold box for CHL I \cite{Ganni2002}. 
The warm helium compression system consists of six oil-flooded screw compressor skids: three 800~horsepower (HP) (597~kW) low pressure stages, one 800 HP medium pressure stage, one 2500~HP (1864~kW) high pressure stage, and one 2500 HP swing compressor. The swing compressor can be configured as a low, medium, or high stage and increases system reliability by taking the place of any one of the other machines during routine maintenance or recovery from an unexpected failure. Several key design requirements, particularly a wide operating pressure range and good efficiency, are addressed by the novel design of the oil management systems on the compressor skid \cite{Knudsen2015}. Figure~\ref{fig:cmpeff} illustrates operational efficiency (isothermal and volumetric) of the CHL~II compressors across a wide range of operating pressures necessary to fully utilize the Floating Pressure Process.

% Option 1 - 3 rows / 2 columns
%\begin{figure*}
%    \centering
%    \includegraphics[width=3.4in]{cmpeff_LPvol.jpg}\hfill
%    \includegraphics[width=3.4in]{cmpeff_top.jpg}
%    \includegraphics[width=3.4in]{cmpeff_MPvol.jpg}\hfill
%    \includegraphics[width=3.4in]{cmpeff_mid.jpg}
%    \includegraphics[width=3.4in]{cmpeff_HPvol.jpg}\hfill
%    \includegraphics[width=3.4in]{cmpeff_bot.jpg}
%    \caption{\label{fig:cmpeff} Measured volumetric (left column) and isothermal (right column) efficiency as a function of pressure ratio and built-in volume ratio for the CHL II low (top row), medium (center row), and high (bottom row) stage compressors, demonstrating good efficiency across a wide operating envelope. (From \cite{Knudsen2015})}
%\end{figure*}

% Option 2 - 2 rows / 3 columns

% Option 3 - Original [NOTE: Change preceding text if this option is preferred... "Figure 24 illustrates operational isothermal efficiency of the CHL II compressors..."]
% \begin{figure}
%\center{\includegraphics[width=3.4in]{cmpeff_top.jpg}% Here is how to import EPS
%
%\includegraphics[width=3.4in]{cmpeff_mid.jpg}
%
%\includegraphics[width=3.4in]{cmpeff_bot.jpg}
%\caption{\label{fig:cmpeff} Measured isothermal efficiency as a function of pressure ratio and built-in volume ratio for the CHL II low (top), medium (middle), and high (bottom) stage compressors, demonstrating good efficiency across a wide operating envelope. (From %\cite{Knudsen2015})}}
%\end{figure}

\begin{figure}[b]
\includegraphics[width=3.4in]{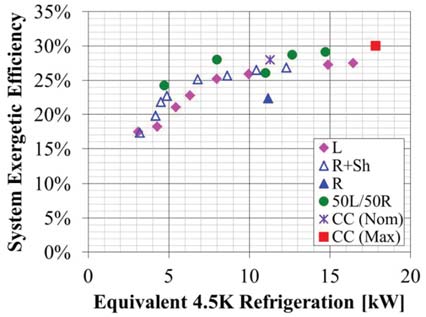}% Here is how to import EPS art
\caption{\label{fig:cbxeff} CHL II cold box performance testing results demonstrate a wide turndown range with remarkably little effect on the overall system efficiency. Test conditions are L (4.5 K liquefaction load) R (4.5 K refrigeration load) and 50L/50R (equal mixed load). Data points for cold compressor (CC) at nominal and max load. (From \cite{Knudsen2017})}
\end{figure}

The first of the two cold boxes, the upper cold box, spans 300 K to 60 K and incorporates several brazed aluminum plate-fin heat exchangers, a liquid nitrogen pre-cooler, and two 80 K purifiers. The other, lower, cold box spans 60 K to 4.5 K and also incorporates several heat exchangers as well as four turbo-expander stages, a 20 K purifier and a 4.5 K subcooler heat exchanger. Efficiency is optimized by designing each expansion stage with an equal temperature ratio, or Carnot step \cite{Knudsen2006}, and compatibility with the Floating Pressure Process is inherent to the design \cite{Knudsen2017}. The CHL II system can be turned down to match load conditions significantly below its maximum capacity, and as shown in Figure~\ref{fig:cbxeff}, exhibits remarkably little loss of efficiency in the process. This turndown is achieved by varying the helium supply pressure from the warm compressors to the cold box between 19.5 and 6.5 bar, without throttling the turbines and with little to no operator intervention \cite{Knudsen2017}. Due to the successful and efficient operation of CHL II, the design has been adopted for the MSU FRIB \cite{Ganni2014,Dixon2015,Knudsen2020} and SLAC LCLS-II \cite{Ravindranath2017} helium refrigeration systems.
 
\subsection{Magnets\label{magnetrefsec}}

The initial CEBAF magnet designs were based on a 4 GeV electron energy
requirement, with an additional goal that the 2,200 magnets in the accelerator
eventually achieve 6 GeV \cite{KB1}. The magnet complement was
measured to support operating the accelerator at 6 GeV beam energy.
In order to operate at 12 GeV
most of the dipoles and quadrupoles in the machine needed to operate beyond
their existing field maps.
In particular, to double the magnetic field required at 12 GeV,
most of the dipoles would become saturated.  Consequently, the bulk material in many magnets needed to be modified
to avoid saturation and all of the magnets are re-measured to magnetic fields
up to the 12 GeV specification. This section will describe the modifications and characterization required
to support a model driven 12 GeV accelerator.

\begin{table}[b]%The best place to locate the table environment is directly after its first reference in text
\caption{\label{tab:magtable}%
Magnets modified and/or remeasured
for 12 GeV beam operations.
}
\begin{ruledtabular}
\begin{tabular}{lcc}
Type&Location&Number\\
\colrule
Dipole&Arcs 1-10&288\\
Dipole&East Spreader&22\\
Dipole&East Recombiner&17\\
Dipole&West Spreader&17\\
Dipole&West Recombiner&17\\
Dipole&Transport Recombiner&17\\
Dipole&Hall Transport&26\\
Quadrupole&Throughout&114\\
Corrector&Throughout&120\\
\end{tabular}
\end{ruledtabular}
\end{table} 

\subsubsection{Dipoles\label{Dipolech}} 

\begin{figure}[b]
\includegraphics[width=3.4in]{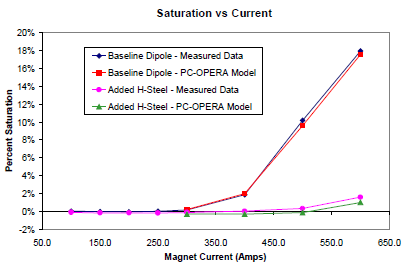}% Here is how to import EPS art
\caption{\label{fig:KBFYYY} Saturation Plot for Converted Dipole. (From \cite{KB1})}
\end{figure}

\begin{figure}[h]\centering
  \includegraphics[width=0.552\columnwidth]{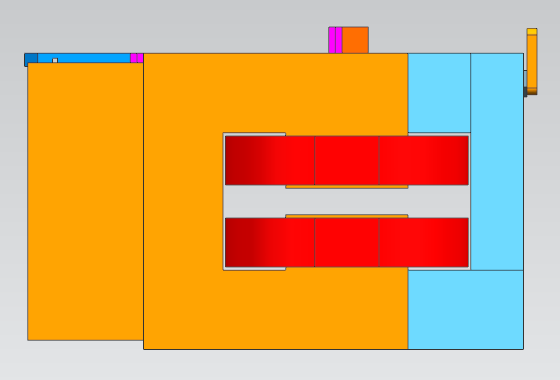}\thinspace\thinspace\includegraphics[width=0.43\columnwidth]{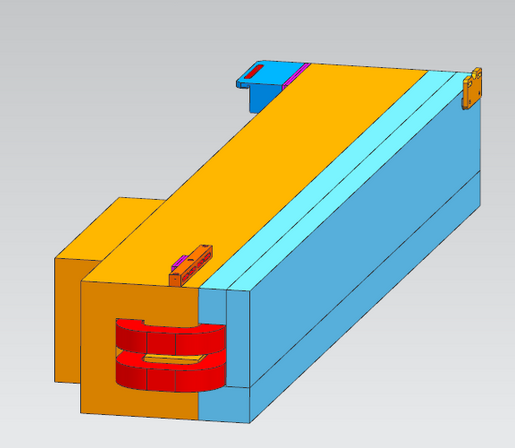}
\caption{\label{fig:KBFXXX1} 6 GeV arc dipole flux return and pole (orange), coil (red), and H-Steel addition
enabling 12 GeV capability (sky blue).}
\end{figure}

Prior to the upgrade project a 2 meter arc dipole magnet was both modeled and measured to understand the saturation
effects resulting from the higher current needed for 12 GeV. The PC-OPERA 2D finite element package was
used for modelling. A plot of the percent saturation is shown in Figure~\ref{fig:KBFYYY} for the measured and
PC-OPERA calculations for an unmodified dipole, and for a dipole with an additional steel return leg.
Due to the good agreement, modelling was then performed to
determine the minimal additional steel needed to reduce saturation effects to acceptable
levels. Such considerations result in an ``H-Steel'' design solution where three additional plates were added to the
existing dipoles to provide sufficient return paths for the magnetic flux generated at 12 GeV operating
currents as seen in Figure~\ref{fig:KBFXXX1}.  The H-Steel plates were fabricated and added to the test dipole so
that magnetic measurements could verify performance.  Figure~\ref{fig:KBFYYY} shows results of the measurements
\cite{KB1}. Additional testing was completed to verify the field quality and thermal integrity of
the magnets at 12~GeV currents. Based on the results of the modelling and tests, the H-Steel
design modification was adopted for the 12 GeV Upgrade.

 
\subsubsection{12 GeV Magnetic Measurement} 

Field integrals and field quality are measured using a combination of stretched wire and
hall probe grids.  Stretched wire measurements provided a simple, fast and accurate
measurement method to use for the dipole mapping.  Because all the arc dipoles had been
mapped at 6 GeV, there was no cause for concern with respect to voids in the steel.
All dipoles are mapped using a single stretched wire method.  Additionally, ~10\% of the dipoles
are mapped using a hall probe grid to ensure measurement integrity and provide detailed mapping information.   

An analysis routine is developed to evaluate the field quality and integrated strength for 12 GeV
dipole magnets mapped with the hall probes.  The analysis calculates results based on the curved
trajectories the beam follows as it moves through the bending dipoles. Data points are analyzed
by a program developed at Jefferson Lab. Field integrals are computed by interpolating between measured
data points to create points on the beam trajectory.  These field values are then integrated
along the curved beam trajectory to calculate the field integrals.  Field quality
is evaluated by comparing the ratio
$B'L/BL$ to the specification where
 $B'L= \int (\partial B/\partial r)dz$
\cite{KB2}. 

\begin{figure}[b]
\includegraphics[width=3.4in]{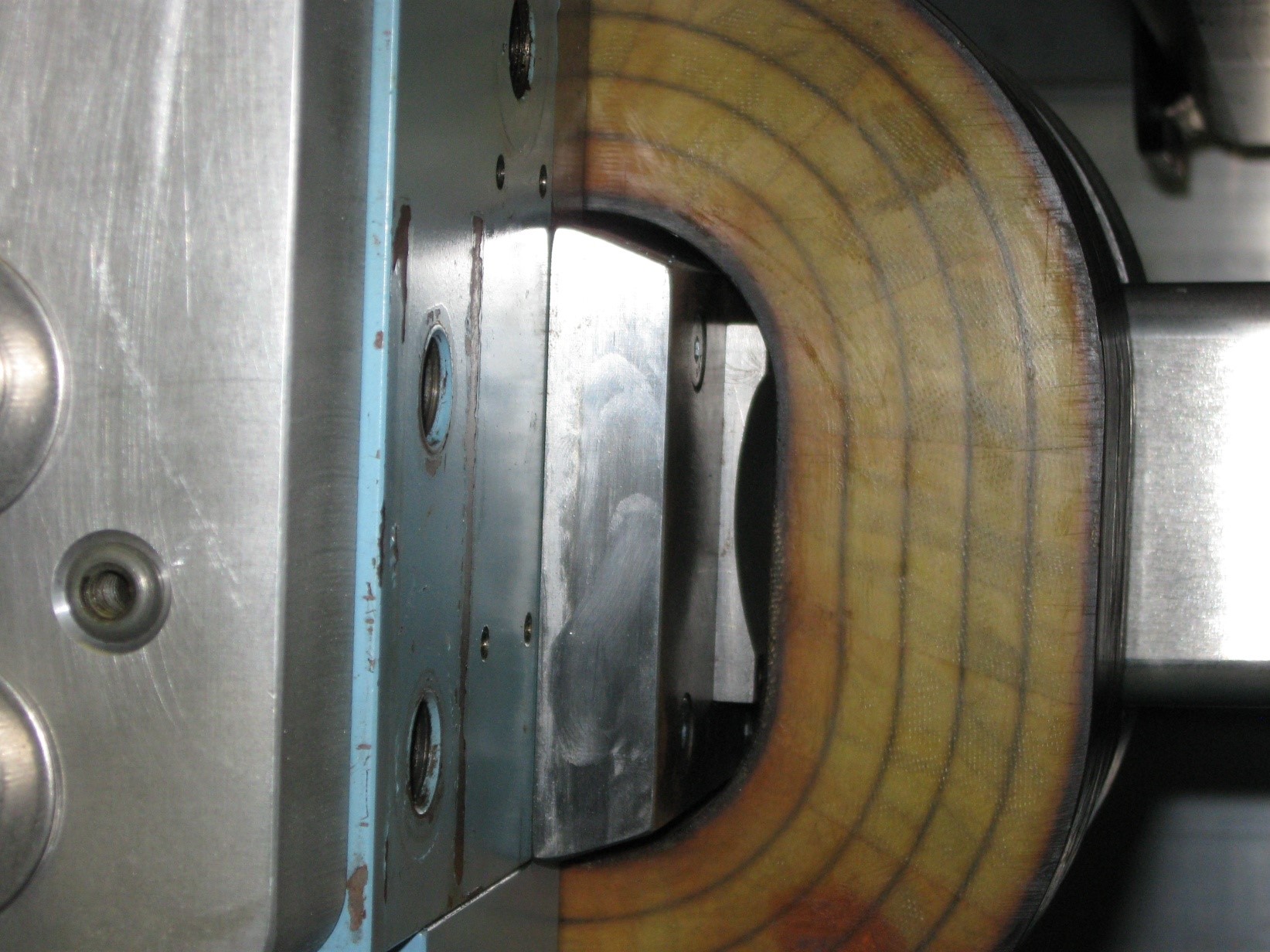}% Here is how to import EPS art
\caption{\label{fig:KBFxyz1} Photograph of Magnet Shim Correcting Field Flatness Installed in a 12 GeV CEBAF Dipole.}
\end{figure}

Measurements identified some integrated strength inconsistencies among arc magnets.  Strength matching
of arc dipoles is required because each arc is powered from a single power supply and dipoles do
not use individual shunts. To solve this issue, field lengthening shims are added to some dipoles
to meet the matching specification.  After correction, gradient measurement results show
acceptable field profiles for all arc magnets and most spreader, recombiner, and extraction magnets. When needed,
field shaping shims are added to correct gradient errors on non-arc dipoles by using a parabolic shim shape to add length
to the off-center field integrals along the horizontal axis as shown in Fig.~\ref{fig:KBFxyz1}.

Measurement values have been compared against model values. Because the detailed breakdown
of the multipole components is not available a method was developed to
compare TOSCA predictions with the measurements from the magnet test stand \cite{KB5}.
Agreement is good when taking into account TOSCA models do not incorporate misalignment or construction errors.

\subsubsection{Quadrupoles} 

The 6 GeV CEBAF experience is used to develop specifications for and to model the
12 GeV quadrupoles. It is found that many existing quadrupoles could
be powered to higher currents to meet the design requirements. Twenty A power
supplies are used in place of pre-existing 10 A supplies in several locations to
increase the focusing strength of those magnets. A second quadrupole and 20 A
power supply is also added at a few locations \cite{KB4}.  
Two new quadrupoles are required for the upgrade. Their designs were based on
existing CEBAF ``QA'' quadrupole designs and required both magnets to fit within the same space
along the beam line and to mount onto existing girders. This requirement eliminates
the need to modify or design new girder parts and assemblies \cite{KB3}. The pole tip designs on these magnets are scaled from the QA quadrupole design.
Pole root saturation and harmonic effects are studied and optimized using Vector Fields
OPERA-2d simulation software.  Each new quadrupole is measured in the Magnet Measurement
Facility.  Rotating coil measurements are used to define magnet strength, multipoles, and
the quadrupole centers.   

\begin{figure}[b]
\includegraphics[width=3.4in]{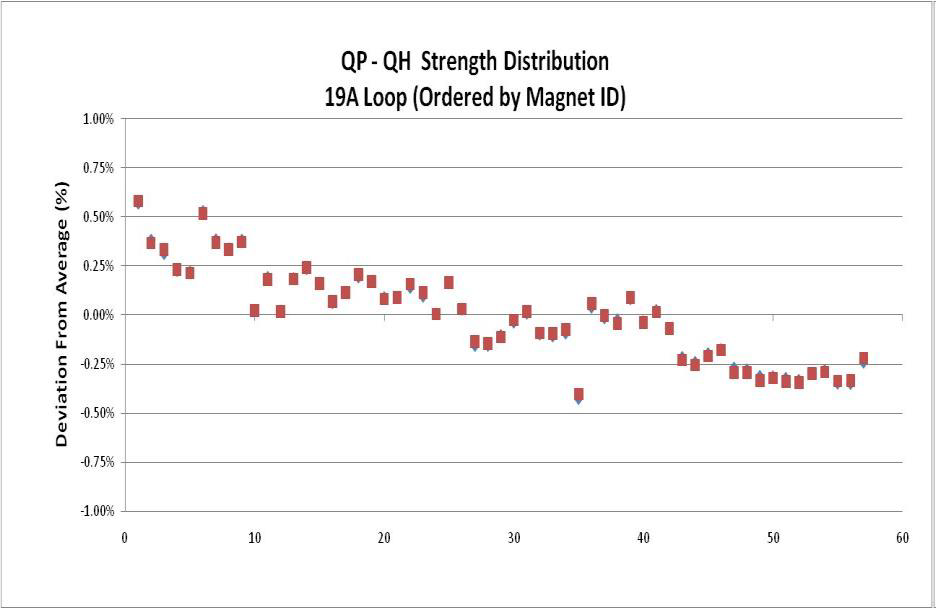}% Here is how to import EPS art
\caption{\label{fig:KBF7} Measured relative magnetic strength deviation from average for QP and QH quadrupoles.}
\end{figure}

\begin{figure}[b]
\includegraphics[width=3.4in]{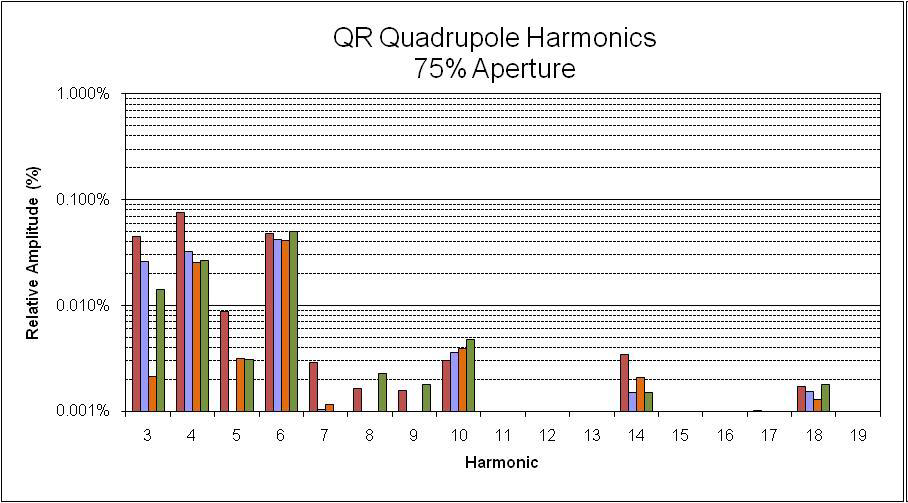}% Here is how to import EPS art
\caption{\label{fig:KBF8} Rotating Coil Field Harmonics Measurement for Four QR Style Quadrupoles at 12 GeV Field Setting}
\end{figure}

\begin{figure}[b]
\includegraphics[width=3.4in]{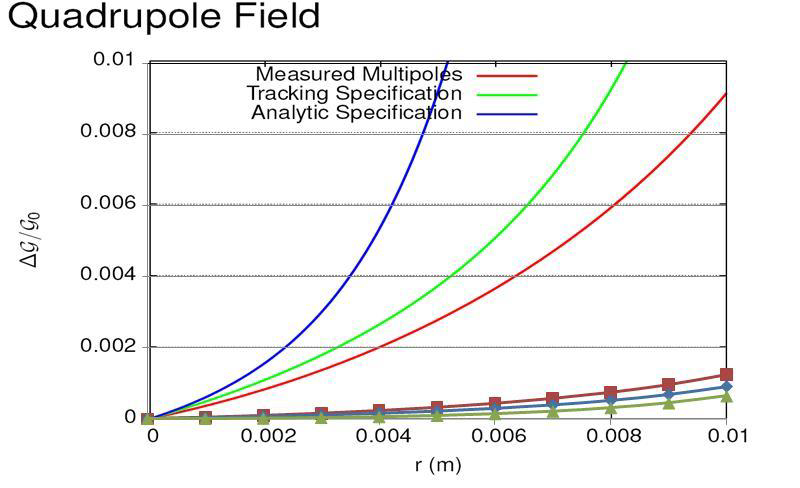}% Here is how to import EPS art
\caption{\label{fig:KBF9} Comparison between beam dynamics requirements (solid curves) and measured performance for a selected quadrupole including several multipoles. }
\end{figure}

Measurement results showed well matched quadrupole performance. Measurements are done on a
rotating probe measurement stand using a MetroLab PDI measurement system along with a
printed circuit board rotating coil.  Magnet strength measurements showed all magnets
within each new family are equal within $\pm$0.5\% as seen in Figure~\ref{fig:KBF7}.  Harmonic content is measured and evaluated
in two ways.  First, individual multipoles are verified to ensure no significant
fabrication errors existed.  To compare with the defined 12 GeV specifications, the
sum of error multipoles relative to radial position are calculated.  An example showing multipole measurements is shown in Figure~\ref{fig:KBF8}.
All new quadrupoles
are measured and shown to meet 12 GeV specifications.
Figure~\ref{fig:KBF9} shows a comparison between the measured quadrupole gradient error and specifications needed from beam dynamics calculations.    
Operating CEBAF at 12~GeV with beam optics settings largely downloaded from an energy-scaled design has proved the acceptable performance for the existing quadrupoles even when
driven at higher currents, and for the new quadrupoles.

\subsubsection{Pathlength Chicane Dipole Upgrades}\label{S:pathlengthChicanes}

The 6 GeV machine circumference (``pathlength'') was adjusted via three-dipole chicanes
in the ``extraction'' regions upstream from each recirculation arc. The physical layout of
each supported a 1 cm span of incremental path (design value +/- 5 mm),
constrained by the installed dipole and power supply capacity. 
Regular measurement throughout the operating experience
at 6 GeV provided values for expected seasonal variation
in pathlength and in variable pass-to-pass pathlength \cite{Teifenback1,Tiefenback2}.

For three-dipole chicanes
path correction is an inverse quadratic function of beam momentum for constant magnetic field.
Nominally, doubling CEBAF energy would reduce the available path compensation by a factor of four
for magnet-limited systems.  The observed
variable circuit-to-circuit path compensation through years of operation
exceeded the anticipated power supply limits.  In order to preserve operational efficiency, an upgrade
of the chicanes is needed. As summarized in Table~\ref{tab:tieftab}, the drive current capacity
in each of the chicane dipoles is increased by a factor of two, and the length of the
dipole magnets in the fourth chicane
(after the second beam pass of the South Linac) is increased by 60\%. Even though the net
path change
capacity is therefore decreased by a factor-of-two (a factor of 40\% for the fourth chicane),
using the modified system plus fine adjustment
of the fundamental operating frequency (at the level of 10s of kHz) provides sufficient control of the
pathlength in the 12 GeV era. Interestingly, operating scenarios have been found
without the need for a tenth chicane in the new recirculation arc beam line leading to the sixth
pass through the North Linac.

\begin{table}[b]%The best place to locate the table environment is directly after its first reference in text
\caption{\label{tab:tieftab}%
Changes to the Pathlength Dipoles (``Dog Legs'') for Operating
with 12 GeV Beam Energy.}
\begin{ruledtabular}
\begin{tabular}{lcc}
Chicane&Current Capacity&Dipole Change\\
\colrule

1&270 A&\\
2&270 to 450 A&\\
3&270 to 450 A&Coil Area Increased\\
4&270 to 450 A&60\% Length Increase\\
5&270 to 450 A&\\
6&270 to 450 A&\\
7&270 to 450 A&\\
8&270 to 450 A&\\
9&270 to 450 A&\\
\end{tabular}
\end{ruledtabular}
\end{table} 

\subsection{Hall D/Arc 10} 

\begin{figure}[hb]
\includegraphics[width=3.4in]{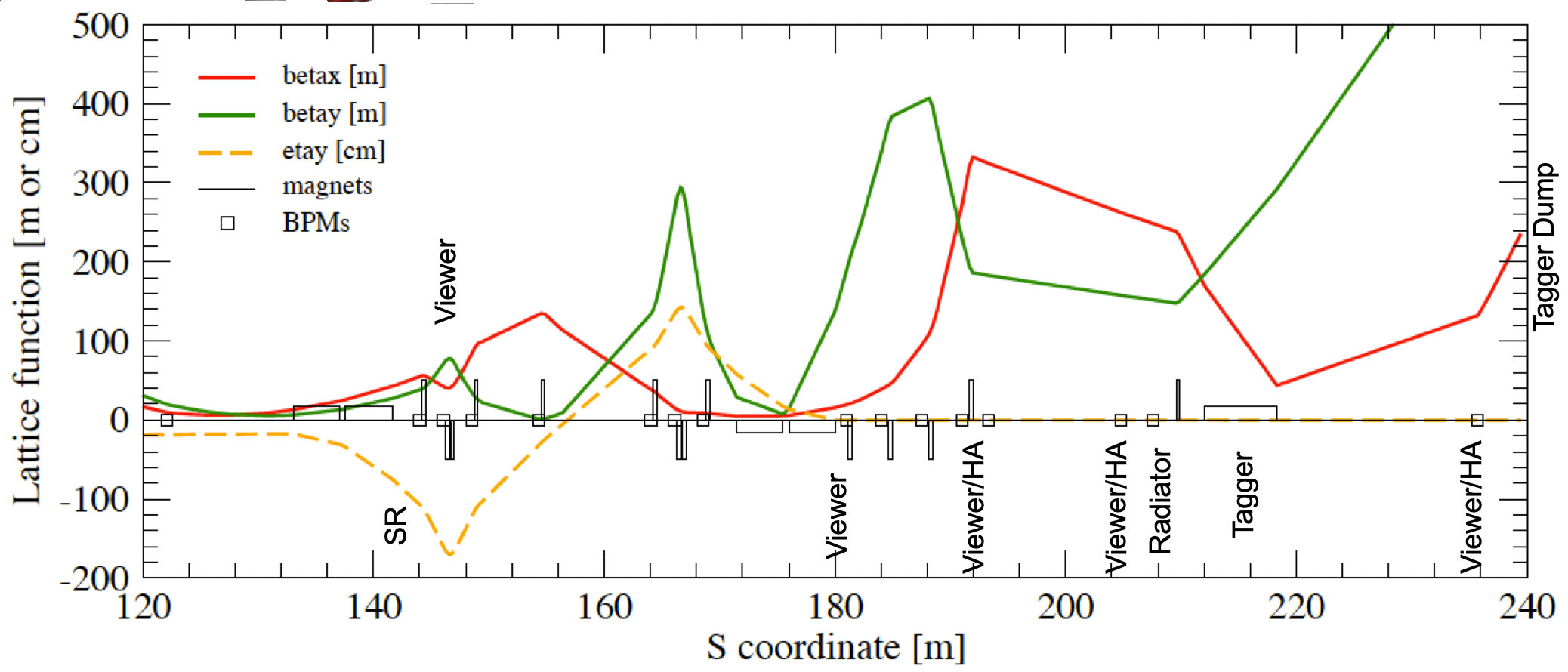}
\caption{\label{fig:hallDoptics} Hall D transport line optics through the vertical ramp to the tagger dump. Locations of synchrotron radiation monitors (SR), viewers, and profile harps (HA) are indicated. Taller vertical rectangles are horizontally focusing (positive) or vertically focusing (negative) quadrupoles.}
\end{figure}  

The original CEBAF magnet transport included nine recirculation arcs (arcs 1 through 9) to support ten linac acceleration passes. Odd-numbered arcs are the ``east'' arcs on the right side of Figure~\ref{fig:scm12GeV}, while even-numbered arcs are the ``west'' arcs on the left side of Figure~\ref{fig:scm12GeV}. To support the addition of Hall D and an additional linac pass, a new arc (Arc 10) is installed below the existing (upgraded) west arcs
as part of the 12\,GeV Upgrade. Corresponding modifications are made to the spreader and recombiner regions at each end of the west arcs, including additional septa, to incorporate Arc 10. These modifications are needed to allow RF beam separation between 5-pass beams to halls A--C and beam into Arc 10, and to include the recombiner merge of the new Arc 10 beam into the North Linac. As in the remainder of CEBAF, all quadrupoles and corrector magnets are independently powered in Arc 10.

Arc 10 is composed of four super-periods. Analogously to the lower arcs, the optics provides second-order achromaticity and linear isochronicity. Arc 10 $M_{56}$ is tunable via quadrupoles throughout the arc optics, though $M_{56}$ is less critical in Arc 10 than in the low-energy arcs. The 32 main Arc 10 dipoles are all on the same main bus and have the same 4\,m length. Dipole trim windings are added to all Arc 10 dipoles to correct for synchrotron radiation beam energy losses; these windings have been tested during commissioning but found to be unnecessary to maintain beam transport quality. Arc 10 has no separate pathlength chicane (see section~\ref{S:pathlengthChicanes}), as sufficient pathlength modifications can be implemented in the arc proper.

The North Linac FODO optics are designed to provide 120 degree phase advance per cell for the lowest-energy (first pass) beam. The higher-energy Hall D beam is thus under-focused and nearly ballistic, leading to tight tolerances on beam optics (particularly divergence) at the exit of the Arc 10 recombiner. Design beta functions are over 200\,m at the end of this pass of the North Linac, entering the east spreader for separation to the Hall D transport line.

The east spreader is modified to add new septum magnets for extraction to Hall D, and to adjust separation geometry to accommodate space for these magnets. Corresponding changes are made to the east recombiner to preserve spreader and recombiner symmetry.  A triplet, quadruplet, and triplet are used to transport and focus beam to a small tuning beam dump $\sim$125\,m downstream of separation before entering a vertical ramp towards the Hall D tagger enclosure. Vertical dispersion from the vertical separation is not corrected until the vertical ramp, which adds sensitivity of optics corrections to energy fluctuations.

The vertical ramp section starts and ends with antisymmetric vertical dipoles (bending a total of 7.8~degrees) to create a +5.2~m vertical dogleg, with seven quadrupoles between the dipoles arranged in a triplet/singlet/triplet configuration. The central quadrupole is located near vertical dispersion zero-crossing and at a vertical beta waist to provide an independent degree of freedom for horizontal beam size and convergence. This is followed by four quadrupoles after the last dipole to provide the other degrees of freedom necessary for control of beam size and convergence in both planes onto the Hall D radiator. The optics of the latter half of the Hall D beam transport are shown in Figure~\ref{fig:hallDoptics}.

The initial Hall D experimental program is dominated by the GlueX collaboration and detector\cite{GlueX,McCaug1}, which uses a polarized photon beam generated by coherent Bremsstrahlung from the passage of 12\,GeV electrons through a diamond radiator. The photon beam convergence and polarization are strongly correlated to the electron beam convergence at the radiator. A collimator 75~m downstream of the radiator is used to select out 40$\%$ polarized photons for use in GlueX experiments. After interaction with the radiator, the electron beam is bent through a large tagger dipole, where a tagger hodoscope correlates low-energy electron events with corresponding photon events in the detector for accurate energy reconstruction. The primary electron beam continues through the tagger dipole and is delivered to the Hall D beam dump. The Hall D transport optics are designed, iterated, and documented in a set of Jefferson Lab Technical Notes \cite{HallDOptics2000,HallDOptics2001,HallDOptics2003}.

Two substantial changes are made in the Hall C beamline
supporting 12 GeV operation. First,
the bending capacity of the dipoles taking the beam to Hall C
is increased as discussed in Section~\ref{Dipolech}. Second, to be able
to bend the beam as required in the Compton polarimeter \cite{JBRef1}, the difference in height between a straight path and the bent path is reduced and a pair of one meter dipoles are added after the M$\o$ller polarimeter to strengthen vertical bending.  The beam line from the shield wall to the diagnostic girder was at a small angle in the 6 GeV era. This offset is eliminated moving all the steering correction to the hall.
 None of the quadrupoles
in the Hall C beamline are replaced in the transition to 12 GeV. Likewise, the dipoles taking the beam to Hall A are upgraded for the enhanced energy to be delivered, but no further changes are needed to the quadrupoles in this beamline.

\subsection{Accelerator Physics}

  At the onset of the project, it has been determined that the main beam physics drivers for the design are the impact of the synchrotron radiation on emittance growth, halo formation and radiation heating. Other issues such as the heating in the accelerator tunnel due to the magnets being operated at higher currents are also quantified.  The first issue may be addressed by a judicious choice of optics combined with a more stringent set of requirements on the magnet field quality. Field specifications are derived from these considerations and utilized in the design of the modified dipoles and the new ones in ARC10. 
  
  \subsubsection{Aperture requirements}
  
  In order to keep the beam synchronized with the RF acceleration in the linacs, a combination of dogleg chicanes, changing the RF frequency of the cavities, and offsetting the orbits in the arcs is employed. The later is necessary because there is no dogleg pathlength adjustment chicane beyond ARC9. Instead, the path length is adjusted by shifting the orbit in the arc. This led to developing a specification for the aperture requirements including beam size, steering allowance, and for ARC10, path length orbit shift.

\subsubsection{Halo Specifications}
Various factors can contribute to the formation of beam halo in particle accelerators. Chief amongst them is the amount of synchrotron radiation, which increases significantly in the 12 GeV machine compared to the 6 GeV machine. This effect causes emittance growth, which in turn leads to larger beam sizes that can sample nonlinear magnetic fields more in the 12 GeV machine. 

Other factors that can contribute to halo formation include the RMS of the beam orbit centroid relative to the magnet center, mismatched beam optics, and scattering off residual beam gas. 

Nonlinear particle tracking simulations are needed to study halo formation. The amount of beam halo due to residual beam gas scattering is expected to be roughly 1/4 of the beam halo in the 6 GeV machine. The 12~GeV design requires careful attention to these factors to contain emittance growth and minimize halo formation.

With the exception of the synchrotron radiation, all these other effects were present in the 6 GeV machine and we can use our past experience and measurements as a benchmark. In particular, the experiments in Hall~B which typically require an electron beam in the range of 1 to 100 nA are very sensitive to halo due to the high luminosity 4$\pi$ detector. Measurements are routinely performed using wire scanners equipped with photo-multipliers for picking up the secondary electron emission generated by the wire going through the beam. This measurement provides dynamic range of over six orders of magnitude. 

The experimental Hall D, which will see an electron beam of 12 GeV and feature a full acceptance detector similar to that in Hall B, sets the maximum allowable halo to be at least six orders of magnitude less than the core of the beam assumed to be Gaussian. 

In order to include the effect of synchrotron radiation and nonlinear mismatches, a set of simulations are performed where 
we generated three representative orbits that simulate a real machine. We introduce random mis-alignments and mis-powering to the  magnets and multipole components. Finally, we apply a global steering of the beamline using the same algorithms one would use in the real machine. This procedure creates three orbits with standard deviations of 0.3 mm, 0.6 mm, and 1 mm, respectively, each consisting of over a 100 million particles. The magnitude of the halo is quantified using the aforementioned measure. 

 From these studies, we conclude that if the orbit RMS is less than 1 mm, we expect a halo to be at least six orders of magnitude less than the signal. 

 The halo due to Mott scattering on the residual gas in the beampipes is evaluated separately and found to be negligible for
 the existing vacuum, thanks to the fact it is inversely proportional to the square of the beam energy.

\subsubsection{Optical Matching and Implications on Emittance Growth}
   The matching specifications for the CEBAF lattice were derived from the requirements that the invariant ellipse distortion resulting from linear errors be exactly compensated by quadrupoles located in the spreader regions.

   The amount of emittance dilution arising from a mismatched beam propagating through a lattice with higher order multipoles and in the presence of synchrotron radiation is estimated separately and used to draw specifications on the allowed magnitude of these multipoles and the amount of mismatch. That mismatch is quantified by the ratio of the area between the design ellipse and the mismatched new ellipse obtained after rematching the optics \cite{roblintn08042}. 

   The process by which one performs these arc by arc corrections is described in \ref{beamtuning}.

\subsubsection{Control of Betatron Envelope in Higher Linac Passes}

  The upgrade of CEBAF from 6 to 12 GeV led to doubling the acceleration in the linacs, adding an extra recirculation arc and modifying the spreaders and recombiners to accommodate these changes. The spreaders and recombiners are two step achromatic vertical bend systems. The nature of this design is such that the peak beta functions in the spreaders are high. The problem is magnified at higher passes where the linac focusing is essentially non-existent. 
  
  A mismatch error in the upper passes will result in a loss of control of the beam envelope as well as a significant emittance growth.  Unlike the 6 GeV CEBAF for which the linac and spreader/recombiner optics were optimized for the first pass, a global approach was chosen in order to minimize the peak betas in the higher passes by trading it with a slightly worse betatron profile at lower passes where it is not significantly impacting beam envelope and emittance growth.  All five passes are optimized together to find the best envelope profile \cite{ChaoNLopti}. The gradient distribution in the linacs is also investigated. 
  
  Emittance growth is impacted by several things, namely the non-linearities in the magnets due to their multipole contents which drove the steering specifications as well as the linac accelerating profile.
  
  The design is iterated several times until we found a satisfactory combination of gradient distribution as well as multipole and steering specifications. This led to installing the new C100 modules at the end of the linacs for practical and budgetary reasons even though the smallest emittance growth is achieved by having these five cryomodules at the start of the linac.

  \subsubsection{Synchrotron Radiation Heating}

Estimates of the synchrotron radiation power deposition are shown in table \ref{tab:srheating}.  The peak deposition occurs in ARC9 and results in a line load of about $0.2$ W/cm. Outgassing is estimated to be well within the existing pumping capacity.

\begin{table}[htbp]
\caption{\label{tab:srheating}Synchrotron radiation heating}
\begin{tabular}{c|c|c|c|c}
& Dipole& Beam& Beam &Radiation\\
            Beamline&Length&Energy (GeV)&Current (\textmu A)& Power\\
\colrule
      Arc6 &  2 m& 6.7 & 90 &\hfil 509 W\\
      Arc7 & 3 m& 7.8 & 90 &\hfil 587 W\\
      Arc8 & 3 m& 8.8 & 90 &\hfil 1029 W\\
      Arc9 & 3 m& 9.9 & 90 &\hfil 1502 W\\
      Arc10 & 4 m& 11.0 & 5 &\hfil 109 W\\
      HallD & 4 m& 12.1 & 5 &\hfil 9 W
\end{tabular}
\end{table}

\subsection{12 GeV CEBAF Optics Design}
Originally, most of the longitudinal bunching in the injector occurred prior to the acceleration up to the final injection energy. The injector experienced difficulties transporting the beam, because of small tails in the longitudinal beam profile. Particles in these tails were not on the crest of the accelerating wave and were ultimately lost preventing machine operation at high beam current. To alleviate this problem the optics in the injection chicane has been re-designed to create additional bunch compression at 123 MeV. A significant advantage of high energy bunching is that the bunching is not affected by the beam space charge. To facilitate this change, a new non-isochronous optics with a negative $M_{56}$ of about $-24$ cm is designed and loaded in the injection chicane magnets. To perform the bunch compression one needs to shift the RF phase of the main injector linac by about 8~degrees. To avoid problems with focusing changes at the beginning of the linac, only the second of the two injector cryomodules is shifted in phase. The new configuration significantly improves machine reliability for high current operation.

For 12 GeV CEBAF the synchrotron radiation effects on beam motion become rather significant in the higher arcs with energies above 6 GeV. Emissions of individual photons excite spurious betatron oscillations; the resulting energy `drop' perturbs the electron trajectory causing its amplitudes to grow leading to cumulative emittance increase. Details of the single particle dynamics are given by M.~Sands \cite{SandsRef}. 

\begin{figure}
\includegraphics[width=\columnwidth]{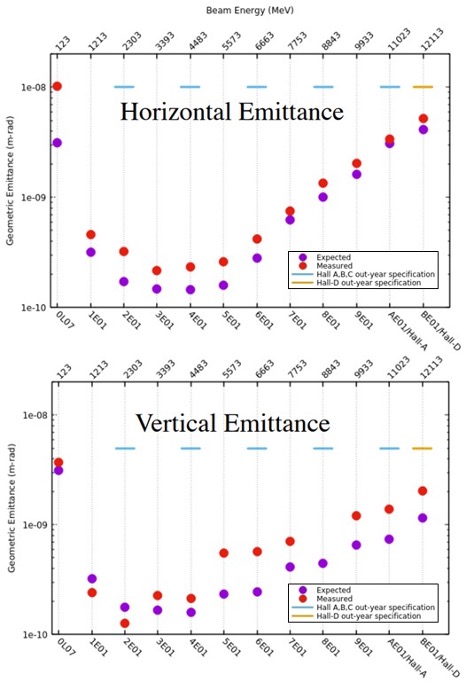}
\caption{\label{CebafEmittances12GeV} CEBAF 12 GeV horizontal and vertical emittances measured at the injector, each arc and at the entrance to Hall D during 12\,GeV commissioning (from \cite{CPPREF4}).}
\end{figure}

In order to limit emittance dilution due to the synchrotron radiation, several options have been explored. Alternative beam optics are proposed for the higher arcs to limit emittance dilution due to quantum excitations \cite{AlexBRef2}. The optics can be implemented within the 6~GeV physical layout of the arcs (baseline design); producing the new optics only involved changes in quadrupole magnet settings. The effect of synchrotron radiation has been suppressed through careful lattice redesign, by appropriately organizing the Twiss functions and their derivatives inside the bending magnets. A Double Bend Achromat (DBA) cell variety using a triplet rather then a singlet to suppress dispersion is chosen as a `building block' for the arc optics. The lattice provides significantly suppressed emittance dilution while offering superior lattice tunability and compactness. The lattices for Arcs 6-10 are reworked based on the above DBA structure. The resulting emittance growth is suppressed by factor of 0.64 compared to the `Standard' Arc 6-10 optics \cite{AlexBRef3}. Figure~\ref{CebafEmittances12GeV} shows that measured CEBAF beam emittances during commissioning were well below specifications, and closely match design expectations~\cite{emittanceevolution,CPPREF4}.

	Responding to the need for diagnosing the beam energy spread, the optics of arcs 1 and 2 have been redesigned and synchrotron light monitors are installed to resolve the beam energy spread with high resolution. The optics goal is to increase the horizontal dispersion by a factor of three and to decrease the horizontal beta function in the middle of both arcs at the location where the new monitors are installed. The new arc optics, with a mirror-symmetric horizontal dispersion pattern, is designed so that it greatly enhances resolution of the beam energy spread measurement without limiting the energy aperture of the beamline. To preserve tunability of the new optics one needs to allow for independent correction of both the horizontal dispersion and $M_{56}$. This is accomplished by appropriate tailoring of the horizontal betatron phase advance inside the arc to provide two pairs of orthogonal ``knobs" (quadrupoles): for dispersion and momentum compaction adjustments. Furthermore, a betatron wave excited by the first tuning quad, which propagates with twice the betatron frequency, is cancelled by the second wave launched by the remaining quad in the pair, so the net betatron wave is confined to the tuning region and subsequently the tuning process does not affect the betatron match outside the arc.

\subsection{Extraction System/4 Hall Operations\label{fourhall}}

\begin{figure*}
\includegraphics[width=6.8in]{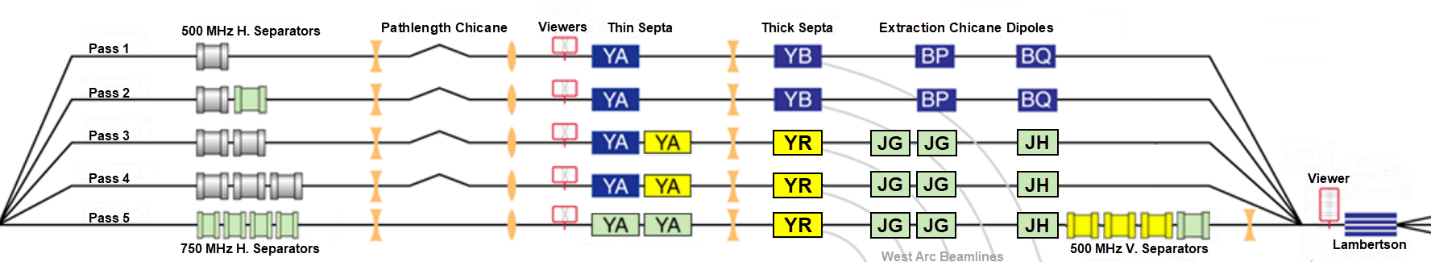}
\caption{\label{SpataFig1} An elevation view of the overall 12 GeV CEBAF extraction system. Gray cavities and blue magnets are in their original locations. Yellow elements were relocated from other 6 GeV CEBAF locations. Green elements were newly constructed for the 12 GeV accelerator.}
\end{figure*}

The extraction system is designed to selectively transport 249.5 MHz and 499 MHz interleaved bunch trains provided by the polarized source to the proper pass and experimental hall needed for the physics program. Figure~\ref{SpataFig1} shows an elevation view of the final configuration for the 12 GeV extraction upgrade. The approximately 170 m long segment represented here starts at the entrance of the southwest spreader and ends at the exit of the Lambertson magnet at the entrance to the Beam Switch Yard. The key elements of the system are 499~MHz and 749.5~MHz RF separators, focusing and defocusing quadrupoles, thin and thick septa magnets, extraction chicane dipoles, beam position monitors, beam viewers, and the Lambertson magnet. In the following sections, a detailed description of these elements and their role in the extraction system is described.

The original 6 GeV CEBAF configuration used ten 499 MHz RF separators with one in the first and second passes, two in the third pass, three in the fourth pass and a set of three cavities oriented for vertical deflection located approximately three meters past the exit of the single YA stack. For the first four passes, the cavities were phased to provide peak deflection to the left for the beam intended to be extracted to Halls A, B or C. This puts the other beam(s) on half power points, 120$^\circ$ from peak phase, to be deflected to the right and sent into CEBAF for recirculation to higher passes. The long drift from the RF Separators to the entrance of the YA stack along with the horizontally defocusing quadrupole between them provided 16.5 mm of separation as measured by precision wires mounted on the beam viewers and the adjacent beam position monitor located in front of the thin septa. Horizontal correctors at the entrance of the RF Separators were used to place the recirculated beam(s) at +5 mm and the extracted beam at -11.0 mm.

The one-meter long YA magnet had a 5 mm wide septum that is protected by a water-cooled molybdenum nose-piece. The magnets were aligned so that the right edge of the septum is at the nominal zero coordinate in the $x$-plane. Beams to the right of the septa enter a field free region while beams to the left saw the full field of the magnet and are kicked to the left. The recirculated and extracted beams continue to drift apart as they were transported to the entrance of the one-meter long YB stack. The defocusing quadrupole combined with the long drift provided 4.5 cm of separation at the entrance of the YB. The recirculated beam(s) continued to drift through the field-free region and arrive at the West Arc point of tangency at the center of the first arc quadrupole for another pass around CEBAF. The one-meter YB, two-meter BP and one-meter BQ magnets comprised an extraction dipole chicane that is used to physically avoid the first stack of west arc quadrupoles next to the BP magnet and to place the extracted beam on zero position and angle at the center of the first quadrupole of the transport recombiner beamline segment. The 2.3 m long Lambertson magnet had an upper and lower set of magnet coils that were independently powered to have a field oriented in the negative y-axis for the upper coil and in the positive $y$-axis for the lower coil. The magnet had three separate vacuum chambers with Hall A in the upper chamber kicked to the right, Hall B in the field-free region between the coils and Hall C in the lower chamber kicked to the left. The Hall A and C beamlines were 2.2 cm above and below the Hall B beamline respectively. Vertical correctors in front of the Lambertson magnet were used to position the beam at the proper elevation for the relevant Hall.

The fifth pass of the 6 GeV extraction system used three RF separator cavities oriented for vertical deflection. These cavities were phased to put the Hall B beam on zero-crossing while kicking the 120$^\circ$ phase delayed Hall A beam up and the Hall C beam down. A pair of YA septa magnets amplified the kick for the Hall A and C beams while leaving the Hall B beam undeflected. Empirical settings were adjusted to place the beam(s) at the proper elevation(s) at the entrance to the Lambertson magnet.
\subsubsection{6 GeV to 12 GeV Layout Changes}
The overall operational paradigm of the 12 GeV extraction upgrade is largely unchanged from that presented above. The following changes are needed to manage the higher energy beams and to accommodate the fourth experimental Hall D.
\vspace{5mm}

\centerline{RF Separators}

\begin{enumerate}
\item The operating power of existing 499 MHz separators is increased.
\item An additional 499 MHz separator is added to the second beam pass.
\item Four horizontal 749.5 MHz separators are installed in the fifth pass beam line beneath the existing 499 MHz cavities.
\item The output coupler for one of four existing 499 MHz RF power amplifiers is modified to operate at 749.5 MHz.
\item The vertical RF Separators are relocated to the transport section beyond the west arc and an additional cavity is added.
\item New 10~kW solid state amplifiers are installed to power the relocated vertical RF Separators.
\end{enumerate}

\centerline{Magnets}
\begin{enumerate}
\item The two fifth pass vertical YA magnets are relocated to the third and fourth passes.
\item Two new YA magnets for the fifth pass are fabricated.
\item The third and fourth pass one-meter YB septa are replaced with existing two-meter YR magnets and a third existing YR has been relocated to the fifth pass.
\item The third and fourth pass two-meter BP dipole magnets are replaced with pairs of a new two-meter JG magnet and another JG pair is added to the fifth pass beamline.
\item The third and fourth pass one-meter BQ dipole magnets are replaced with a new two-meter JH magnet design and another JH magnet has been added to the fifth pass beamline.
\item The vertical correctors, used in the first through fourth pass for setting the elevation of the beams at the entrance of the Lambertson, are replaced with higher field magnets.
\end{enumerate}

\subsubsection{RF Separators}
\begin{figure}
\includegraphics[width=3.4in]{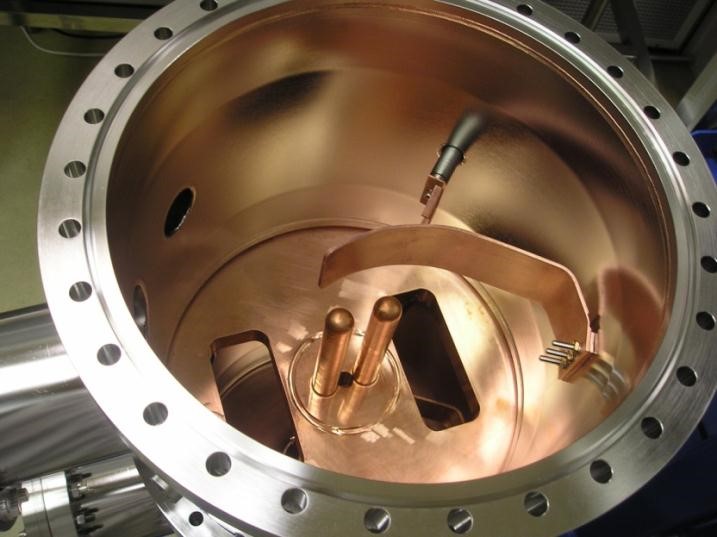}
\includegraphics[width=3.4in]{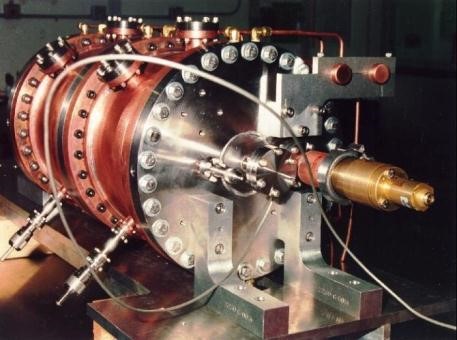}
\caption{\label{SpataFig2} Interior and fully assembled views of a 499 MHz RF Separator cavity.}
\end{figure}

The RF separator cavities for CEBAF were conceived of and designed in the early 1990s \cite{Yao1} with the first proof of principle experiment conducted in the CEBAF injector using 45 MeV beam in 1992 \cite{SpataRef1}. Each separator cavity is a two-cell warm copper structure with each cell containing four co-planar copper rods to concentrate the TEM dipole mode along the central axis of the cavity. A pair of copper rods can be seen in the interior view of Figure~\ref{SpataFig2} along with a mechanically actuated tuning plate, coupling holes for the adjacent cell, and the 15 mm beam aperture between the rods. The unattached 14” end flange holds the other pair of copper rods. Water channels in the end flanges and center flange deliver coolant to the rods that are fitted with internal septum plates. The interior of the cavity bodies are copper plated and water-cooled.

Power is delivered through a critically coupled inductive copper loop mounted on a 1-5/8” coaxial adapter and the field is measured through an under coupled loop probe. A fully assembled 499 MHz cavity is shown in Figure~\ref{SpataFig2}.
To allow for simultaneous beam delivery in all four CEBAF experimental Halls a concept using 249.5~MHz electron bunches and 749.5 MHz RF separators was proposed in 2012 \cite{SpataRef2,SpataRef3}. Electromagnetic and thermal analysis studies for the shorter structure began in 2014 and a prototype cavity was fabricated for bench testing. The CST electromagnetic design simulations center on an optimization of the high-power input coupler position, loop size and rotation, and the tuner paddle position to optimize frequency and field flatness \cite{SpataRef4}. Four production cavities have been produced, installed in CEBAF, and then commissioned with beam in 2018 \cite{SpataRef5}.

The nominal reference values and calculated power requirements are shown in Table~\ref{tab:spatab1} and Table~\ref{tab:spatab2}. For both the 499 MHz and 749.5 MHz cavity designs the modeled shunt impedances have been experimentally verified through beam-based measurements of the horizontal beam position at the entrance to the down stream YA septa as a function of cavity power with the beam phase at $\pi$/2 relative to zero-crossing phase.

\begin{table}[b]%The best place to locate the table environment is directly after its first reference in text
\caption{\label{tab:spatab1}%
Shunt impedance for 499 MHz and 749.5 MHz cavity designs and required cavity phase relative to zero-crossing.}
\begin{ruledtabular}
\begin{tabular}{lc}
Parameter&Value\\
\colrule
499 MHz Cavity Shunt Impedance ($\Omega$)&2.10E+08\\
749.5 MHz Cavity Shunt Impedance ($\Omega$)&1.04E+08\\
1-5 Pass Horizontal Beam Phase (radians)&$\pi$/2\\
5th Pass Vertical Beam Phase (radians)&$\pi$/3\\
\end{tabular}
\end{ruledtabular}
\end{table}

\begin{table}[b]%The best place to locate the table environment is directly after its first reference in text
\caption{\label{tab:spatab2}%
Power requirements for RF Separators based on beam energy, deflection angle and the number of cavities per pass.}
\begin{ruledtabular}
\begin{tabular}{lcccccc}
&Number&Angle per&Total&&Power at&Cavity\\
&of&Cavity&Angle&Energy&Cavities&Power\\
Pass&Cavities&(\textmu rad)&(\textmu rad)&(MeV)&(W)&(W)\\
\colrule

1&1&221&221&2303&1238&1238\\
2&2&116&232&4483&2553&1276\\
3&2&116&232&6663&5695&2848\\
4&3&81&243&8843&7305&2435\\
5&4&40&158&11023&7291&1823\\
5V&4&62&248&11023&11861&2961\\
\end{tabular}
\end{ruledtabular}
\end{table}

\begin{figure}
\includegraphics[width=3.4in]{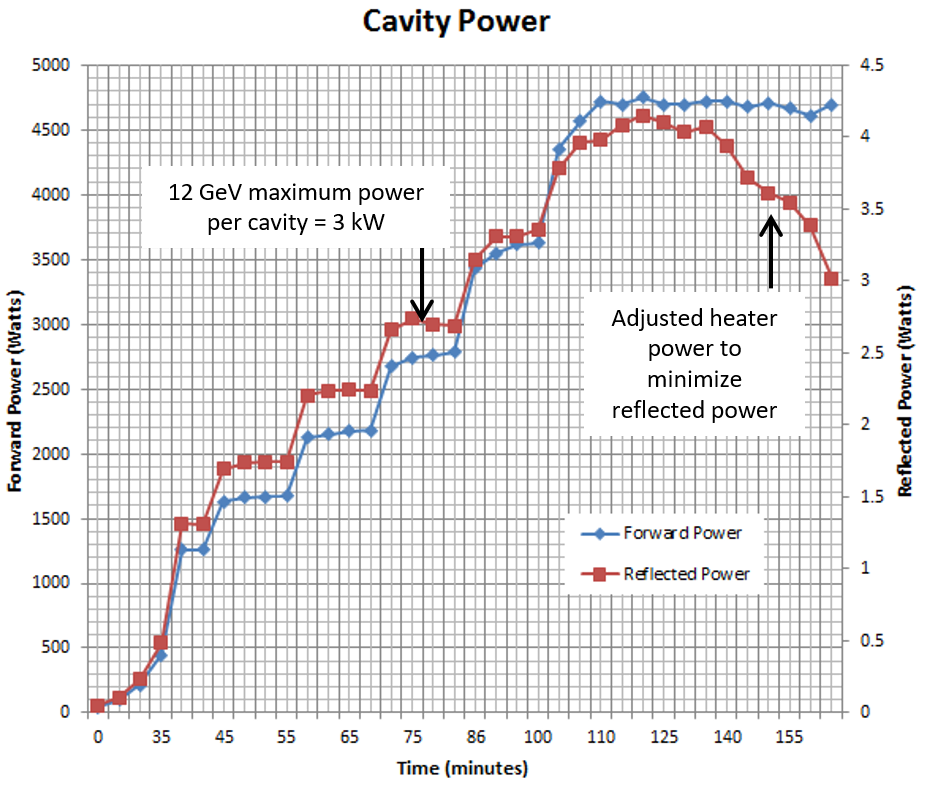}
\includegraphics[width=3.4in]{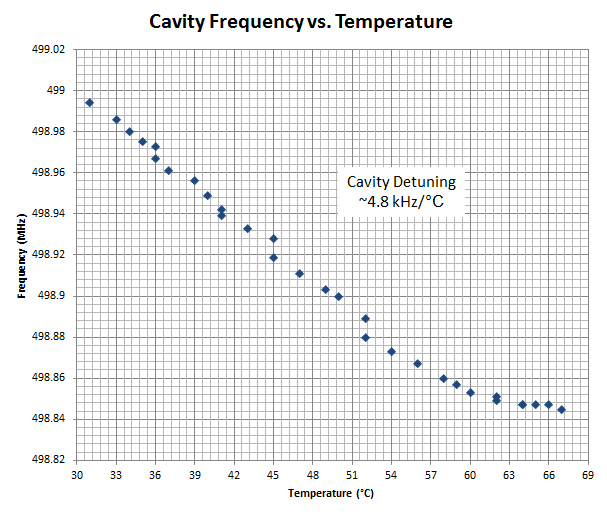}
\caption{\label{SpataFig3} Power ramp ending in a one-hour high-power run at 4.7 kW (upper) and cavity frequency vs.~temperature curve (lower).}
\end{figure}

The peak power per cavity from Table~\ref{tab:spatab2} is just under 3 kW. To verify thermal modeling and that cavity frequency shifts as a function of temperature are within range of the heater-based resonance control system a high power test has been conducted on a 499 MHz cavity. The results for both are shown in Fig.~\ref{SpataFig3}.

\begin{figure}
\includegraphics[width=3.4in]{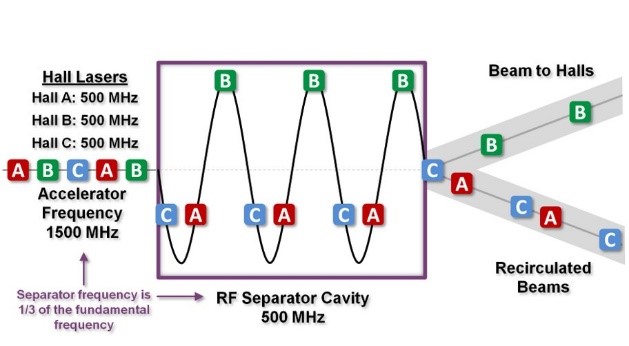}
\includegraphics[width=3.4in]{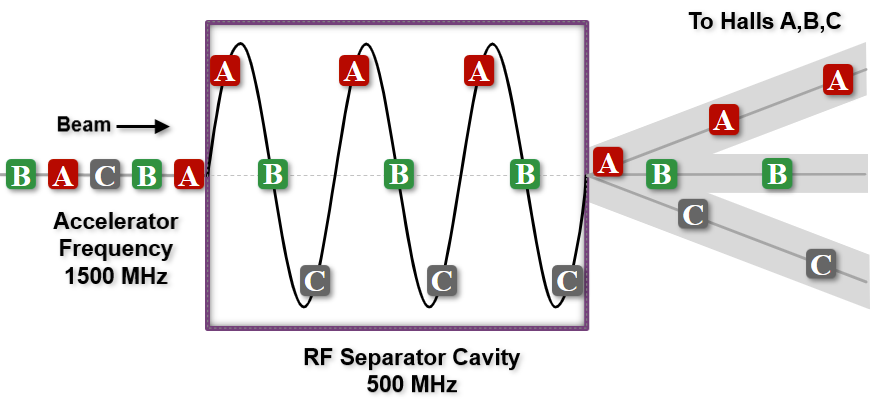}
\caption{\label{SpataFig4} Schematic showing how beams are interleaved, separated and delivered to each experiment hall for first through fourth pass horizontal extraction and for fifth pass vertical extraction. Configuration when Hall D not receiving beam.}
\end{figure}

\begin{figure}
\includegraphics[width=3.4in]{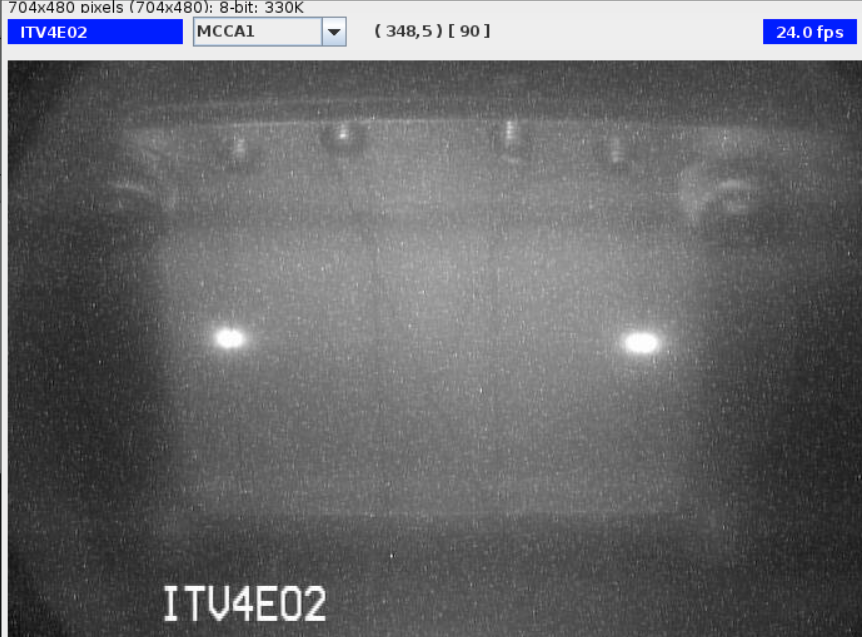}
\includegraphics[width=3.4in]{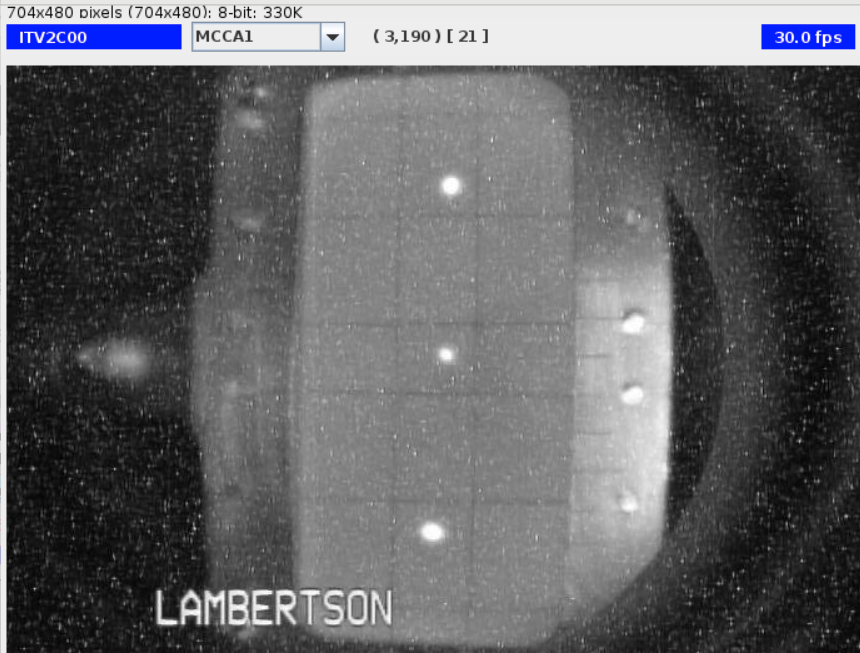}
\caption{\label{SpataFig5} Viewers showing 16.5 mm of separation in front of the 2nd pass YA thin septum (above) and three beams at their proper elevation at the entrance to the Lambertson magnet for Halls A, B and C (below).}
\end{figure}

The required phase relationship for the 499 MHz horizontal and vertical extraction systems is shown in Fig.~\ref{SpataFig4}. The 749.5 MHz phase relationship has been shown earlier in Figure~\ref{fig:mpfig6}. Typical horizontal and vertical separation on beam viewers is shown in Fig.~\ref{SpataFig5}.

\vspace{5mm}
\centerline{RF Amplifiers}
\vspace{5mm}
The RF power solution for the initial 4 GeV CEBAF installation consisted of six 499 MHz modular solid state amplifiers (SSA) each with their own low-level RF control module and capable of delivering 1 kW. The amplifiers were connected to the relevant ten cavities on the beamline that were required to support the pass configuration of the scheduled physics program. The maximum number of amplifiers needed at any one time was five allowing the sixth amplifier to serve as a hot spare. The connections were made through a patch panel system with each cavity being powered by a single amplifier.

\begin{figure}
\includegraphics[width=3.4in]{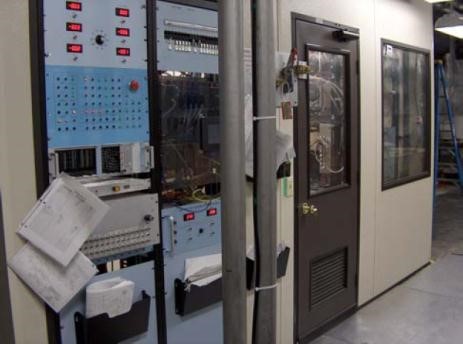}
\includegraphics[width=3.4in]{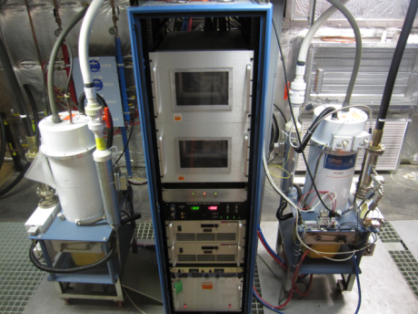}
\caption{\label{SpataFig6} The 20 kV High Voltage enclosure and controls above and a pair of 10 kW IOTs with their controls below.}
\end{figure}

To address obsolescence issues with the original 1 kW SSA systems and to support the higher power beam during the 6 GeV era, four 499 MHz 10 kW Inductive Output Tube (IOT) systems, floating on a 20 kV DC high voltage deck, were installed. An example is shown in Fig.~\ref{SpataFig6}. These UHF RF transmitters provide power for the second through fifth pass cavities while components of the aging original solid-state system were retained to power the first pass separator cavity. The output coupler for one of the 499 MHz IOTs was modified to operate at 749.5 MHz for the fifth pass horizontal system.

\begin{figure}
\includegraphics[width=3.4in]{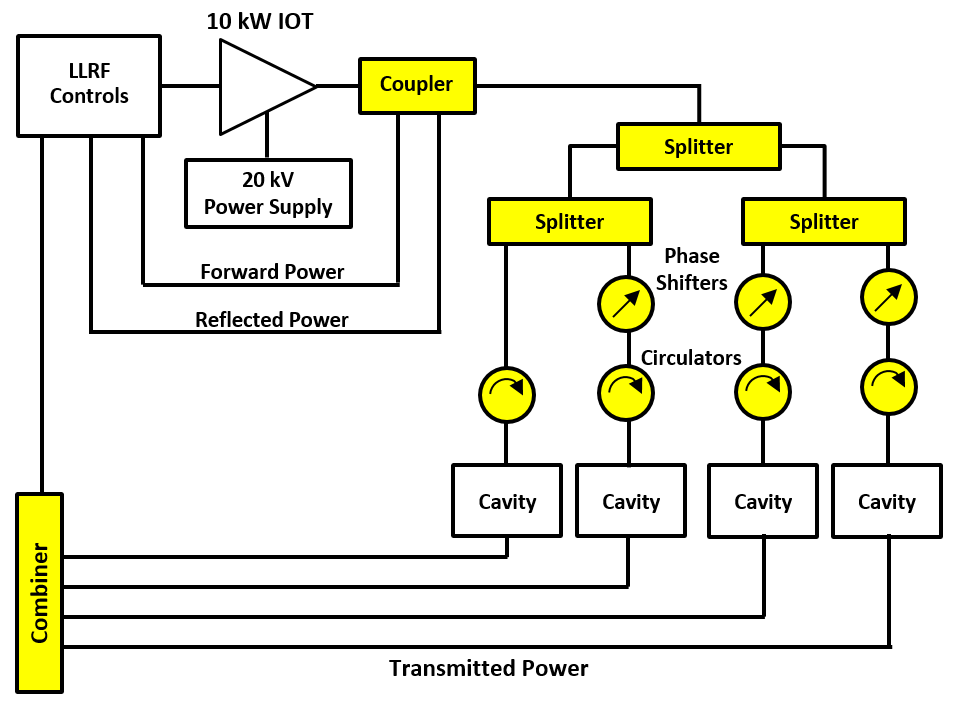}
\caption{\label{SpataFig7} RF distribution system for the 12 GeV Upgrade IOT systems.}
\end{figure}

The IOTs are each connected to the multiple cavities of a relevant pass through a network of high-power splitters, phase shifters, circulators and combiners as shown in Figure~\ref{SpataFig7}.  A procedure to optimize the phase of each cavity of the combined system relative to the beam has been developed. The system is first powered with cavities 2, 3 and 4 terminated into water-cooled loads. The LLRF controls are then used to find the zero crossing phase that provides a rightward beam deflection with positive changes in the phase of the LLRF control module. The phase is determined by beam position monitors and viewers at the entrance of the YA thin septa for first through fifth pass horizontal systems and at the entrance of the Lambertson magnet for the fifth pass vertical system. Each of the remaining cavities is then incrementally reconnected to the system with their phase shifter used to return the system to the same zero crossing. The mechanical phase shifters have a limited range of 175$^\circ$ at 499 MHz and 225$^\circ$ at 749.5 MHz. If the proper setting is out of reach, a piece of hardline is inserted to center the phase-shifter zero crossing response. 
There have been operational challenges with the 749.5~MHz RF system related to thermal management in the RF distribution system as well as long-term amplitude and phase drift in the power delivered to each cavity.

Two additional upgrades for the RF power systems have been accomplished to complete the overall program. Funding from the American Recovery and Reinvestment Act (ARRA), secured in 2013, was used to build an addition on an existing service building and to procure and install four 499 MHz 10 kW solid-state amplifiers for the fifth pass vertical RF Separators. An additional unit was purchased in 2018 to power the first pass horizontal system and to retire the original 1 kW solid-state amplifiers. In addition, ARRA funding allowed a 499 MHz superconducting deflector to be built and cold tested \cite{suba1,suba2}.
The cavity achieved a transverse voltage of 3.3 MV with peak surface fields of 32 MV/m and 49 mT \cite{suba3}.

\subsubsection{Magnets}

The extraction system magnet upgrade has been accomplished through the reuse of existing YA/YB septa and BP/BQ dipole magnets. The YA and YR magnets are relocated and outfitted with modified vacuum chambers, two new YA magnets have been fabricated, and new JG and JH dipole magnet have been developed and fabricated for the extraction chicanes. All new magnets are measured for field integral and field quality as described in Section~\ref{magnetrefsec}. In addition, a long-standing error in the $\int BdL$ of the BP and BQ magnets have been corrected through the addition of 1” shims to lengthen the magnets. The magnet transitions for third through fifth pass, design bend angles and integral field strength are shown in Table~\ref{tab:msextab3}.

\begin{table*}[b]%The best place to locate the table environment is directly after its first reference in text
\caption{\label{tab:msextab3}%
Physical parameters, design bend angles and the integral 
field strength for the 12 GeV Extraction magnets.}
\begin{ruledtabular}
\begin{tabular}{l|c|c|c|c|c|c|c|c}
Pass&Energy (MeV)&4 GeV/6 GeV&12 GeV&Magnet Type&Length (m)&Septa (mm)&Total Angle (mrad)&Total B-dL (G-cm)\\
\colrule
3&6600.67&YA&2 YA&Septa&1&5.0&-1.876&-41694\\
&&YB&YR&Septa&2&31.5&-39.675&-881791\\
&&BP&2 JG&Dipole&2&NA&81.335&1807710\\
&&BQ&JH&Dipole&1&NA&-37.767&-838389\\
\colrule
4&8760.40&YA&2 YA&Septa&1&5.0&-1.876&-55336\\
&&YB&YR&Septa&2&31.5&-39.675&-1170444\\
&&BP&2 JG&Dipole&2&NA&81.335&2399202\\
&&BQ&JH&Dipole&1&NA&-37.767&-1113989\\
\colrule
5&10920.13&YA&2 YA&Septa&1&5.0&-1.701&-62536\\
&&YB&YR&Septa&2&31.5&-39.20&-1441336\\
&&BP&2 JG&Dipole&2&NA&81.335&2988452\\
&&BQ&JH&Dipole&1&NA&-37.767&-1388522\\
\end{tabular}
\end{ruledtabular}
\end{table*}

\subsection{Beam Diagnostics\label{beamdiagref}}

The beam diagnostics upgrade consists of adding devices to the new beamlines in Arc 10 and Hall D and to the modified spreaders and recombiners. The solutions deployed are a mix of replicating existing hardware, using existing solutions with some upgrades to beamline devices and electronics, and developing new components. The details for each system are discussed in the following sections. Table~\ref{tab:msdiagtab1} lists the total number of devices and the locations where they have been installed for the upgraded facility. Table~\ref{msdiagtab2} lists the operational range, precision, and accuracy specifications for the various diagnostic systems. 

\begin{table*}[b]%The best place to locate the table environment is directly after its first reference in text
\caption{\label{tab:msdiagtab1}%
Added beam diagnostics inventory and installed locations for the 12 GeV CEBAF Upgrade.}
\begin{ruledtabular}
\begin{tabular}{lccccc}
Diagnostic System&Injector&Arcs \&&Spreader and&Hall D&Total\\
&&Extraction&Recombiner&&\\
\colrule
Antenna BPM&2&38&22&0&62\\
Stripline BPM&0&0&0&26&26\\
Nano-Amp Cavity BPM&0&0&0&2&2\\
Wire Scanner&0&2&0&4&6\\
Beam Viewer&0&5&6&6&17\\
Synchrotron Light Monitor&0&1&0&2&3\\
\end{tabular}
\end{ruledtabular}
\end{table*}

\begin{table*}[b]%The best place to locate the table environment is directly after its first reference in text
\caption{\label{msdiagtab2}%
Specifications for CEBAF Diagnostics.}
\begin{ruledtabular}
\begin{tabular}{lll}
Device Type&Operating Range&Precision/Accuracy\\
\colrule
Antenna-Style&Position: -8 mm $<$ x/y $<$ 8 mm&30 \textmu m / 100 \textmu m\\
Beam Position Monitor&Transport Style: Current: 50 nA $<$ I $<$ 200 \textmu A&\\
&Linac Style: Current: 1 \textmu A $<$ I $<$ 2 mA&\\
\colrule
Stripline&Position: -8 mm $<$ x/y $<$ 8 mm&30 \textmu m / 100 \textmu m\\
Beam Position Monitor&Current: 10 nA $<$ I $<$ 200 \textmu A\\
\colrule
RF Cavity nA&Position: -12 mm $<$ x/y $<$ 12 mm&100 \textmu m / 300 \textmu m\\
Beam Position Monitor&Current: 100 pA $<$ I $<$ 1 \textmu A\\
&Beam Size: $\sigma_{x/y}$ $<$ 4 mm &\\
\colrule
RF Cavity nA&Position: -12 mm $<$ x/y $<$ 12 mm&100 nA / 1 \textmu A\\
Beam Current Monitor&Current: 60 nA $<$ I $<$ 1 mA&\\
&Beam Size: $\sigma_{x/y}$ $<$ 4 mm &\\
\colrule
Wire Scanner&Position: -10 mm $<$ x/y $<$ 10 mm&10 \textmu m / 10 \textmu m\\
&Current: 2 \textmu A $<$ I $<$ 50 \textmu A&(both apply to sigma)\\
&Beam Size: 25 \textmu m $<$ $\sigma_{x/y}$ $<$ 4 mm &\\
&RMS width: 25 \textmu m $<$ width $<$ 4 mm &\\
\colrule
Fluorescent Screen&Position: -12 mm $<$ x/y $<$ 12 mm&500 \textmu m / 1 mm\\
Beam Viewers&Current: 100 pA $<$ I $<$ 50 \textmu A&\\
&Beam Size: $\sigma_{x/y}$ $<$ 4 mm &\\
\colrule
Synchrotron&Position: -12 mm $<$ x/y $<$ 12 mm&10 \textmu m / 10 \textmu m\\
Light Monitor&Current: 1 nA $<$ I $<$ 1 mA&(both apply to sigma)\\
&Beam Size: 100 \textmu m $<$ $\sigma_{x/y}$ $<$ 4 mm &\\
\end{tabular}
\end{ruledtabular}
\end{table*}

\subsubsection{Beam Position Monitors} 

\begin{figure*}
\includegraphics[width=6.8in]{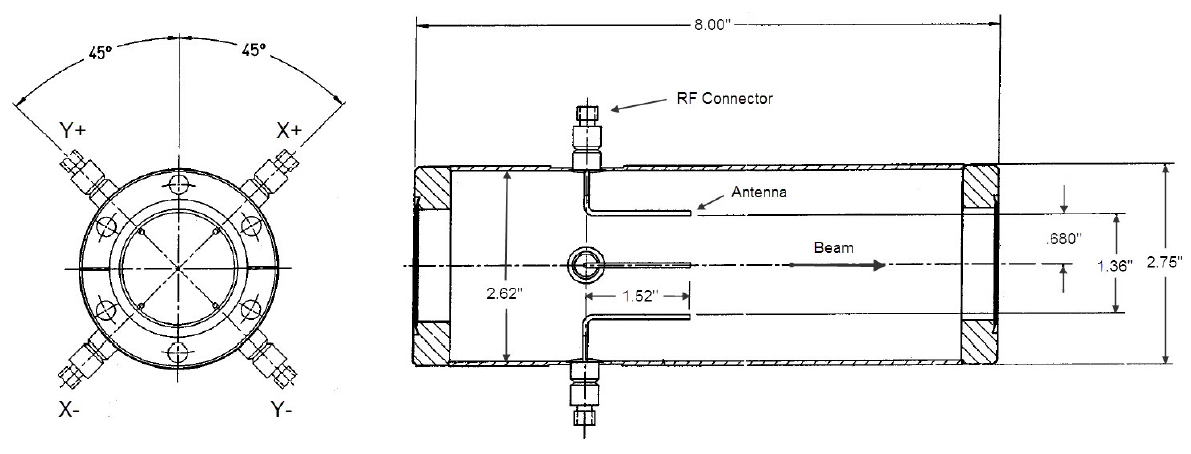}
\caption{\label{msdiagf1} Schematic for M15 antenna-style beam position monitor. M20 BPMs have a larger diameter with the same length.}
\end{figure*}

The 6 GeV CEBAF beam position monitor (BPM) configuration included 450 antenna-style beam position monitors consisting of two different types of a similar design \cite{MSdiagr1}. They had four thin quarter-wave antennae symmetrically placed around the beam and oriented at 45 degrees from the normal x-y axes to avoid false signals being induced from synchrotron radiation in the bend planes. A schematic representation of an M15 BPM can is shown in Figure~\ref{msdiagf1}. The majority of BPMs were installed on girder assemblies with the BPM located immediately upstream of a quadrupole. The M15 style was installed in all locations with the exception of the first two recirculation arcs, the extraction regions, and in spreaders and recombiners. In these locations, an M20 BPM with increased bore was used to accommodate the larger beam tubes in these dispersive sections of the accelerator. The M20 was mounted to a pair of 4-5/8” conflat flanges and has an inner bore of 1.87” compared to the 2-3/4” flanges and 1.36” inner bore of the M15 design. 

The original requirements for the BPM system were to detect beam currents from 1 \textmu A to 200 \textmu A for CW beams as well as 60 Hz tune mode pulses as short as 100 \textmu s with an average current of ~10 \textmu A. The position resolution specification was 100 microns. The first generation of electronics implemented to meet these specifications was a CAMAC based heterodyne solution referred to as the 4-channel or transport BPM system \cite{MSdiagr1}. To first order, the performance of the 4-channel system was sufficient in the early years of commissioning and operating CEBAF. Correcting some limitations in the original system and the need to add new features led to the development of the VME-based Switched Electrode Electronics (SEE) system \cite{MSdiagr2}. 

For example, the 4-channel system suffered from drift in the gain between plus and minus channels (see Figure~\ref{msdiagf1}) for each rotated plane. The total 5-pass linac beam current specification is from 1-1000 \textmu A and the 4-channel system lacked the required dynamic range for these beam currents. There was a need to distinguish the beam orbit for each of the linac passes during tune-mode operations. Finally, there was an emerging need to incorporate a high-speed data acquisition system suitable for use as a time domain diagnostic and to support the development of feedback systems for correcting AC line harmonics. 

\begin{figure*}
\includegraphics[width=6.8in]{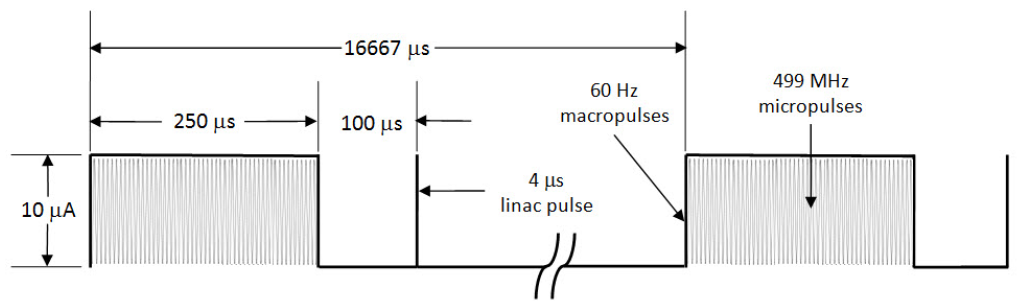}
\caption{\label{msdiagf2} Tune mode 250 \textmu s 60 Hz macropulse  current structure followed by the linac 4 \textmu s ``snake” pulse.}
\end{figure*}

To mitigate gain drift between plus and minus channels the SEE system switches between the pairs of antenna at 120 kHz and uses a single electronics circuit for detection. Figure~\ref{msdiagf2} provides a representation of the CEBAF 60 Hz tune-mode current structure, showing a 250 \textmu s macropulse and a 4 \textmu s linac ``snake'' pulse after a 100 \textmu s delay. Most of the beam diagnostic systems are triggered to take data 65 \textmu s after the leading edge of the 60 Hz macropulse. The one-pass transit time around CEBAF is 4.237 \textmu s. To allow for independent pass position measurement using the successive 4 \textmu s current pulses as they snake through the linacs, the system timing of the linac-style SEE electronics is tuned to take readings delayed by the machine recirculation period for each successive pass.  The transport style systems include a multiplexer connecting the in-tunnel RF modules of vertical stacks of ARC BPMs to a service building VME chassis. Specifications for both linac and transport style electronics are shown in Table~\ref{msdiagtab2}. The SEE BPM system was designed in 1994 and then implemented in the accelerator segments as shown in Table~\ref{msdiagtab3}.  

\begin{table}[b]%The best place to locate the table environment is directly after its first reference in text
\caption{\label{msdiagtab3}%
Regions where SEE BPM electronics are installed and the installation periods.}
\begin{ruledtabular}
\begin{tabular}{cl}
SEE Install Year&Regions Upgraded\\
\colrule
1995&North and South Linac\\
&Spreader and Orbit Locks for\\
&East Arcs (1,3,5,7,9), Hall A\\
1998&Injector and Spreader\\
&Orbit Locks for West \\
&Arcs (2,4,6,8), Halls B and C\\
Mid to late&Transport Recombiner\\
2000s&Remaining segments in West Arcs\\
\end{tabular}
\end{ruledtabular}
\end{table}

The 12 GeV BPM upgrade encompassed two main technical approaches. First, a mix of M15 and M20 style BPMs are added to the existing SEE electronics systems. The majority of these systems are located in the new Arc 10 and in the modified Spreaders and Recombiners.  Two more BPMs are added in the Injector segment. There are 62 total BPMs installed in the locations summarized in Table~\ref{tab:msdiagtab1}. 

\begin{figure}
\centering
\includegraphics[width=3.4in]{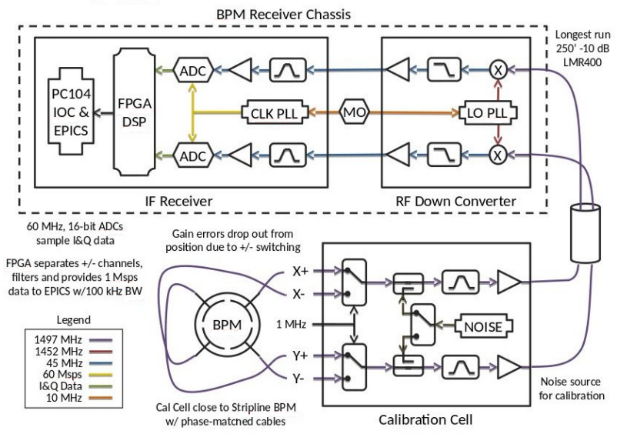}
\includegraphics[width=3.4in]{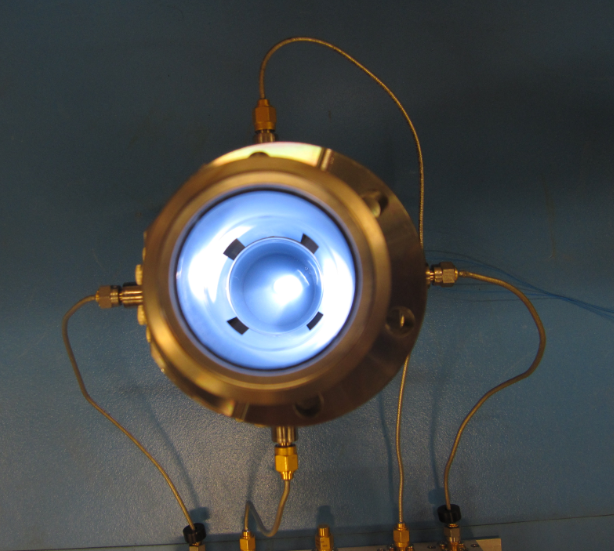}
\caption{Schematic view of the digital Diagnostic Receiver (DR) and an image of a stripline beam position monitor developed for the Hall D beamline.. \label{msdiagf3}} 
\end{figure}

Second, as shown in Fig.~\ref{msdiagf3}, a new Diagnostics Receiver (DR) and Stripline BPM have been developed to meet the low beam current specification of less than 1 \textmu A up to 5 \textmu A for the Hall D beamline. The three main blocks of the system are: (1) a calibration cell that includes a multiplexer to switch between pairs of wires at 1 MHz, a pre-amplifier block to amplify the 1497 MHz signals before sending them to the service building, and a noise source to calibrate the system; (2) an RF downconverter, filter, and amplifier block to lower the signal frequency to 45 MHz and condition it for the next stage; and (3) a digital IF section, which filters, samples, de-multiplexes, demodulates, and performs CEBAF-specific functions related to beam delivery, including a PC-104-based IOC tied to the EPICS control system. The 1452 (= 1499 - 45) MHz LO is tied to the 10 MHz CEBAF Master Oscillator. 

The new stripline BPM is a precision-machined component that registers the pickups more accurately as compared to the M15 and M20 antenna designs. Twenty-six of these devices and their diagnostics receiver (DR) chassis are installed in the Hall D beamline. The system is capable of providing position readbacks with good signal-to-noise ratio starting around 10 nA. The Hall D physics program requires beam currents below this threshold for production running as well as for calibration of their total absorption counter (TAC) at a few nanoamps of beam current. For these conditions two nA-style BPMs, described in the next section, have been implemented. 

\subsubsection{Nano-Amp BPMs} 

A system of low intensity beam position and current monitors capable of operating in the current range of 1~nA – 1000~nA was developed in 1996 for Hall B \cite{MSdiagr4}. Initial attempts to extend the dynamic range of the stripline BPM system by a factor of thirty for this application were unsuccessful due to limitations in the electronics architecture. A cavity BPM design was adopted that is conceptually similar to the RF Separator design described earlier. 

Each nA BPM system consists of a pair of position sensitive pillbox cavities with field perturbing rods operating in a dipole mode oriented for horizontal and vertical measurement and a simple current sensitive pillbox cavity to normalize the signals from the position sensitive cavity pair. The requirements for the system are listed in Table~\ref{msdiagtab4}. The mechanical design for the 1497 MHz position sensitive cavities is shown in Fig.~\ref{msdiagf5}. The cavity is coarsely tuned by adjusting the end plate spacing and a plunger system is then used for fine-tuning the input coupling. Each cavity has an output coupling loop and a test probe for independently measuring field strength. The four rods are spaced 3 cm apart transverse to the beam axis. The current-sensitive cavities are a similar design without the field-perturbing rods. The three-cavity system is installed as a unit in a temperature-stabilized enclosure as shown in Fig.~\ref{msdiagf5} (lid and thermal blanket not shown).  

 \begin{table}[b]%The best place to locate the table environment is directly after its first reference in text
\caption{\label{msdiagtab4}%
nA BPM Cavity Specifications.}
\begin{ruledtabular}
\begin{tabular}{ll}
Parameter&Specification\\
\colrule
Operating Range&1 nA - 1000 nA\\
BPM Resolution&70 pV/m at 1 nA \\
Position Measuring Range&$|x|$, $|y|$ $\le$ 5 mm\\
Resonant Frequency&1497 MHz\\
Loaded {\it Q}&3500\\
Beam Line Aperture&3.0 cm\\
Diameter&19.0 cm\\
Depth&9.5 cm\\
Rod Gap&3.0 cm\\
Material&Copper Plated Stainless\\
\end{tabular}
\end{ruledtabular}
\end{table}

\begin{figure}
\centering
\includegraphics[width=3.4in]{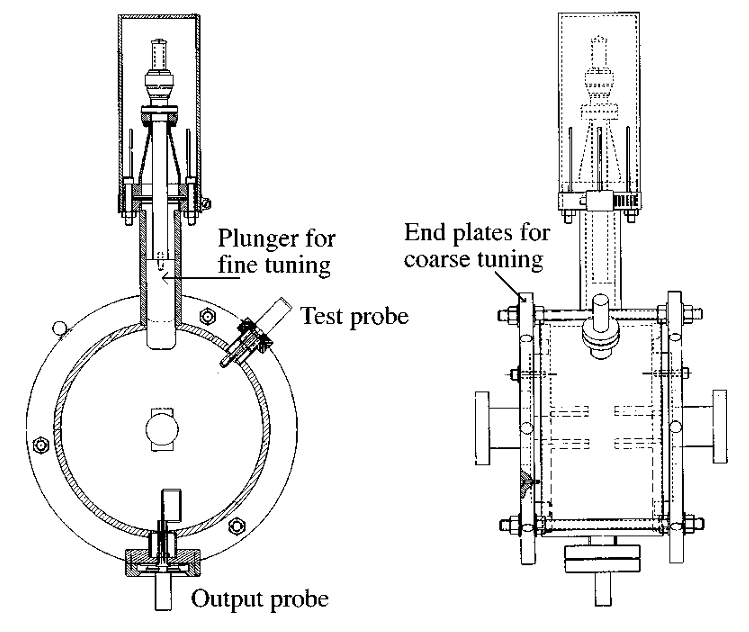}
\caption{Schematic of the nA BPM Position Cavity with 3 cm rod gap. . \label{msdiagf4}} 
\end{figure}

\begin{figure}
\centering
\includegraphics[width=3.4in]{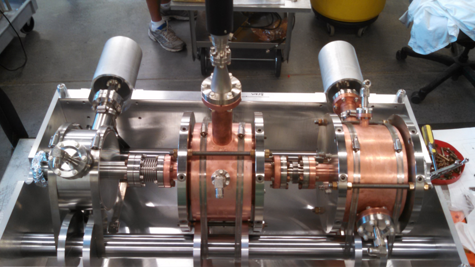}
\caption{3-cavity nA BPM system mounted in its thermal enclosure. \label{msdiagf5}} 
\end{figure}

In the electronics the 1497 MHz signal from each cavity is amplified, mixed with a 1497.1 MHz local oscillator (LO) to down convert to 100 kHz and then processed by a lock-in amplifier. The 100 kHz reference, LO and 1497~MHz test signal are all generated in an external reference module, which is tied to the CEBAF Master Oscillator. 

For the 12 GeV Upgrade, two of these nA BPM systems are installed at the top of the ramp in Hall D, upstream of the photon radiator and Tagger Magnet to optimize the position and angle of the electron beam before hitting the photon radiator. 

\subsubsection{Wire Scanners} 

The use of wire scanners to measure transverse beam size and absolute position, and thereby to infer beam emittances, energy, and energy spread, has been in place at CEBAF since the initial operations \cite{MSdiagr5,MSdiagr6}. The majority of the early systems in the main accelerator used CAMAC architecture for controlling stepper motors and reading back beam induced wire signals. During the 6 GeV era, the controls began migrating towards VME solutions, retiring the CAMAC systems with some remaining at the start of the 12 GeV Upgrade. 

Wire scanners in the main accelerator are primarily used to match the beam to the design optics at three different energies in the Injector, in the matching section preceding each recirculation arc, and at the entrance of each Hall beamline as described in Chapter~\ref{beamdelref}. In addition wire scanners are used upstream of targets in the halls to optimize the beam spot size and convergence for the experiments. 

\begin{figure}
\centering
\includegraphics[width=3.4in]{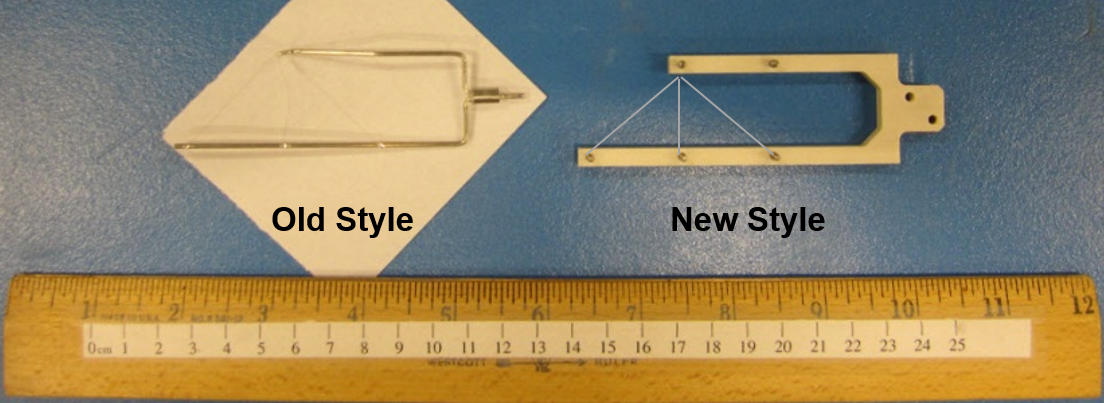}
\caption{Wire scanner with metal fork and glued wires and the upgraded polymer fork using retaining screws for stability. \label{msdiagf6}} 
\end{figure}

The 12 GeV Upgrade for wire scanners consisted of three parts: 
(1) Developing a more robust fork assembly that uses retaining screws as opposed to gluing the wires to a metal frame [Fig.~\ref{msdiagf6}], (2) building and installing 6 new wire scanner assemblies, (3) developing an upgraded electronics package for Hall D and propagating that solution around the accelerator to replace existing CAMAC and VME systems \cite{MSdiagr7}. The new assemblies were deployed to match the beam at the entrance to
Arc~10, measure the spot size and position at the end of the northeast spreader beamline for Hall D, match the beam at the entrance of the Hall D ramp. Two additional systems are deployed for matching the beam at the top of the Hall D ramp and for projecting the convergence of the photon beam onto the physics target in the Hall and for measuring the spot size at the Hall D beam dump.

The new electronics chassis is capable of driving up to eight wire scanners and includes stepper motor drivers, power for pre-amplifiers and data acquisition for wire signals from either digital encoders or analog linear potentiometers. The FPGA based main controller board includes a PC/104 computer running EPICS as a local IOC.

\subsubsection{Viewers} 

\begin{figure}
\centering
\includegraphics[width=3.4in]{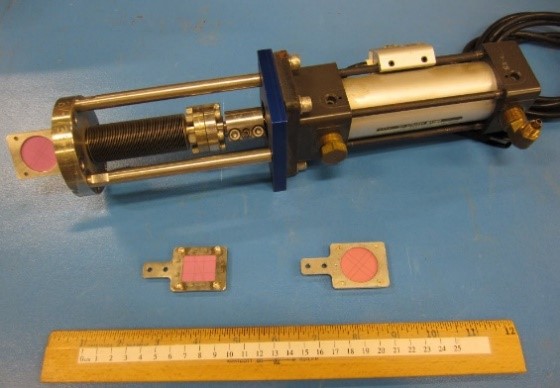}
\caption{Pneumatic viewer assembly and an extraction and standard viewer flag. \label{msdiagf8}} 
\end{figure}

The CEBAF beam viewers are fluorescent screens that emit optical light, which is then focused onto CCD cameras. Beginning in the 6 GeV era the material used was Chromox-6, an alumina doped ceramic material from Morgan Technical Ceramics. Prior to the upgrade, there were ~125 viewers in CEBAF. All but the extraction viewers are 28.45 mm diameter x 0.25 mm thick discs mounted to a 25.4 mm diameter frame. In each linac, six of these were installed on every fourth girder between cryomodules. They were modified to have a hole in the center used for threading lower pass beam to allow imaging of the next pass on the viewers. The extraction viewers were shown in Figure~\ref{SpataFig5} of Chapter~\ref{beamdiagref}. Shown in Figure~\ref{msdiagf8} is a standard pneumatic viewer assembly and viewer flags. 

For the 12 GeV Upgrade 17 new viewers have been installed as follows: 
one extraction-style viewer in front of the 5th pass YA magnets, five viewers at zero dispersion points in the new Arc 10 beamline, one at the Arc 10 recombiner, two in the northeast Spreader Hall D beamline, five in the new Hall D beamline, one at the Hall D beam dump, and finally, two in the transport recombiner after the fifth pass.

\begin{figure}
\centering
\includegraphics[width=3.4in]{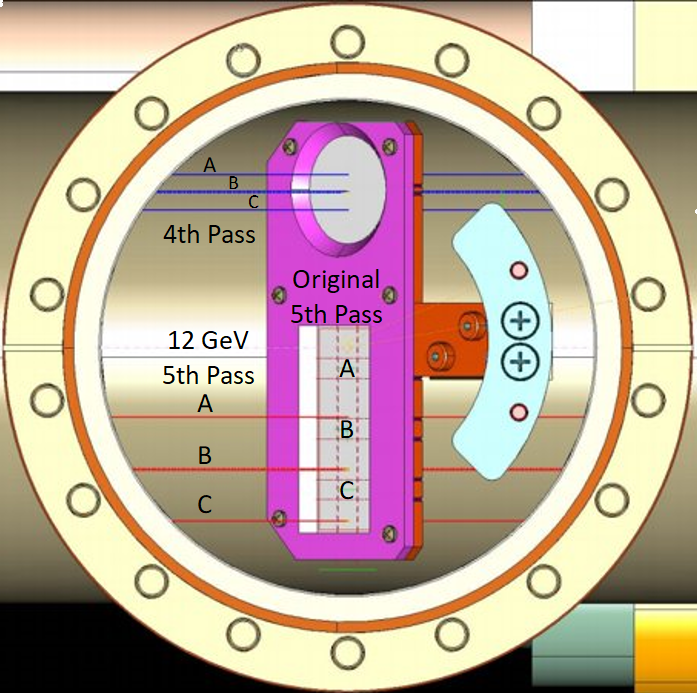}
\caption{6 GeV Transport Recombiner viewer diagram with shifted 12 GeV trajectories shown for fifth pass. \label{msdiagf9}} 
\end{figure}

The 6 GeV spreader and recombiner design used pairs of 2-meter YR septa magnets to separate 4th and 5th pass beams. In the transport recombiner a viewer system was designed to show both 4th and 5th pass beam trajectories between the septa pair. The 12 GeV design requires 3-meter ZA septa with a larger separation between 4th and 5th pass trajectories. Figure~\ref{msdiagf9} shows the original viewer flag with nominal beam positions and the new design trajectories for 5th pass. The system has been modified for the 12 GeV beam paths.

\subsubsection{Synchrotron Light Monitors} 

Since 1986, there were plans to use synchrotron light monitors in CEBAF \cite{MSdiagr8}. Throughout the 6 GeV era, we have routinely used these devices to monitor energy stability, energy spread, and intrinsic spot size at multiple locations. Leading into the 12 GeV Upgrade there were three synchrotron light monitors in the CEBAF main accelerator that were in routine use; at the exit of the injector chicane dipole 0R02, at the high dispersion point in Arc 1 after dipole 1A09 and at the high dispersion point in Arc 2 after dipole 2A09. The dispersion at these locations was 1.26 m for the Injector location and 6.5 m in the Arcs. 

Using synchrotron light monitors to measure Twiss parameters was also considered in the early days of CEBAF \cite{MSdiagr10} and realized with the installation of systems at four homologous points in Arc 7 \cite{MSdiagr11}.

\begin{figure}
\centering
\includegraphics[width=3.4in]{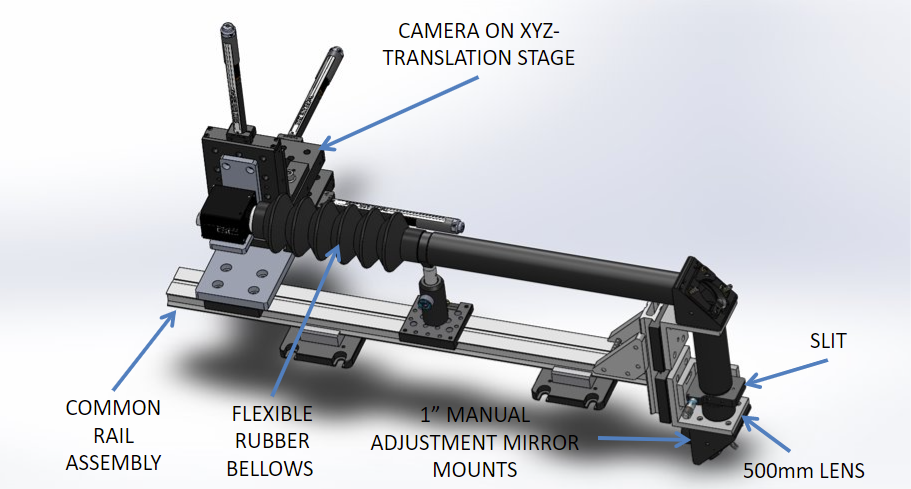}
\includegraphics[width=3.4in]{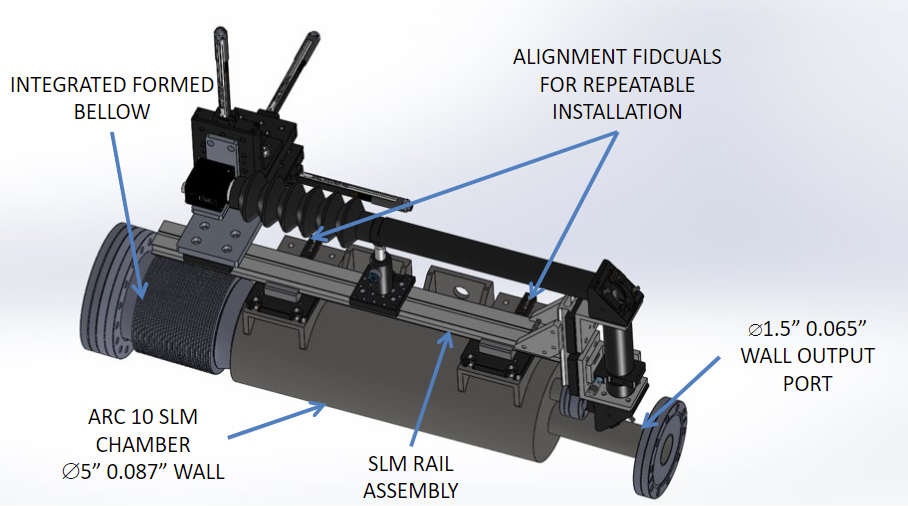}
\caption{Modular SLM assembly on rail system and mounted to the Arc 10 Vacuum Chamber. \label{msdiagf10}} 
\end{figure}

These systems worked reasonably well but the 12 GeV Upgrade provides an opportunity to develop a new modular design that could drop into any location around the accelerator. The images in Fig.~\ref{msdiagf10} capture the details for the new design. The new system is lightweight, low cost, rugged, easy to fiducialize offline and then align in the tunnel, and can be installed without modifying any existing dipole vacuum chambers \cite{MSdiagr9}.

For the 12 GeV Upgrade three new synchrotron light monitor assemblies are installed in Arc 10 at the exit of the 4 m dipole, in the northeast spreader at the exit of MXLBS08, and after the second vertical-bend dipoles at the bottom of the ramp into Hall D. The same design replaces the three existing 6 GeV systems. All of these devices are in dispersive locations and provide energy stability and energy spread information.

\subsection{Machine Protection System}

The Machine Protection System (MPS) is designed to protect beam line and beam line components from damage
due to beam strike events or other equipment failure, which may result in costly damage to machine
components or radio-activation. The two main components of the MPS are the MPS Beam Containment system
and the Fast Shutdown (FSD) system. The MPS Beam Containment system detects
both acute and chronic beam loss events and is made up of a collection of beam monitoring devices
strategically placed around the machine. The FSD is a network collecting and evaluating beam loss
monitoring signals and
triggering fast beam shutdown devices to terminate the beam.

\subsubsection{Beam Containment System}

The main function of the Beam Containment system is to prevent beam damage to machine components
due to acute beam loss. Secondarily, the system supports minimizing radioactivation of
beamline components by reducing low-levels of beam loss.
The beam containment system employs a variety of detectors and subsystems to monitor and react to
beam loss around the CEBAF facility. These include: (1) the Beam Loss Accounting (BLA) System, (2) Machine Protection
Beam Loss Monitors (MPS-BLM), (3) Diagnostic Beam Loss Monitors (DIAG-BLM), and
(4) Beam Loss Ion Chambers (BLIC).
It is important to note that the Beam Containment System is layered.
The BLA system is able to detect a gross beam loss greater than 2 \textmu A, while
BLMs protect the accelerator beam line and its components from low-level beam loss in the range from
10 nA to 2 \textmu A.

\it{Beam Loss Accounting}\rm:
In the BLA system, the average current out of the injector is measured by a RF cavity
current monitor, and compared to the current measured similarly in each of the experiment
Halls. When the summed total current
measured in the Halls is 2 \textmu A less than the injector measurement
the beam is shut down.
The original design of the BLA system was suitable to the requirements for 12 GeV operation, and only
needed expanding to cover the new Hall D point of beam delivery.

\begin{figure}
\centering
\includegraphics[width=3.4in]{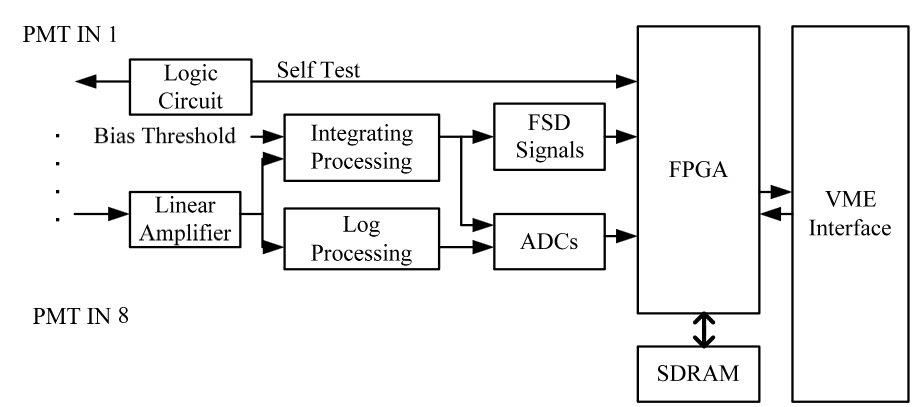}
\caption{The system block diagram of the BLM board. (From \cite{JKRef1})\label{fig:figjk1}} 
\end{figure}

\it{Beam Loss Monitors}\rm:
Beam loss is detected throughout the CEBAF site by promptly
detecting the radiation generated by a beam strike using
photomultiplier tubes.
The addition of new machine segments and the increase of beam energy triggered extensive review of
the Beam Loss Monitor network. As a result of the analysis, the location and quantity of both
FSD-interlocked BLM tubes and beam diagnostic BLMs changed around the machine.
The count of FSD-interlocked BLMs
increased from 45 to 70, while the count of diagnostic BLMs decreased from 111 to 94. The overall
machine coverage provided by the BLMs significantly improved through this effort. At the same time newly
redesigned BLM cards were installed that added flexibility in machine protection configuration and
diagnostic capabilities \cite{JKRef1}. In developing the new BLM hardware Jefferson Lab took the opportunity to
migrate to a VME-based system using FPGAs as shown in Figure~\ref{fig:figjk1}.

\it{Beam Loss Ion Chambers}\rm:
Ion Chambers are used to protect areas with a high ambient radiation environment like High Power
Dumps and Target Systems. While there were no significant changes in the type and number of Ion
Chambers due to 12 GeV upgrade, it is worth noting the use of Ion Chambers for protecting the High
Power Dump Diffusers in Halls A and C.
In each location there are two dedicated Ion Chambers monitoring the backscatter of radiation
from the High Power Dump Diffusers. With the help of specialized FSD cards, these Ion Chambers
cause the FSD system to trip if radiation levels are lower than expected
as this condition indicates burn-through of the dump's diffusers and turning
the beam off protects the dump from catastrophic
failure. This mode is the opposite of the ``trip high'' condition normally associated with Ion Chambers.

\it{Beam Envelope Limit System}\rm:
Beam Envelope Limit System (BELS) is a high reliability PLC based system \cite{JKRef2}, which ensures that
CEBAF runs within accelerator operations and safety power limits. For 12 GeV CEBAF the Operations Envelope 
(the maximum beam power in normal operations), is 1.1 MW and the Safety Envelope (the maximum power which
if exceeded causes significant administrative burden), is 1.3 MW.
This system provides a tiered reaction to exceeding 1100 kW utilizing different beam shutdown methods for each:
(1) Operator warning after 1 minute, (2) Control system shutdown after 5 minutes, (3) MPS shutdown after 10 minutes,
and finally, (4) the personnel safety system will terminate the beam after 15 minutes.

The calculated total CEBAF power is the sum of power delivered to individual beam
destination segments as calculated based on the beam energy and actual beam current delivered to these segments.
Since Hall D is a low power beam destination (with an FSD-monitored power limit of 60 kW), as is Hall B (55 kW),
the power delivered to Hall D is not monitored by the BELS system and the addition of Hall D did not trigger
significant modification of the system. The increased beam energy delivered to Halls A and C required only the
modification of BELS software configuration parameters to allow operating at 12 GeV.

\subsubsection{Fast Shutdown System}
The FSD system is a network of electronic cards (nodes) strategically located throughout the CEBAF facility.
The nodes form a tree structure with the cards aggregating input signals and propagating them to the top-level
FSD node in the Injector segment, which controls the beam shut off. When the FSD system is triggered by a
beam loss event the system responds by shutting off the electron beam in less than 50 microseconds.
Control software provides the ability to mask the FSD input signals, allowing for easy and flexible
but reliable FSD system configuration according to changing beam delivery destinations or changed conditions
in the segmented CEBAF structure.

For the 12 GeV upgrade the existing FSD infrastructure has been expanded to integrate new areas but did not
change significantly beyond that. The upgrade did not require any new FSD input types, and response time
of the existing FSD cards and their network is sufficient to meet the beam shutoff requirements with
the higher energy electron beam.

The new digital Low Level RF controls for the C100 cryomodules are designed
to produce summary output FSD signals covering Quench, Arc, IR, and Vacuum (waveguide and beam line) fault detection.
These fiber optic 5 MHz FSD summary signals feed directly from each new RF zone into dedicated FSD input cards
installed respectively in the North and South Linacs. Summary of FSD signals from the Linacs further travels
to the Master FSD card in the Injector segment.

Protection of the beamline and new hardware installed in the Hall D and Hall D Tagger segments
required extension of the FSD system. This new installation utilized VME based FSD cards.
Similarly to other segments, there is a local master FSD card aggregating FSD signals from all
local sources within Hall D, and the aggregated summary is sent to the Master FSD card in the Injector segment.

\subsection{Site Cooling and Power Upgrades}
 
To support the 12 GeV upgrade, new and upgraded cooling and electrical systems are needed
to meet the 12~GeV project requirements. Included are modifying the existing Low Conductivity
Water (LCW) Systems and the CHL Condenser Water System, installing a new Passive Chilled Beam System in
the accelerator tunnel, providing new utilities for the new experimental hall (Hall D), and
extensively upgrading and improving CEBAF's electrical power system. Table~\ref{tab:tablecw}
summarizes the new total design
values after the upgrade.

\begin{table*}[b]%The best place to locate the table environment is directly after its first reference in text
\caption{\label{tab:tablecw}%
CEBAF cooling and power systems requirements after 12 GeV upgrade. gpm stands for gallons per minute water flow rate.
}
\begin{ruledtabular}
\begin{tabular}{lcccc}
Location&LCW System&Condenser Water&Chilled Water&Power - New Unit\\
&(gpm)&(gpm)&(gpm)&Substations\\
\colrule
&&&&3 MVA (4 ea)\\
Accelerator&5600&4900&200&2 MVA (1 ea)\\
&&&&1.5 MVA (1 ea)\\
\colrule
CHL&&3100&&5 MVA (2 ea)\\
\colrule
Hall D&385&1120&600&2 MVA (1 ea)\\
&&&&1 MVA (1 ea)\\
\end{tabular}
\end{ruledtabular}
\end{table*}

There are four Low Conductivity Water (LCW) Systems that were upgraded to support 
the 12 GeV project. The systems have been expanded to provide cooling for new and upgraded magnets,
the additional RF zones, and the additional power supplies.  Table~\ref{tab:tablecw2} shows
the 6 GeV operational
capacities and the new 12 GeV design flow capacity. All LCW systems provide 2 M$\Omega$ water at 95~$^\circ$F.
\begin{table*}[b]%The best place to locate the table environment is directly after its first reference in text
\caption{\label{tab:tablecw2}%
LCW Systems in the 6 GeV and 12 GeV eras
}
\begin{ruledtabular}
\begin{tabular}{lcccc}
Load&West Arc&North Linac&East Arc&South Linac\\
&LCW System&LCW System&LCW System&LCW System\\
&(gpm)&(gpm)&(gpm)&(gmp)\\
\colrule
6 GeV Flow&660&955&453&987\\
\colrule
12 GeV Magnets&212&&34&\\
\colrule
12 GeV RF Zones&&903&&891\\
and Power Supplies&&&&\\
\colrule
Total 12 GeV gpm&&&&\\
Required&872&1858&487&1878\\
\bf{Design Capacity}&\bf{1000}&\bf{2000}&\bf{600}&\bf{2000}\\
\end{tabular}
\end{ruledtabular}
\end{table*}

\begin{figure}[b]
\includegraphics[width=3.4in, angle=-90]{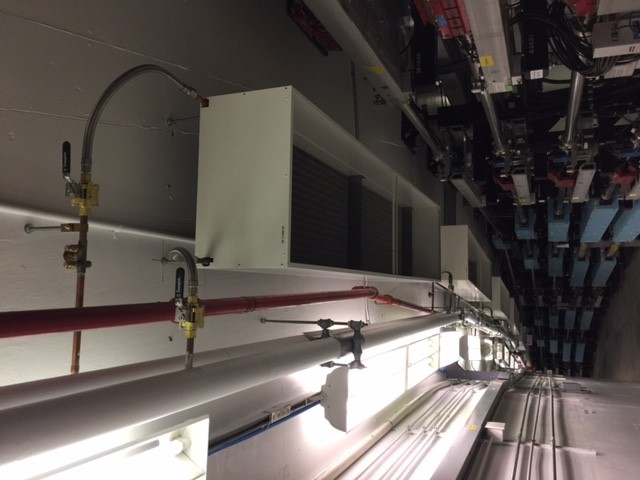}% Here is how to import EPS art
\caption{\label{fig:figcw1} Photograph of chilled beams of the air conditioning system next to the East Arc
magnets
}
\end{figure}

The 12 GeV upgrade requires four times more cooling at the bending magnets than was required
for 6~GeV operations.  The heat generated from these magnets must be removed to allow personnel
access to make equipment repairs within one hour of interrupting magnet operations.
Ninety-five $^\circ$F ambient air temperature is specified; without air conditioning tunnel temperatures
could exceed 135 $^\circ$F. 
The new air conditioner is a natural convection non-condensing cooling system. It consists of chilled water
systems (located above ground) providing chilled water to 132 chilled beams (cooling radiators)
mated with required automated controls, and provides tunnel cooling without producing condensation.
Each chilled beam requires 1.5 gpm of chilled water and is 96''x20''x12'' as shown in Figure~\ref{fig:figcw1}. Each of
the A/C systems for each arc produces 60 tons of aggregrate
cooling and has been in operation for several years. The chilled
water system maintains tunnel air temperature contributing to overall
accelerator stability and worker safety.

\begin{figure}[b]
\includegraphics[width=3.4in]{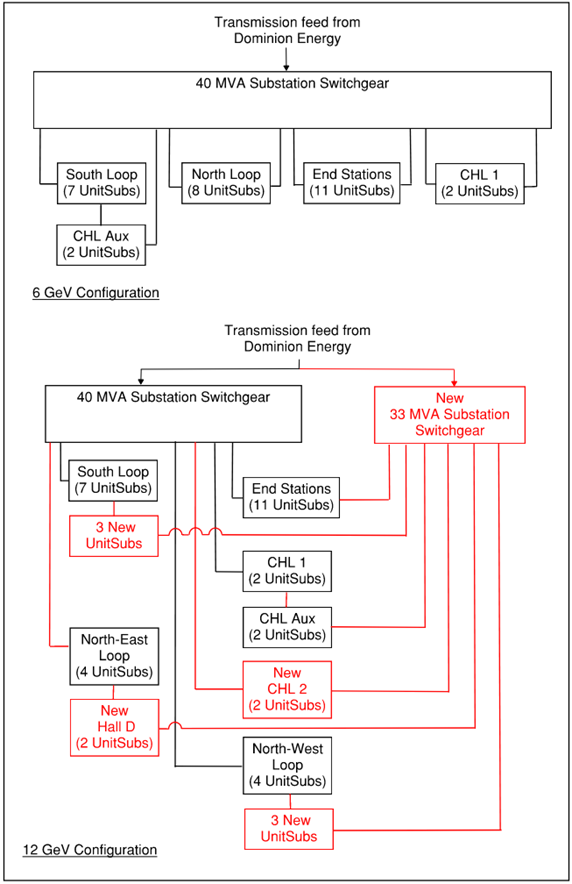}% Here is how to import EPS art
\caption{\label{fig:figcw2} Electrical Power Distribution comparing 6 GeV and 12 GeV CEBAF}
\end{figure}

Figure~\ref{fig:figcw2} shows a one line diagram of the electrical distribution system at 6 GeV (shown in black) with
the modifications for 12 GeV shown in red.  An additional 33 MVA substation switchgear was added to the
existing Dominion Energy (formerly Dominion Virginia Power) overhead transmission line feed to accommodate
the additional power requirements for the 12 GeV upgrade as well as to provide a more reliable and robust
15 kV distribution system.  During the 6 GeV era, all the power for the accelerator site was fed from
the 40 MVA Primary Substation Switchgear through four 15 kV loops; the South Loop, the North Loop, the CHL Loop,
and the End-Station Loop.  The new 12 GeV site power distribution included adding six new unit 
substations to the existing North and South loops. The north loop was split into a northeast and northwest
loop with the two new unit substations for the Hall D complex added to the north east loop. The CHL Aux unit
substations were moved to the CHL1 loop, and the CHL2 loop was created with the 2 new 5 MVA unit substations.

Additional electrical power was added for the new high power amplifiers (HPA's) and the
additional RF zones in each of the North and South Linac Buildings as outlined in Table~\ref{tab:tablecw}.
A new 1.5 MVA unit substation was provided at the east end of the North Linac Building to
account for this power need. A new 2 MVA unit substation was provided at the west end of the
South Linac Building to meet the additional power needs as well as providing additional box
power supply power in the supporting west arc service building (W2) with a standard operating
headroom of 30\%.  New feeders run from each unit substation to a new switchboard in both the
North and South Linac Buildings.

All magnet power supplies were re-fed to the newly installed unit substations and indoor
switchboards. The existing switchboards that previously supported the power supplies
during 6 GeV operations were rewired and used to power the upgraded LCW equipment, cooling
towers, and chilled water systems.

\section{Operating CEBAF at 12 GeV}

In this major section, we summarize three significant aspects of 12 GeV CEBAF project performance: results for the newly installed cavities; effects of particulate movement around the CEBAF accelerator and its results on cavity performance; and recent enhancements of operations procedures and software that have led to significantly improved accelerator tuning and overall reliability.

\subsection{Installed C100 Cavity Performance}

After each C100 cryomodule was installed in
CEBAF it was commissioned. SRF
Commissioning consists of a set of tests designed to quantify the
performance aspects of the cavities that are most important in an
operational setting.  Commissioning tests are focused on determining
maximum stable operating gradients and measuring field emission, dynamic
heat loads ($Q_0$), and microphonics.  

\subsubsection{Determining the Maximum Gradient}\label{SRFCOM}
The 12 GeV specification states that a C100 cryomodule must be capable
of delivering a stable energy gain of 108 MV.  Therefore, each cavity
in a C100 cryomodule cavity must deliver, on average, a usable
gradient of at least 19.2 MV/m.  The first step in the commissioning
process, once cavities have been mechanically tuned, is to determine
the highest stable gradient available from each cavity.

The first step in gradient determination is to quantify the RF cable
losses in order to calibrate RF power levels.  Then, while running
pulsed RF into the cavity, the gradient is calculated from the emitted
power.  The loaded $Q$ ($Q_L$) is calculated at this time as well.  The
field probe calibration is then set so that gradient as calculated
from the field probe power level is equal to the gradient as
calculated from the emitted power.  From this point on, the gradient
derived from field probe power is used as the relevant gradient
measure.

Once the gradient is calibrated, pulsed RF power is increased in small
steps.  C100 cavities will frequently go through a series of
non-repeating quenches as the gradient is increased.  The process
continues until the cavity reaches a limiting condition.

Potential gradient limitations include quenching, high dynamic heat
loads, warm RF window temperatures, vacuum degradation in either the
beamline or the waveguide guard vacuums, arcing in the guard vacuum,
or finally the administrative limit of 25 MV/m.  For the majority of C100
cavities, the final limitation is a repeatable quench.  Most of the
remaining cavities will be limited by RF heat load or by the
administrative limit of 25 MV/m.  This administrative limit is meant
to protect the cavities from new field emitter creation.  The current
controls and available RF power would limit normal operation of these
cavities to gradients lower than 25 MV/m.

\begin{figure}
\centering
\includegraphics[width=3.4in]{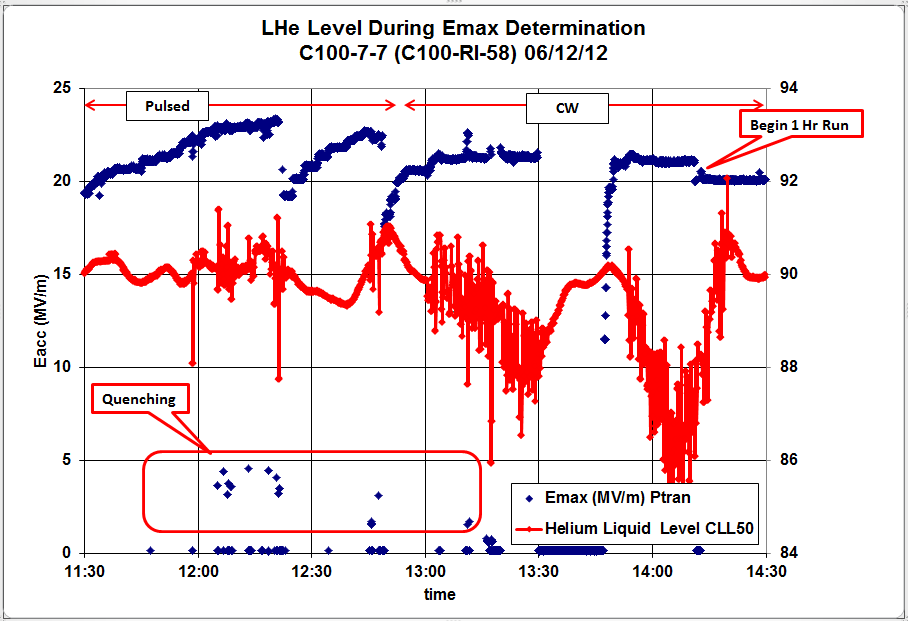}
\caption{Sample Individual Cavity $E_{max}$ Determination (From \cite{MD1}). The blue curves give the cavity gradient and the red curves give the liquid helium level during the test. \label{fig:MD_C100_fig1}} 
\end{figure}

Once the maximum gradient is defined, the limit is then tested using
CW RF.  When the absolute maximum gradient is known, the next step is
to determine the maximum stable operating gradient.  The maximum stable operating gradient is found by
lowering the gradient below the maximum just enough to avoid fault
conditions over the course of running the cavity CW at least an hour.  This
procedure provides an opportunity for the helium circuit and the beamline and
waveguide guard vacuums to settle.  Figure~\ref{fig:MD_C100_fig1}
shows the process of 
raising the gradient to determine the maximum gradient.  In this
example, the gradient has already been increased to roughly 20 MV/m.
Over the next several hours, the cavity is pushed through a series of
quenches in pulsed mode until a final maximum of approximately 23 MV/m
is reached.  Then, while running CW RF, it is determined that the RF
heat load is too high above 20 MV/m and the gradient is lowered to a
point where it will run stably.  The red trace shows the 2K helium
liquid level.  Periods where the RF heat load exceeds the capabilities
of the helium vessel's plumbing show large oscillations in the liquid
level.  During the last 15 minutes shown in the figure, the gradient
has been turned down enough that the helium bath begins to stabilize
and the one hour run begins at 20.1 MV/m.

%insert figure 1 here

\begin{figure}
\centering
\includegraphics[width=3.4in]{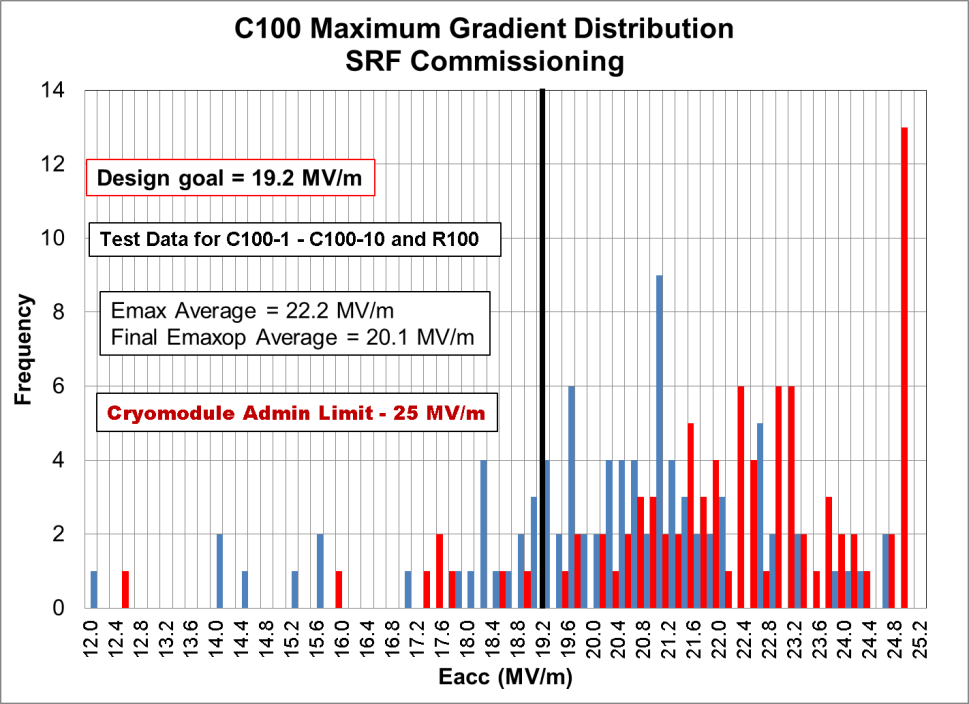}
\caption{Maximum operating gradients $E_{maxop}$ both before (blue) and after (red) the optimization procedure.  \label{fig:MD_C100_fig2}} 
\end{figure}

Finally, once $Q_0$ measurements of all the cavities in a cryomodule have been completed and the static and
dynamic heat loads are known, a further optimization of gradients is
completed.  The optimization takes into account the one hour run gradients along
with heat load information and provides a set of gradients that allow
for all eight cavities to operate at the highest stable gradients,
while staying within the dynamic heat load budget of 240 W.
Figure~\ref{fig:MD_C100_fig2} 
shows the distribution of the absolute maximum gradients (red) and compares
that with the distribution of final maximum operating gradients
($E_{maxop}$) after the optimization is completed (blue). The final
maximum operating gradients are entered into machine operations software as a maximum operating gradient permitted for that superconducting cavity.

%insert figure 2 here

\subsubsection{Field Emission}
\begin{figure}
\centering
\includegraphics[width=3.4in]{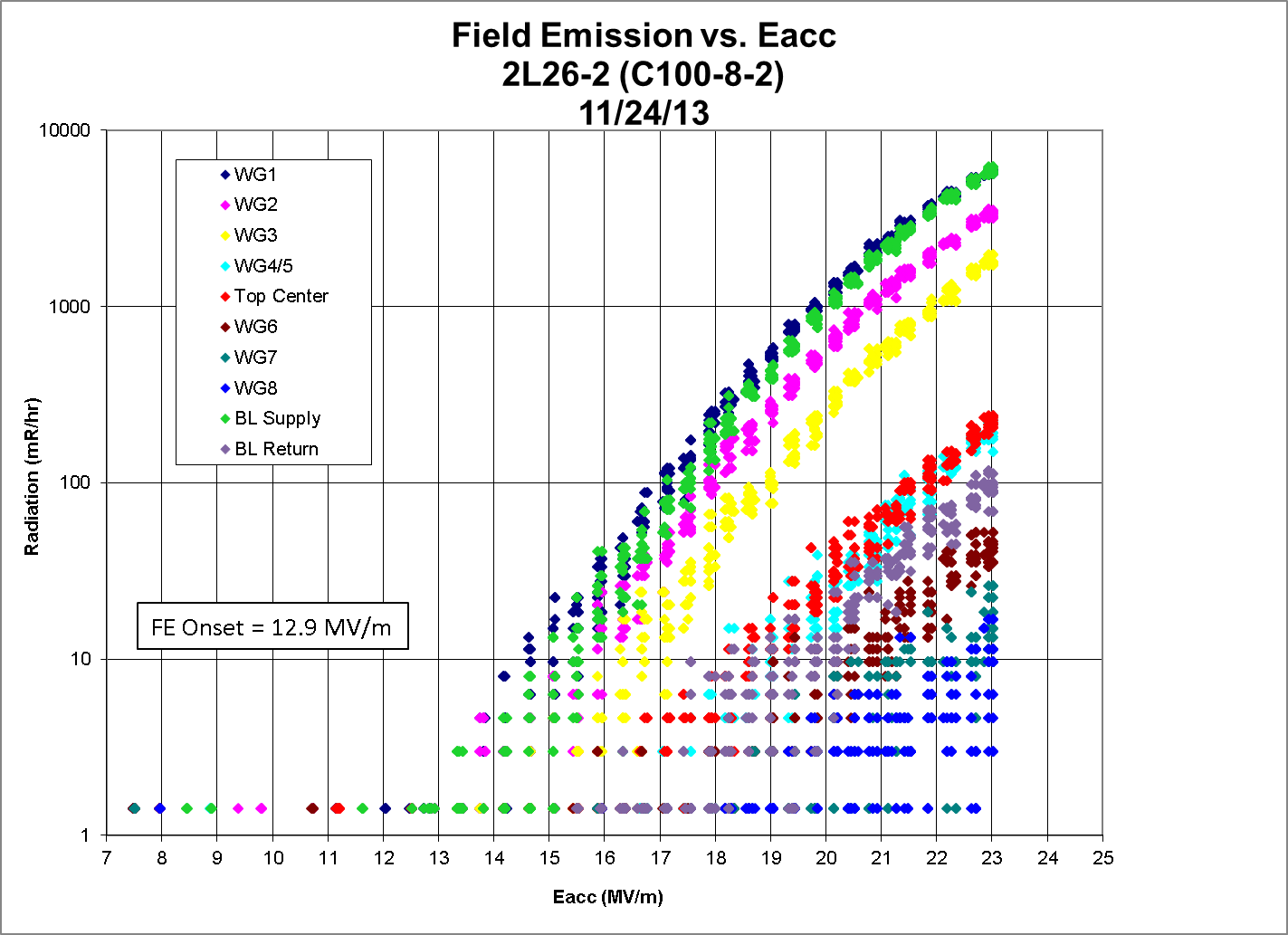}
\caption{Measured Field Emission Count Rate as a Function of Accelerating Gradient for a typical cavity. \label{fig:MD_C100_fig3}} 
\end{figure}

After the $E_{maxop}$ extended run is completed, measurements of x-rays
produced by field emission as a function of gradient are made.  A set
of 10 Geiger---Mueller (GM) tubes are placed on the cryomodule at
several locations, including the beamline at either end of the
cryomodule, and at the Fundamental Power Couplers (FPC's) \cite{MD1}.
Figure~\ref{fig:MD_C100_fig3}  shows a set of measurements for a
typical cavity.    

%insert Figure 3 here

\begin{figure}
\centering
\includegraphics[width=3.4in]{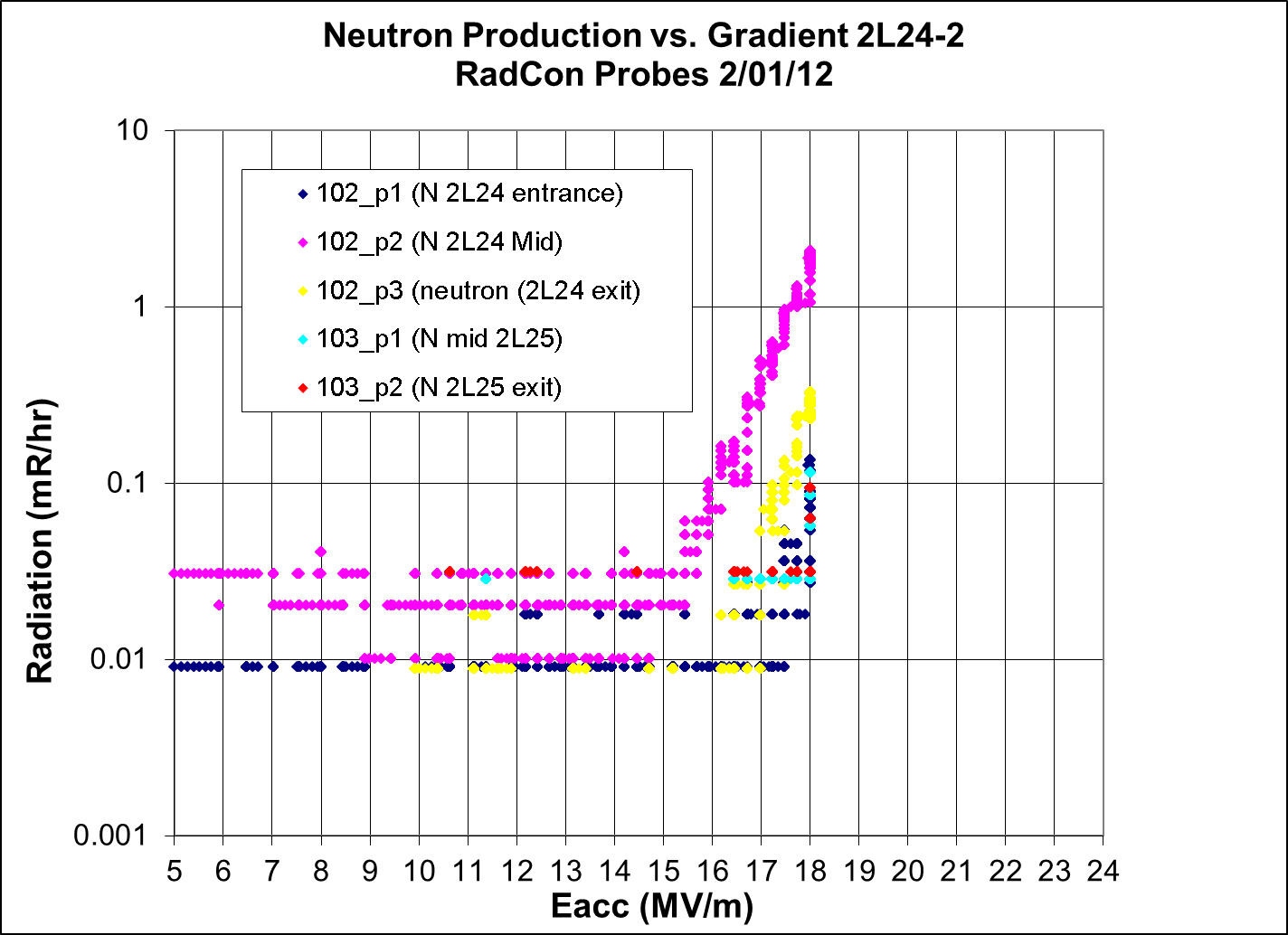}
\caption{Measured Neutron Production as a Function of Accelerating Gradient for a typical cavity \label{fig:MD_C100_fig4}} 
\end{figure}

Neutron production was measured during commissioning of the first two
cryomodules that were installed.  This, however, has not been a
routine measurement on all of the C100 style cryomodules as the
necessary instrumentation was not always available.
Figure~\ref{fig:MD_C100_fig4} shows an 
example of neutron production.  

% insert figure 4 here

\begin{figure}
\centering
\includegraphics[width=3.4in]{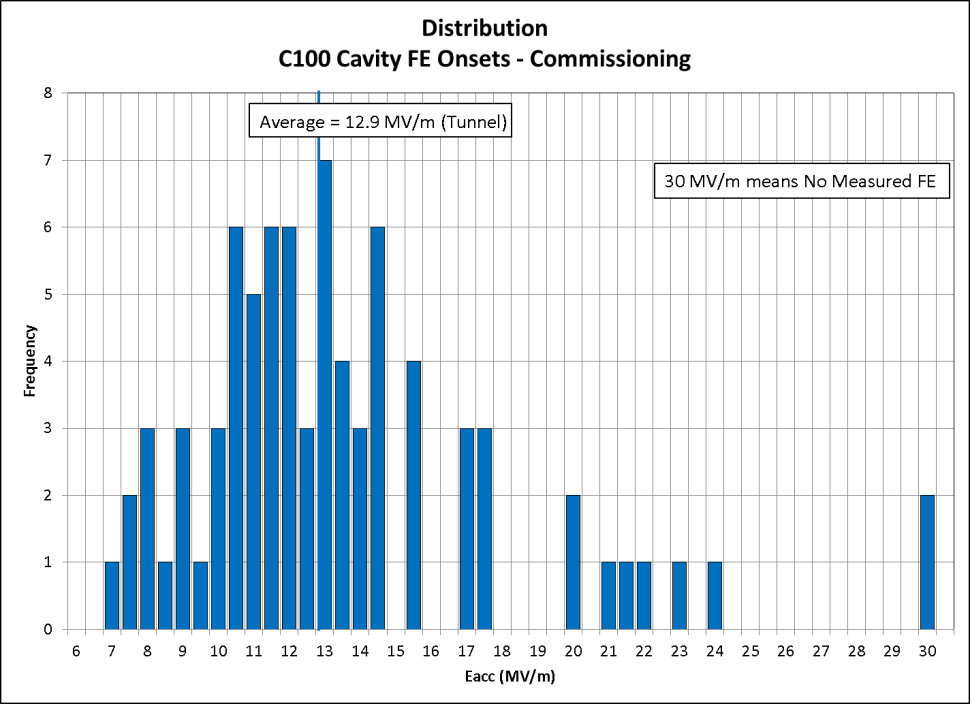}
\caption{Field Emission Onset Gradients for 79 Cavities in the Initial Complement of C100 Cryomodules, as Determined by Measured Radiation at a Level of 1 mR/hour. \label{fig:MD_C100_fig5}} 
\end{figure}

Figure~\ref{fig:MD_C100_fig5} shows the distribution of field
emission onset 
gradients for the C100 cryomodules as measured during initial commissioning.
The average across all of the C100's and the R100 was 12.9 MV/m.  The
detector resolution of the system in use to measure the radiation meant that the criterion for onset gradient would be defined as the lowest gradient at which
any of the channels measured a value of about 1 mR/hr. 

% insert figure 5 here

\subsubsection{$Q_0$ and Heat Load}
After the maximum gradients of stable operation for the individual cavities have been established, $Q_0$s are measured.  The $Q_0$s are calculated from a calorimetric measurement of
the power dissipated by the cavity into the helium bath.  This is
accomplished by isolating the cryomodule from the helium transfer
lines and measuring the rate of rise of helium pressure with RF off,
with a known heater power, and finally with RF on.  This method can
resolve power dissipation as low as 1 Watt \cite{MD1}.

\begin{figure}
\centering
\includegraphics[width=3.4in]{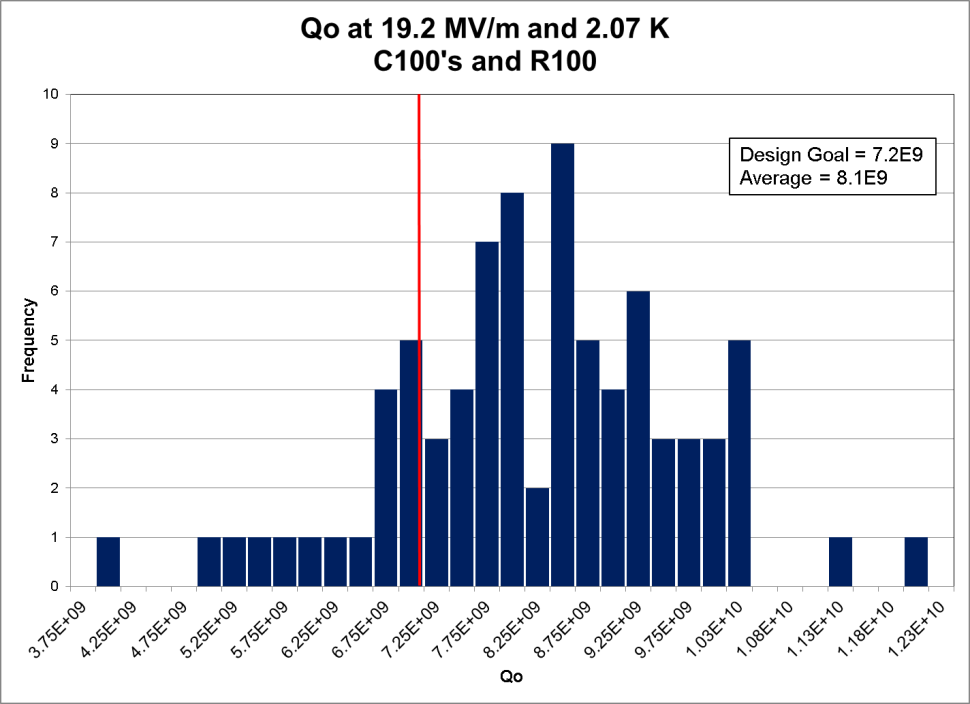}
\caption{$Q_0$ distribution for each C100 cavity at the lesser of $E_{maxop}$ or 19.2\,MV/m. The design goal was $7.2\times 10^9$ and the average was $8.1\times 10^9$.\label{fig:MD_C100_fig6}}
\end{figure}

Figure~\ref{fig:MD_C100_fig6} shows the distribution of measured $Q_0$ values at 19.2\,MV/m for
all C100's and the R100.  Roughly 25\% of the cavities could not be
measured at 19.2\,MV/m due to
 gradient limitations. The average over all the cavities clearly exceeds the 12 GeV project requirements.

% insert figure 6 here

\begin{figure}
\centering
\includegraphics[width=3.4in]{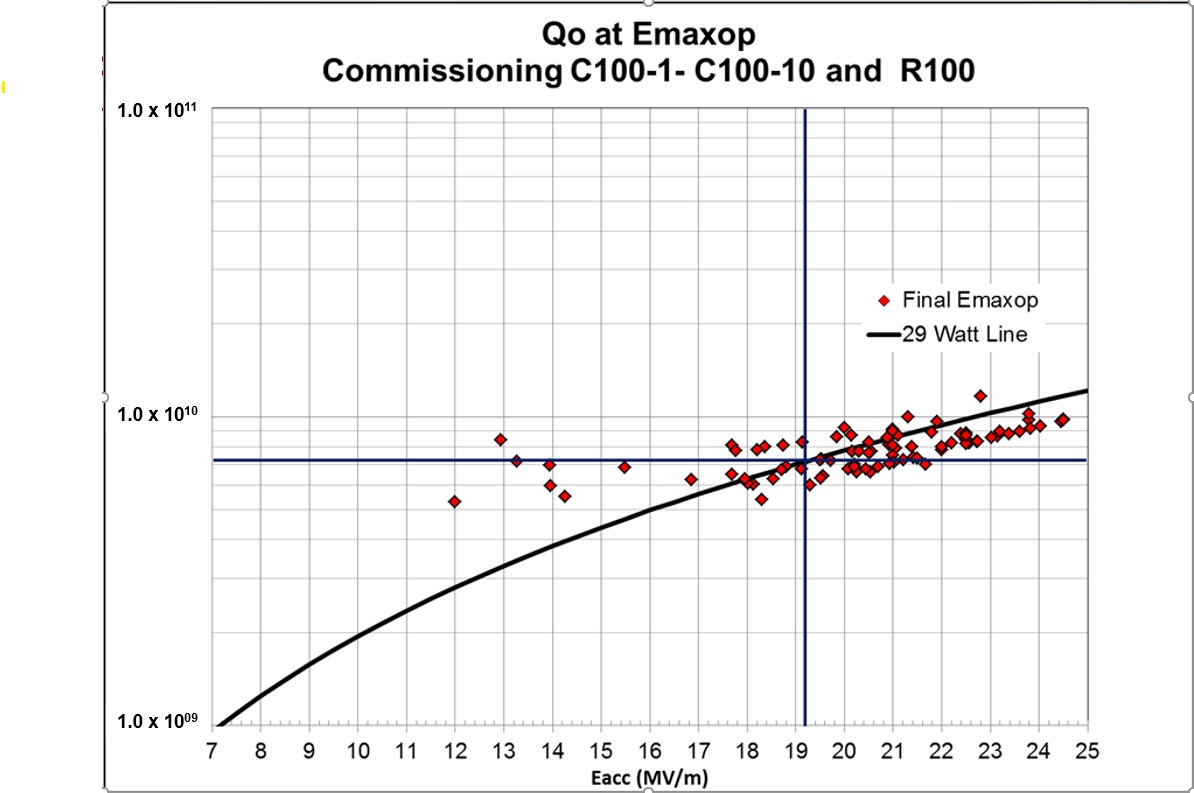}
\caption{C100 cavity $Q_0$ at maximum operating gradient. The vertical line is the upgrade project specification. \label{fig:MD_C100_fig7}} 
\end{figure}

After the $Q_0$ vs. $E_{acc}$ data has been measured for all eight
cavities in 
a cryomodule, an optimal set of maximum gradients can be defined that
take into consideration the extended run gradients and the heat loads
measured at various gradients.  This optimum is calculated within a constrained maximum
allowable heat load per cavity of 35 W and a total heat load for all
eight cavities of 240 W.  Figure~\ref{fig:MD_C100_fig7} shows the $Q_0$
for each cavity at the 
final $E_{maxop}$ gradient.  The black curve on this graph denotes the $Q_0$
that is equivalent to 29 W of dynamic heat load across a range of
gradients.  The crossed lines indicate the gradient and $Q_0$
specifications. 

%insert figure 7 here

A final step in the maximum gradient determination is to turn on all
eight cavities at the final $E_{maxop}$ gradients.  The cavities are then
run for at least an hour in this configuration.  Should the heat load be
too high, a run of more than a few minutes is not possible.
Table~\ref{tab:MD_C100_1} shows the integrated voltages at which these
eight cavity runs 
were accomplished for each new cryomodule. Only two cryomodules fell slightly short of the 108 MV goal and the average performance is comfortably above the 108\,MeV 12\,GeV project requirement.

%insert table 1 here
\begin{table}
\centering
\begin{tabular}{lc} \\ \hline
Cryomodules   & Voltage \\ 
              & (MV)  \\ \hline
C100-1    &   104 \\
C100-2    &   120  \\
C100-3    &   124  \\
C100-4    &   105  \\
C100-5    &   110 \\
C100-6    &   113 \\
C100-7    &   113  \\
C100-8    &   109  \\
C100-9    &   117  \\
C100-10   &   116 \\
R100      &   116  \\ \hline
Average   &    113     \\\hline
\end{tabular}
\caption{The cumulative operating voltage for each C100, all eight
  cavities per C100 operated simultaneously for at least one hour. 
\label{tab:MD_C100_1}}
\end{table}

\subsubsection{Microphonics and Tuning Sensitivity}
The 12 GeV project ``budgeted'' for 25 Hz peak total detuning (4 Hz
static plus 21~Hz dynamic) based on the available klystron power
(13 kW), the design $Q_{\rm ext}$ for the fundamental power couplers  ($3.2\times10^7$),
and maximum beam load (465 \textmu A) [3]. 

The measurement of cavity detuning due to external vibration sources
and the vibrational modes of the cavity/cryomodule structure is
conducted in both the cryomodule test facility and in the tunnel.  The results of these
measurements tend to be location and environment dependent. 

Microphonics testing of the first unit (C100-1) met design goals
marginally, but results were higher than expected based on prototype
testing.  This unexpected result was due at least in part to the low
loss cell shape used for the C100 cavities.  The cell walls are more
vertical as they approach the iris making them more susceptible to
deflection than the original CEBAF cell shape.  Even though the
detuning due to microphonics was lower than the 12 GeV allowance, a
detailed vibration study was initiated and conducted on the first two
C100 style cryomodules, the R100 and C100-1.  This study led to a
simple modification of the pivot plate in the tuner assembly that
reduced the amount of detuning in later cryomodules by an average of
42\% \cite{DPowers1}.

\begin{figure}
\centering
\includegraphics[width=3.4in]{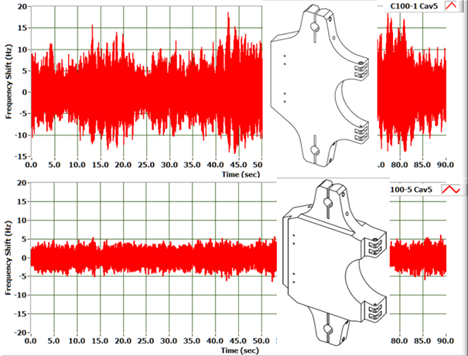}
\caption{90 second microphonic detuning measurement for original and modified pivot plate. Detuning fluctuations are reduced almost 50\%. (From \cite{MD1})\label{fig:MD_C100_fig8}} 
\end{figure}

Figure~\ref{fig:MD_C100_fig8} depicts the frequency shifts due to
microphonics over a 90 
second period in cavities with and without the modified tuner and
shows how the pivot plate was modified.  The cavity with the modified
tuner has had an almost 50\% reduction in detuning.

In addition to reducing sensitivity to microphonics, modifying the tuner assembly led to an average reduction of 35\%
(348 Hz/Torr to 228 Hz/Torr) in the cavity pressure sensitivity
(detuning due to pressure changes).  An average reduction of 25\%, in
the static Lorentz detuning (from -2.16 Hz/(MV/m)$^2$ to -1.62
Hz/(MV/m)$^2$) was measured as well. 

\subsubsection{SRF Commissioning Summary}
Commissioning results show that these cryomodules were able to deliver
an average energy gain of 113 MeV which exceeds the design goal of
108 MV.  The C100 cavities were able to operate at an average maximum
operating gradient of 20.1 MV/m. However, during routine beam operations in the Fall of 2022, nine C100 zones
averaged only 86 MeV, where most had substantial field emission radiation.
Plans for improving C100 operating performance are discussed in Section~\ref{CPP}.

\subsection{Particulate Movement in High-Gradient SRF Linacs}
        
Particulates that have settled on the inner surface of beamline components other than SRF cavities,
such as inter-cryomodule warm sections, pose no harm. However, when they migrate to the RF surface of an SRF cavity
a number of impactful consequences may result. Particulates that have landed on the cavity iris region may
become new field emitters, giving rise to an increased electron field emission at the high required operational
cavity gradient. Some secondary effects induced by enhanced field emission such as rapid beamline
vacuum excursions, frequent charging of components made of insulation materials, and accelerated boiling of
bath liquid helium, have the acute consequence of reducing the linac energy output. Such limits in the
collective operational acceleration by the ensemble of installed cavities, even though individual cavities
are intrinsically capable of higher gradient as demonstrated in their individual qualification testing, need to be avoided.
Other secondary effects such as the field emitted electrons producing
gamma and neutron radiations are chronic effects degrading
and ultimately damaging accelerator components, in turn negatively impacting the operating schedule and
maintenance cost for the accelerator.

\begin{figure}
\centering
\includegraphics[width=0.515\columnwidth]{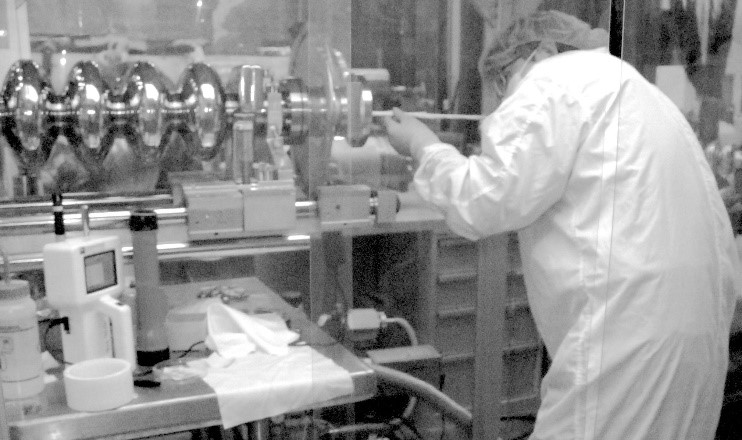}\thinspace\thinspace\includegraphics[width=0.445\columnwidth]{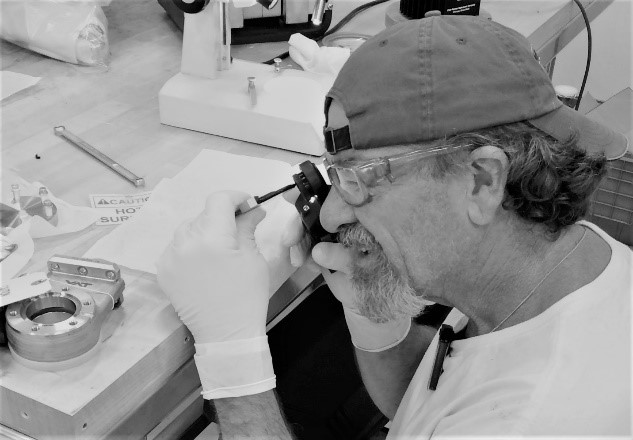}
\caption{Identifying particulate sources. Left: collecting particulates from a 5-cell cavity previously
operated with beam for reconstructing contamination distribution and off-line determination of particulates
sizes and compositions; Right: Inspecting the sealing surface of a beamline gate valve Viton seal aided
with a magnifying glass for large particulates. \label{RLFig1-2}} 
\end{figure}

Particulate movement is currently understood as a driving mechanism behind the apparent loss of energy reach in
CEBAF \cite{RGRef1}. Understanding the controlling variables of particulate movement in accelerator-quality
vacuum with or without CW electron beams is a prerequisite to solving and perhaps ultimately
reversing the slow energy loss problem. Besides, a lasting solution requires the knowledge
of the particulate sources and mechanisms of particulate movement in the entire CEBAF linac beamline systems.
To that end, a fresh effort was started in 2014 \cite{RGRef2,RGRef3,RGRef4,RGRef5,RGRef6}, coinciding the onset of CEBAF 12 GeV era operation, with a 3-pronged strategy: (1) identifying
(see Fig.~\ref{RLFig1-2}) and reducing particulate sources; (2) identifying the particulate transporting mechanisms and
blocking particulate traffic into the cavity space; and (3) developing effective in-situ particulate removal
apparatus and procedures and applying them at scheduled intervals
\cite{McCaug2}.
        
\begin{figure}
\centering
\includegraphics[width=0.480\columnwidth]{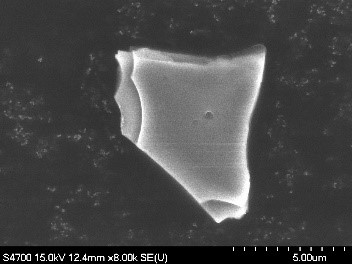}\thinspace\includegraphics[width=0.480\columnwidth]{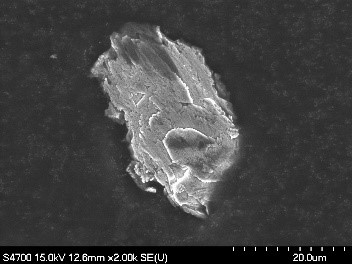}

\includegraphics[width=0.480\columnwidth]{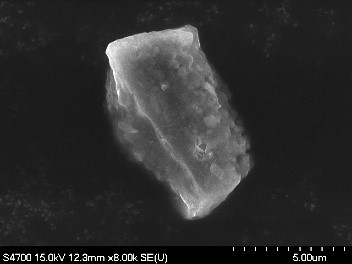}\thinspace\includegraphics[width=0.480\columnwidth]{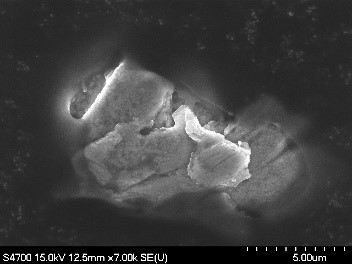}
\caption{Examples of particulates of Ti/Ta, stainless-steel, silicate, and copper (clock-wise starting at upper left), found on the surface of cavities extracted from beamlines. (From Ref.~\cite{RGRef2})  \label{RLFig3-6}} 
\end{figure}

As established in Ref.~\cite{RGRef2}, a critical first step taken at the beginning of this campaign against particulates
in CEBAF was to collect with a suitable method particulate matter from the vacuum surfaces of components including
SRF cavities that had been operated with beam for some time. The collected samples were transferred to carbon tapes
which were then analyzed with an SEM for characterization. Typical examples of particulates found on the surface of
cavities removed from the cryomodule FEL-2, being refurbished into C50-12, are shown in Fig.~\ref{RLFig3-6}. 
    
Through systematic collection and characterization of particulates from cavities and beam pipes in the cryomodule
FEL-2, physical evidence of particulate movement was revealed \cite{RGRef2}. Ti/Ta particulates, a
characteristic material of the differential elements in the cryomodule ion pump (so called B pump in the CEBAF
nomenclature) were detected in all four of the sampled cavities, two being close to the B pump (first two cavities in
the string) and the other two being away from the pump (last two cavities in the string). Stainless-steel and silicate particulates
were observed in abundance. All these observations point to a consistent picture of particulate sources being
outside of SRF cavities and particulate loading, by some movement mechanisms, post cryomodule installation.
Several changes were implemented in CEBAF SRF linac operation and maintenance practices based on the findings
of 2014-2015 particulate collection and identification effort, all targeted at reducing source particulates,
including the implementing ``cavity-quality cleaning'' of adjacent warm girder beamline UHV components of any
future cryomodule extracted from the accelerator tunnel for refurbishment. Furthermore, high voltage conditioning of ion pumps (hi-potting)
has been prohibited over the entire CEBAF linac system and the B pumps are disabled during any planned cryomodule
warm up. Modern NEG/ion pumps have replaced the current conventional or differential ion pumps in
the CEBAF SRF linacs since the summer of 2016 \cite{RGRef7}. From 2016 onward, extracted cryomodules
and warm girder beamline UHV components are further sampled for particulate characterization with an improved
collection method and automated SEM analysis procedure \cite{RGRef8,RGRef9}. This resulted in a growing
catalog of particulates, confirming and reinforcing the extent of particulate contamination and the
need for controlling particulate sources external to SRF cavities.
     
\begin{figure*}
\centering
\includegraphics[width=3.5in]{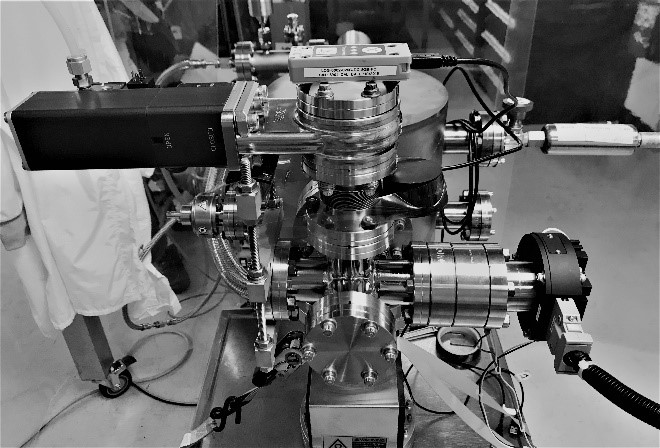}\thinspace
\includegraphics[width=3.5in]{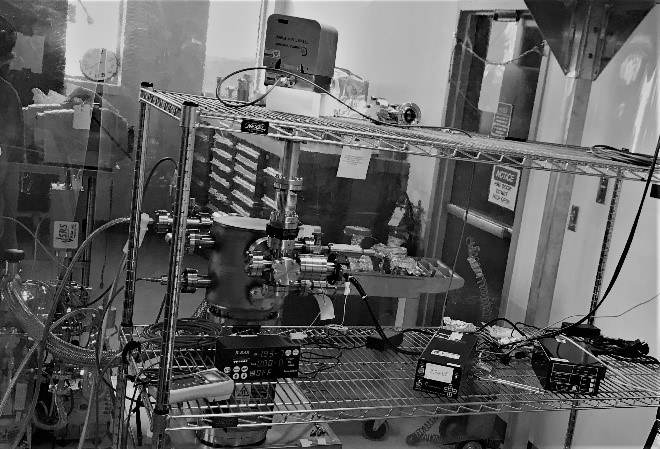}
\caption{Laboratory vacuum test bed instrumented with a vacuum particle counter configured for evaluation of particulate generation attributes of CEBAF beamline gate valves (Left) and ion pumps (Right).
\label{RLFig7-8}} 
\end{figure*}

Recently, particulate source identification efforts moved to evaluating the in-situ particulate generation
of the regular beamline components in their nominal use for beam operation. The current focus
is the cryomodule isolation gate vales (two each for every installed cryomodule) and ion pumps (one each for
every installed cryomodule and one each for every warm girder between adjacent cryomodules). A laboratory test
bed (see Figure~\ref{RLFig7-8}) has been established since May 2019. Preliminary test results have established a correlation
between service life and particulate generation for both the beamline gate valves and ion pumps. No
particulates down to 0.3 micron in size were detected for a freshly in-house rebuilt gate valve with accumulated
open/close cycles up to 1000.  In comparison, particulates up to 2 micron in size were frequently detected for
a gate valve extracted from the CEBAF North Linac. A differential ion pump extracted from the North linac zone
1L23, which was the B pump of a new cryomodule C100-6 installed for the CEBAF 12 GeV upgrade, was tested with
controlled vacuum in the range of 10$^{-7}$ to 10$^{-4}$ torr and varying gas species such as N2, He, and Ar. No particulate
down to the detection limit of the vacuum particle counter was detected regardless of its operating high voltage.
In comparison, particulates up to 2 micron in size were easily detected in a conventional ion pump, which was
extracted from the former JLAB FEL with standard operating high voltage. 
     
Presently, the gate valve and ion pump evaluation is dedicated to determining the onset of particulate
generation as the accumulated service life increases. The outcome of this effort is a set of recommended operational
procedures as well as maintenance schedules of the current beamline gate valves and ion pumps for effective control
of particulate generation. Furthermore, alternative valves and pumps possessing superior particle generation
attributes are to be evaluated as future options for the CEBAF linac SRF systems. 
  
Depending on the nature of moving forces, different modes of particulate movement may be identified. In routine
CEBAF electron beam operation, the intrinsic electrostatic force levitates, suspends and transports
the charged population of particulates that are exposed to or irradiated by various species including
X-rays, gamma rays, and electrons. When excessive charge accumulation reaches a point where the internal
repulsive Coulomb force exceeds the tensile strength of the body material, particulate explosion or
fragmentation results, leading to particulate mobility. In interrupted beam operating conditions, such
as fault-triggered gate valve closures, the mechanical shocking force launches particulates originally at rest
on a given site. Launched particulates then follow ballistic trajectories governed by gravity and ultimately land
at a remote site. In an accident condition when cavity string vacuum is lost, either partially or completely, the
hydrodynamic force arising from gas inrush disperses particulates. The landscape of particulate distribution
on the beamline UHV surfaces over the affected linac section might be profoundly changed in a catastrophic vacuum
loss event. Last but not least, the thermal force, arising from temperature gradients that exist in the beamline
axial direction because of alternating cold and warm components in the CEBAF linacs, acts on suspended particulates,
which then tend to drift adiabatically towards the cold cavity walls.	
        
An interesting observation has recently been made due to microscopic SEM inspection of the sealing surfaces
of Viton seals from beamline gate valves extracted from the CEBAF North Linac. Earlier optical inspection established
that particulates were embedded along these sealing surfaces. Attempts to characterize these particulates
using an ordinary SEM however failed because of severe charging in the non-conducting Viton elastomer.
By using a special SEM at the College of William \& Mary, that specimen charging problem was overcome when the specimen was
measured in the ambient air. The microscopic images revealed the concentration of particulates
captured, with an estimated density of $10^4-10^5$ per~mm$^2$ for particulates 10 micron or smaller in size. The elemental
composition of these particulates has a large overlap with those collected from the beamline UHV surfaces. This
recent observation lead us to conclude that we now have the first physical evidence of the existence of charged
particulates in the CEBAF beamline spaces. Moreover, in view of the outcome from the test bed gate valve evaluation
which shows zero particulate generation from a freshly in-house rebuilt gate valve, we now have a
potential future solution to reduce the particulate input into cavities, namely blocking particulate movement
using the Viton seal as a particulate trap. Alternative particulate traps, such as electrostatic precipitators,
are a potential solution as well. We plan to evaluate these options in conjunction with developing a plan
for scheduled maintenance of the CEBAF beamline gate valves.

An important step in understanding particulate movement, but currently missing, is its direct observation.
Toward that end, a novel particulate detector was invented and patented at Jefferson Lab. The detector is based on the phase and amplitude interruption of a laser
beam interacting with a passing particulate, which is introduced through a window into the accelerator beamline.
The design package was completed in September 2019 and the first
demonstration unit has been built in collaboration with OmniSensing Photonics LLC. Extensive bench testing is on-going, which is to be followed by a field test in CEBAF. We anticipate that by applying such particulate detectors in the CEBAF SRF linacs, detecting and diagnosing particulate movement in real time will become available, providing needed information to
guide solutions for reducing, preserving, and even possibly reversing the problem of slow loss of CEBAF
energy reach.

\subsection{Beam Delivery\label{beamdelref}}

During CEBAF's operational life many procedures
and processes have been developed in order to operate
recirculated linacs efficiently. Many of these
processes have been improved as a result of the
upgrade project. In this section we highlight individual
systems in the accelerator that are important for
accurate and timely beam delivery. These tools are
all by now sufficiently developed that operations
staff routinely utilize them during initial setup
of the accelerator after a long down, to affect a
change of CEBAF configuration needed as part of the
physics program, or to analyze the existing machine
configuration during operations. In particular we discuss
the methods to manage the large number of magnet
settings in the accelerator, our optics verification
tools, the pathlength systems, and the linac energy management system.

\subsubsection{Model Driven Settings}\label{Optcontrol}

For the 12 GeV upgrade, improvements in agreement between the CEBAF
model and machine performance, along with new software tools and
processes, were implemented such that new machine configurations can
be set from the model with less tune time.  Over the course of 12 GeV
CEBAF commissioning, these new tools and processes were tested and
improved upon.  The result was a measurable reduction in necessary
time for new machine configurations.

A CEBAF Modeling Team was formed to
establish tools and procedures for model-driven configuration of
12 GeV.  The Modeling Team chose the accelerator simulation code
\texttt{elegant}~\cite{elegant} to model the machine.  The Modeling
Team established a formal feedback process such that model
discrepancies discovered during commissioning and operation are fed
back to the model, thus providing a path for convergence.  The process
includes a formal audit to verify consistency and correctness.

To address configuration control, the CEBAF Element Database (\texttt{CED}) was
created ~\cite{CED},~\cite{CED2011}.  \texttt{CED} is a relational
database that stores beamline elements and their attributes.  It is
the authoritative source of hardware, control system, and model
information for the accelerator.  It is accessed real-time by control
system software and operator tools.  Operator screens are generated on
the fly from \texttt{CED} so they are always correct and up to date.
A number of high-level software tools were developed, based on
\texttt{elegant} and \texttt{CED}, to provide operators the means to quickly and consistently configure and tune the machine.

For example, elegant Download Tool 
(\texttt{eDT}) is a high-level software tool that generates magnet
design setpoints for various machine energies and pass configurations
based on the modeled \texttt{elegant} values stored in
\texttt{CED}~\cite{eDT}. \texttt{eDT} also compares the present
machine setpoints to the design setpoints and provides a means to
highlight off-design magnets.

\subsubsection{Beam Optics Tuning}
\label{beamtuning}
\centerline{Emittance Measurement and Matching}
\vspace{2mm}

During the 6 GeV era, transverse optics matching was manually performed
using designated tuning knobs while observing differential orbits
produced by diagnostic
kickers~\cite{Lebnim,linearopticsCEBAF}.  The
\texttt{qsUtility} software toolset was developed to perform
transverse emittance measurement and matching for 12 GeV CEBAF in a
more deterministic and reproducible fashion~\cite{qsUtility}.

The \texttt{qsUtility} software toolset automates the measurement of emittance
and Twiss parameters, along with computing quadrupole settings to achieve the
design Twiss parameters at each matchpoint.

The emittance is measured by varying the field strength of one or more quadrupole magnets while measuring the beam size with a downstream wire scanner as described in ~\cite{MintyZimmerman}.  To save time, the beam size measurements are performed with the ``zig-zag'' method described in ~\cite{zigzag}.

The Twiss parameters at the entrance of the quadrupole that was varied during the measurement are determined by solving Equation (\ref{MZ_4.20}) for $\epsilon$, $\beta$, and $\alpha$ using the least squares method outlined in ~\cite{MintyZimmerman}.
\begin{equation}
\label{MZ_4.20}
\begin{pmatrix}
(\sigma_{x}^{(1)})^{2} \\
(\sigma_{x}^{(2)})^{2} \\
(\sigma_{x}^{(3)})^{2} \\
... \\
(\sigma_{x}^{(n)})^{2} \\
\end{pmatrix}
=            
\begin{pmatrix}                      
(R_{11}^{(1)})^{2}&2R_{11}^{(1)}R_{12}^{(1)}&(R_{11}^{(1)})^{2} \\
(R_{11}^{(2)})^{2} & 2R_{11}^{(2)}R_{12}^{(2)} & (R_{11}^{(2)})^{2} \\
(R_{11}^{(3)})^{2} & 2R_{11}^{(3)}R_{12}^{(3)} & (R_{11}^{(3)})^{2} \\
... \\
(R_{11}^{(n)})^{2} & 2R_{11}^{(n)}R_{12}^{(n)} & (R_{11}^{(n)})^{2} \\
\end{pmatrix}
\begin{pmatrix}
\beta(s_{0})\epsilon \\
-\alpha(s_{0})\epsilon \\
\gamma(s_{0})\epsilon \\
\end{pmatrix}
\end{equation}

The superscript indices in parentheses refer to the measurement step, the $\sigma_{x}^{(n)}$ are the beam size measurements, and the $R_{11}^{(n)}$ and $R_{12}^{(n)}$ are the transport matrix elements for the beamline from the varied quadrupole and the wire scanner.

Once the upstream Twiss parameters are determined, a set of quadrupole setpoints to match to the design Twiss parameters at the matchpoint is computed using the built-in optimizer in \texttt{elegant}~\cite{elegant}.

Matching is performed at the exit of the Injector, at each of the ten spreaders, and at the entrance to each of the four experiment halls.  Occasionally, hand tuning in one or more
recombiners is needed using the Courant-Snyder measurement, as described in
\cite{Lebnim,linearopticsCEBAF}, to produce
optics suitable for matching at the downstream spreaders.
\begin{figure}[!htb]
\centering \includegraphics*[width=80mm]{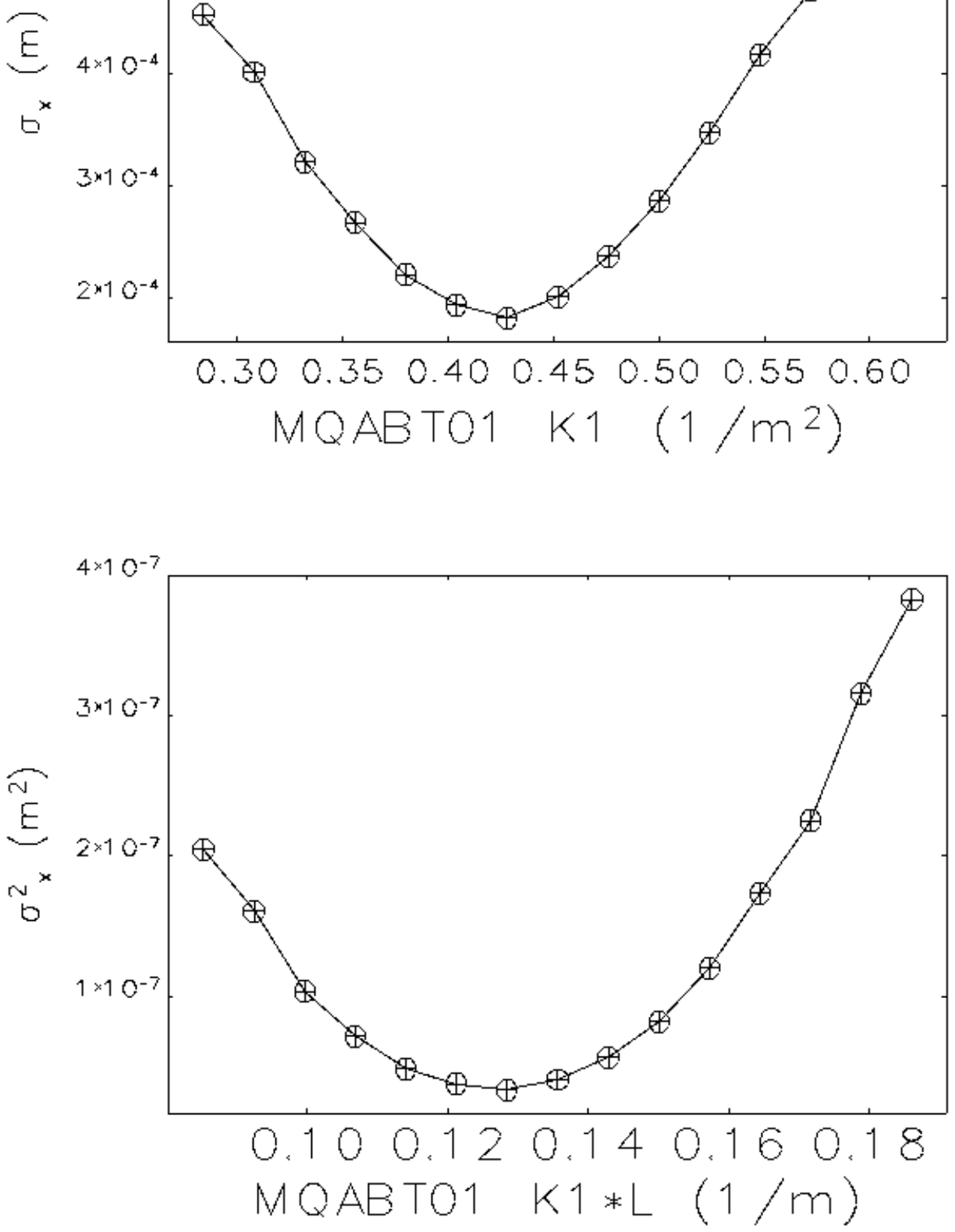}
\caption{Emittance Measurement Example}
\label{WEPOR03_f1}
\end{figure}

The ray-tracing technique described in ~\cite{rayTrace1} and
~\cite{rayTrace2} is being explored for use during machine setup.  The
technique involves injecting a number of orbits into a region of the
accelerator and monitoring the position response to trace out the phase
ellipse of the beam as it travels.  The ray-trace technique measures
the beam optics at multiple locations simultaneously which reduces
setup time and provides a more global understanding of the machine
\vspace{2mm}optics.

\centerline{Dispersion Measurement and Correction}
\vspace{2mm}
Dispersion measurement at CEBAF is performed by modulating the beam
energy and observing the differential beam positions.  The
differential beam positions are proportional to
dispersion~\cite{linearopticsCEBAF}. The final four cavities in the Injector Linac and the eight cavities in the twentieth cryomodule in the North Linac are used for energy modulation~\cite{Lebnim}, with
the North Linac cavities used most commonly.  Dispersion correction is
performed arc by arc by adjusting designated pairs of quadrupole magnets
in each recirculation arc for horizontal dispersion correction, and
designated pairs in each spreader and recombiner for vertical
dispersion correction.  Adjustments are performed while observing the
downstream differential orbits and adjusting designated quads to
cancel the dispersion leakage out of each dispersive region.

A new software tool for displaying dispersion measurements in a more
operator-friendly fashion, along with an automatic dispersion
optimization method to speed up dispersion corrections are being
explored for future use at CEBAF~\cite{dispersionTool}.

\subsubsection{Beam Locks}

The CEBAF beam experiences both slow drifts and fast fluctuations in
beam position and energy.  Slow drifts are due to magnet power supply fluctuations, temperature drifts, ground motion, and the like.  Fast fluctuations are primarily induced by power line frequency interference.  In order to ensure that the beam stays
within the energy and orbit apertures of the machine and within the
users' requirements, a set of feedback locks has been\vspace{3mm} developed.
\centerline{Orbit Locks}
\vspace{0.5mm}

A set of slow orbit locks was implemented to stabilize the beam
against slow orbit drifts at frequencies less than 1~Hz \cite{slowOrbitLocks,empiricalLocks}.  The slow orbit locks maintain beam positions
into the Injector linac, each of the ten recirculation arcs, each of
the 5 extraction regions, and various locations in each of the
experimental hall transport lines.  Each lock uses a pair of
correctors and BPMs for each plane to maintain the required
beam position and angle into the region of interest.  The locks are
calibrated empirically by applying small kicks with the lock correctors
and measuring the resulting BPM positions to produce a response
matrix.  The orbit lock server uses the Control Device (CDEV)
interface layer ~\cite{CDEV} to communicate with instances of the
orbit lock GUI and the EPICS control system.  
%Figure
%~\ref{orbitLockExpanded} %shows an example of the orbit lock that
%maintains the launch into Arc 4.
%\begin{figure}[!htb]
%  \centering \includegraphics*[width=80mm]{orbit_lock}
%  \caption{Orbit Lock Example}
%  \label{orbitLockExpanded}
%\end{figure}

\begin{figure}[!htb]
\centering \includegraphics*[width=80mm]{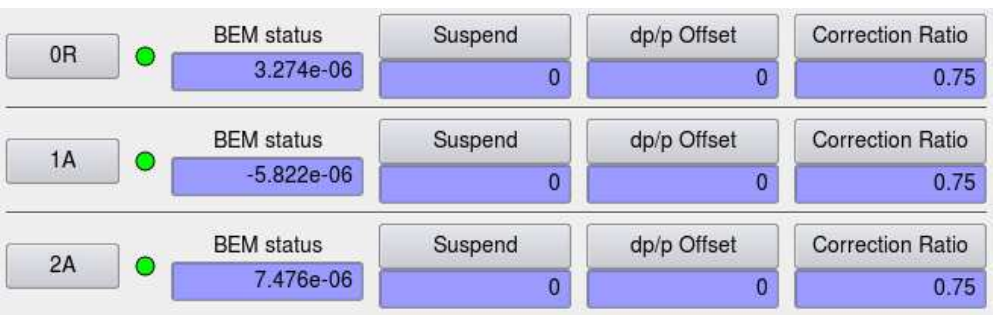}
\caption{Energy Lock GUI}
\label{elock}
\end{figure}

\centerline{Arc Energy Locks}
\vspace{2mm}
A set of slow energy locks was also implemented to stabilize the beam
against similarly slow energy drifts~\cite{elocks}.  The energy locks
adjust the gradient setpoints in selected SRF cavities at or near the
end of each linac to maintain the correct beam energy in the
downstream arcs.  There is a lock to maintain the correct energy from
the Injector linac through the Injector Chicane, from the North Linac
into Arc 1, and from the South Linac into Arc 2 (Fig.~\ref{elock}).

\begin{figure}[!htb]
\centering \includegraphics*[width=80mm]{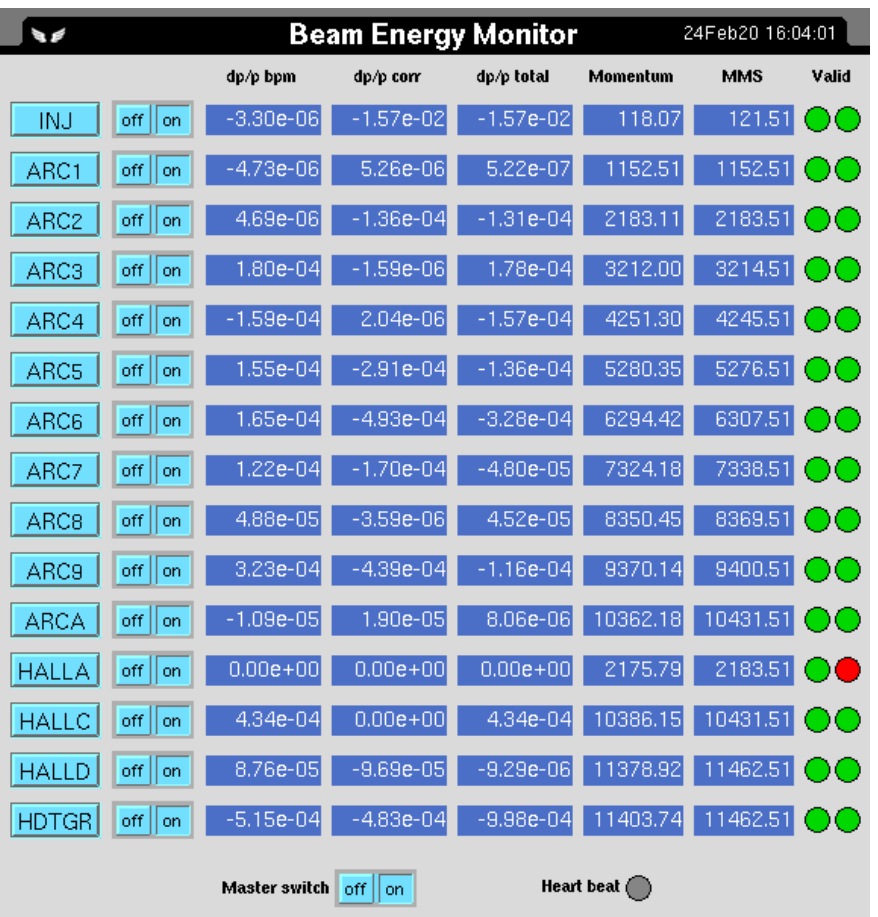}
\caption{Beam Energy Monitor}
\label{bem}
\end{figure}

The Beam Energy Monitor (BEM)~\cite{BEM} provides the energy input to
the arc energy locks.  BEM computes the beam energy in each arc using
the arc magnet power supply setpoint, corrector setpoints, and BPM
position readbacks (Fig.~\ref{bem}).

\begin{figure}[!htb]
\centering \includegraphics*[width=80mm]{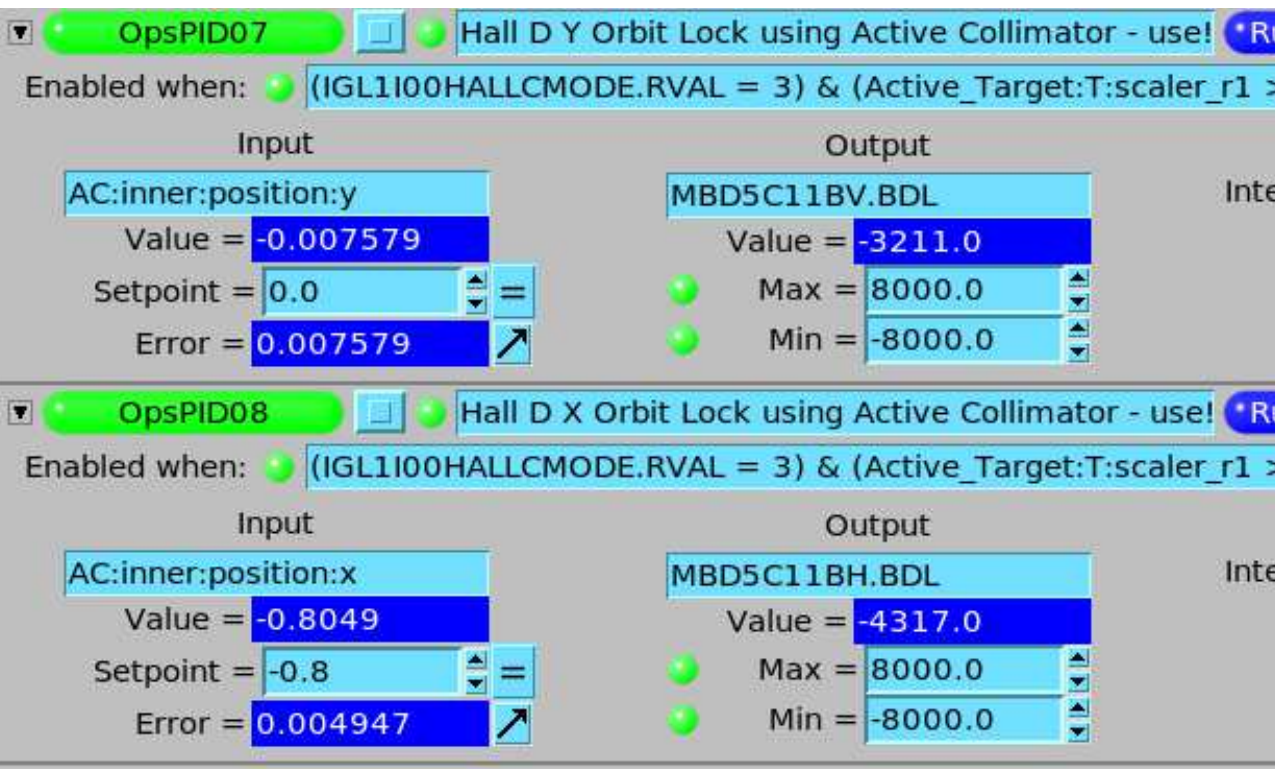}
\caption{Generic Lock Example}
\label{genericLockExample}
\end{figure}

\centerline{Generic Locks}
\vspace{2mm}
In addition to dedicated slow orbit and energy locks, a so-called
``Generic Lock'' architecture was developed to allow operators to
easily implement PID locks between arbitrary process variables
~\cite{genericlocks}.  Examples of Generic Locks include locks to
maintain stable beam current, RF phase locks, and short term
experiment-specific orbit locks.  The Generic Lock tool allows the
operator to specify input and output process variables, PID gains,
expressions to enable or disable the lock based on other process
variable values, etc.  Figure ~\ref{genericLockExample} shows an
example of a Generic Lock to maintain horizontal and vertical beam
positions on the Active Collimator in Hall\vspace{2mm} D.

\centerline{Fast Feedback}
\vspace{2mm}
In addition to slow orbit and energy drifts, the beam experiences fast
fluctuations in beam position and energy.  These fluctuations
primarily occur at harmonics of the power line frequency (60~Hz, 120~Hz,
180~Hz, etc.)~\cite{linefreq}.  A Fast Feedback system was implemented
to squelch these fast fluctuations ~\cite{ffb1,ffb25,ffb3,ffb4}.  The system was originally installed in Halls A and C,
and was expanded to control the beam delivered to Hall D as part of the 12~GeV upgrade ~\cite{ffb5}.

\begin{figure}[!htb]
\centering \includegraphics*[width=80mm]{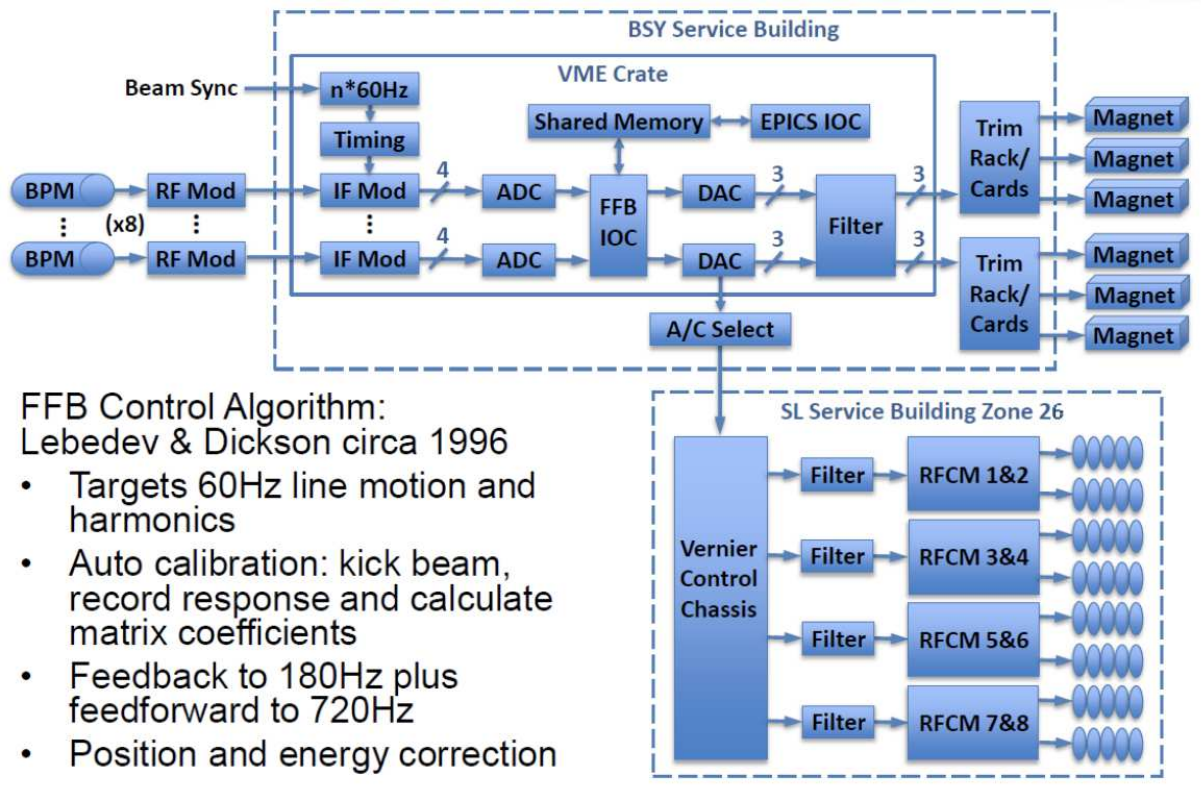}
\caption{Fast Feedback Block Diagram}
\label{ffb_schematic}
\end{figure}

\begin{figure}[!htb]
\centering \includegraphics*[width=80mm]{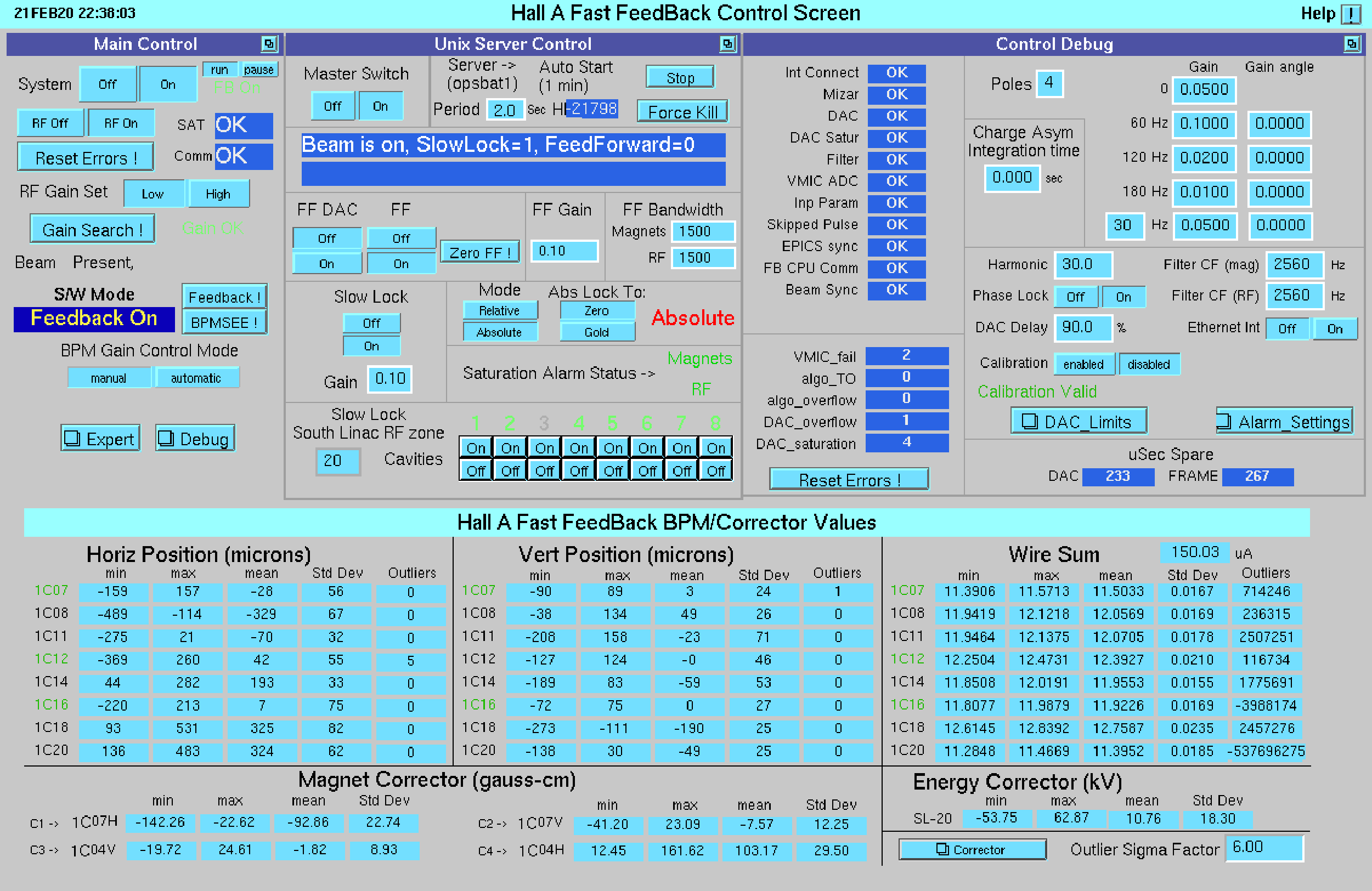}
\caption{Fast Feedback Control Screen}
\label{ffb_control}
\end{figure}

The Fast Feedback system is connected to a set of BPMs which were
modified to provide a high enough frame rate in order to be useful as
an operational feedback system ~\cite{ffb3}.  Outputs from the Fast
Feedback system include a set or air-core correctors for position
control and an RF vernier for energy control~\cite{ffb25}.  Figure
~\ref{ffb_schematic} shows a block diagram of the system.  Figure
~\ref{ffb_control} shows the control screen for the Hall A Fast
Feedback system.

The Fast Feedback system suppresses fluctuations from the first three
power line harmonics.  A Feed Forward system was added to suppress
higher order harmonics (up to twelve).  The Feed Forward system
predicts future beam motion by analyzing BPM and corrector data from
the recent past ~\cite{ffb6}.

The air-core correctors and RF vernier have a limited dynamic range.
A Slow Lock was added to compensate for slow drifts which could drive
the beam outside the range of the air-core correctors and/or the RF vernier
~\cite{ffb6}.

\subsubsection{Pathlength and $M_{56}$ Measurement and Correction}\label{PLControl}

\centerline{Introduction}
\vspace{2mm}
Accurate measurements of pathlength and pathlength change versus
momentum ($M_{56}$) are critical for maintaining minimum beam energy
spread in CEBAF.  Pathlength in CEBAF tends to drift due to seasonal
and diurnal temperature changes and long-term ground motion.
Overall pathlength in CEBAF is measured and controlled as it was during 6 GeV running\vspace{2mm} \cite{KRAFFT2006314,PathLength1,PathLength2,pac2001tief}.

%%Pathlength for each arc is
%% normally corrected by adjusting the excitation of dogleg chicane
%% magnets at the entrance of each arc.  $M_{56}$ for each arc is
%% corrected by adjusting a designated set of quadrupole magnets in each
%% arc.

\centerline{Measurement Devices}
\vspace{2mm}
Pathlength and $M_{56}$ are determined using a precision phase
detector~\cite{PathLength1} measuring the relative arrival time
of the electron bunches at a longitudinal pickup cavity operating at 1497 MHz located at the end
of each linac.  A beam macropulse with a duration of 4 \textmu s
(less than the recirculation time of 4.2 \textmu s) is
established. The output of the cavity comes out in successive 4 \textmu s bursts separated by 0.2 \textmu s, each burst RF phase locked to the beam current of each pass, respectively.  A difference in
pathlength between passes is measured as a phase difference between
the RF from each burst~\cite{PathLength2}. Briefly, the path length is adjusted via the pathlength chicanes so that all the measured phases are identical. Once this is achieved, higher beam passes transit the linac SRF cavities at the same phase they did on the first pass to high precision.

\begin{figure}[!htb]
\centering \includegraphics*[width=86mm]{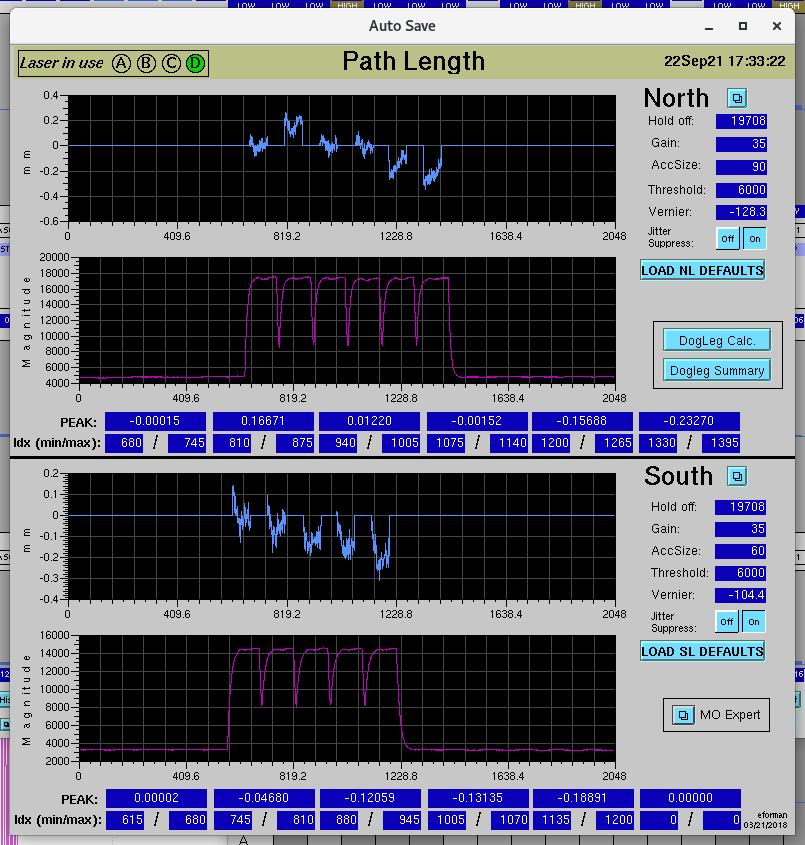}
\caption{Cavity Monitor Waveform Display}
\label{plBefore}
\end{figure}

For 12 GeV CEBAF, functionalities to digitize the waveforms generated by the
cavity monitors and to store them as EPICS waveform database records were
added.  Figure~\ref{plBefore} shows a display of the waveforms for
each cavity monitor.

  The magenta traces are the cavity monitor output amplitude
levels for each pass, 6 for the passes through the North Linac and 5 for the passes through the South Linac.  The output voltages are proportional to beam currents
for each pass, therefore they are a useful indication of beam
transmission. The blue traces in Figure~\ref{plBefore} represent the relative arrival times of each pass as measured by the RF phase from the cavity monitor. In the example shown, the accumulated path length from pass one to pass two through the North Linac is 150 microns long, 0 microns for pass one to passes three and four, 150 microns too short for pass 1 to pass 5 and 200 microns short for pass one to pass six. In this case the South Linac is exhibiting a classic pattern often found when the accelerator contracts: 1-2 is 50 microns short, 1-3 is 100 microns short, 1-4 is 150 microns short, and 1-5 is 200 microns short. Because the North Linac data do not reflect this same type of
pattern, CEBAF did not actually contract between readings.  Rather, the measurement indicates that the South Linac is off crest and requires a phase\vspace{2mm} adjustment.

\centerline{Pathlength Correction}
\vspace{2mm}

For pathlength correction, the main changes in 12 GeV CEBAF are that there are up to 6 beam passes through the North Linac, and there is no Arc 10 pathlength chicane. The pathlength correction is now a three-step process. First, the overall accelerator Master Oscillator (MO) frequency is adjusted to globally correct the overall pathlength, including the Arc 10 beam pass. Once this frequency is established, the ARC 1-9 pathlength chicanes can be adjusted to the proper values by the same process used previously. As discussed in Section \ref{S:pathlengthChicanes}, at 12 GeV the dogleg chicanes now allow the pathlength to be adjusted by up to
$\pm$10 degrees of RF phase, or $\pm$5.6 mm of pathlength.  As the final step, the pathlength of the Arc 10 pass can be fine-tuned by horizontally steering the beam inboard or
outboard in the arc itself using steering correctors, which decreases or
increases the distance of beam travel through the arc. 
The same method
can be used in the other nine arcs in addition to dogleg adjustments
to provide a range of several additional millimeters of pathlength
\vspace{3mm} correction.
\centerline{Dogleg Calculator Tool}
\vspace{1mm}

\begin{figure}[!htb]
\centering \includegraphics*[width=86mm]{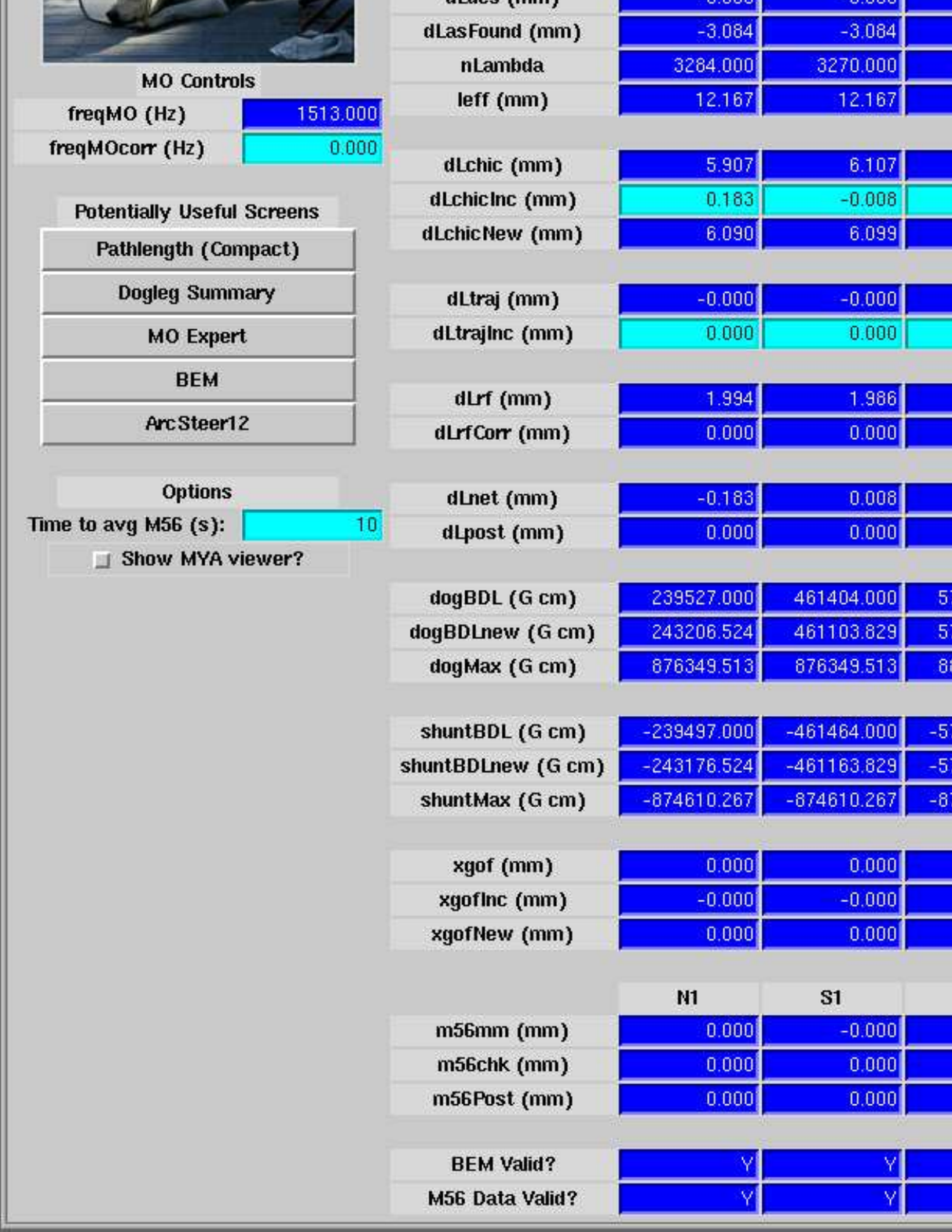}
\caption{Dogleg Calculator Tool}
\label{dogCalc}
\end{figure}

During the 6 GeV era, pathlength was corrected manually by adjusting
dogleg magnets while observing the cavity monitor output traces on a
pair of oscilloscopes.  For 12 GeV, the cavity monitor outputs are
stored as EPICS waveform database records, which allows for automation
of pathlength correction.  A new software tool, called
\texttt{DogCalc12}, was developed to quickly compute and apply new
dogleg chicane setpoints to correct pass-by-pass pathlength in a
single step.  The time to correct pass-by-pass pathlength was reduced
from an order of hours to less than one minute.  Figure~\ref{dogCalc}
is a screen capture of the new tool.

\texttt{DogCalc12} reads the pathlength errors from the EPICS waveform
data and displays the measured pathlength errors along with the
computed dogleg chicane setpoints which will correct the errors.  The
tool will recompute the dogleg setpoints to accommodate pathlength
corrections using MO frequency adjustments or arc orbit\vspace{2mm} offsets.

%% \begin{figure}[!htb]
%% \centering \includegraphics*[width=80mm]{191220645_4}
%% \caption{Pathlength traces after correction}
%% \label{plAfter}
%% \end{figure}

%% \subsection{MOMod}
%% See ~\cite{pac2001tief}

\centerline{$M_{56}$ Correction}
\vspace{2mm}
The pathlength measurement system is also used to measure $M_{56}$ for each arc.
$M_{56}$ is the change in pathlength for a given momentum change~\cite{PathLength2}
\begin{equation}
  \Delta pathlength = M_{56}\frac{\Delta p}{p}.
\end{equation}
$M_{56}$ is measured by applying a small momentum offset and observing the resulting
change in pathlength.  Corrections for each arc are performed by adjusting the settings of
designated quadrupole magnets according to the design beam optics in the arc.

\subsubsection{Linac Energy Management}

Linac energies are set via a software tool called Linac Energy Management (LEM). Given a requested operating beam energy, LEM distributes the accelerating gradient to individual cavities in a way which minimizes overall machine RF fault rate \cite{JB2}. Once the gradient distribution is determined, quadrupoles in the linac are adjusted to the machine model values scaled by the actual beam first pass energy at the quadrupole.

LEM must take into account several aspects of the overall accelerator configuration to complete a setup. Obviously, cavities that are off-line for any reason must be tracked and eliminated from the optimization. In addition, LEM tracks and uses ``operations maximum gradient'' set points, one for each cavity, determined by the running operations history of that particular cavity. Roughly, the operations maximum gradient is the largest gradient set point, determined by the operations staff through use, that the cavity operates reliably. In no case, should this gradient exceed the maximum gradient as determined in SRF commissioning outlined in Section~\ref{SRFCOM}. Included in the LEM optimization is that the operations (and by implication the SRF commissioning) maximum gradient for each cavity is not exceeded. LEM
automatically distributes gradient to all operating cavities using a solution that minimizes the RF trip rate, but adheres to these constraints \cite{JB1}.

In early 12~GeV CEBAF running, the C100 cavities had to be turned down to approximately 80\% of their design value
for energy gain due to field emission. Consequently, the old cavities had to be pushed to higher
gradients and therefore higher fault rates.  The final energy to Hall D was lowered
from 12.0 to 11.6 GeV to mitigate overall fault rates.

\section{Future Plans}

\subsection{Machine Learning}

Following the lead of other scientific disciplines, such as astronomy and high-energy physics, accelerator physics has started to leverage machine learning to address challenging problems. In the U.S., this is largely in response to recent National and Department of Energy (DOE) artificial intelligence (AI) initiatives \cite{BTenn}. We note that despite the terms “machine learning” and “artificial intelligence” often being used interchangeably, machine learning is a subset, albeit a large one, of the more general field of AI. A helpful definition of machine learning is “the field of study that gives computers the ability to learn without being explicitly programmed” \cite{CTenn}. This represents a major paradigm shift from conventional programming where the user inputs data and a set of explicit rules is used to generate the output. Machine learning, on the other hand, takes as its inputs data and the corresponding answers (or labels) and infers the rules. The rules can then be applied to new, unlabeled data. This is an example of supervised machine learning since the data is associated with a label and represents the most common class of machine learning. Unsupervised learning, by contrast, is another category of machine learning which takes unlabeled data as its input and seeks to organize it into clusters or to reduce its dimensionality.

The rise of machine learning – across sectors as diverse as commerce, healthcare and science, among others – is being driven by the confluence of compute power, abundant data, open-source software and theoretical advances in the field. Historically, particle accelerator systems have been a source of enormous amounts of data, not only by users (i.e., experimental beamlines, detectors), but also from machine diagnostics which record data about the beam, hardware components and their various subsystems. With the advent of specialized coprocessors, such as graphical and tensor processing units and cloud-based computing resources, compute power is available to analyze, process, interact and visualize large data sets in ways that were not possible before. The quality of free resources available for learning to build machine learning systems, coupled with the accessibility of open-source software which incorporates the latest algorithmic advances, makes for a low barrier of entry into the field. Where machine learning was once a niche field practiced by subject matter experts and trained machine operators, one can now reproduce state-of-the-art results on a personal computer by following a simple tutorial.

Particle accelerators represent a class of complex scientific instruments which are comprised of many interacting subsystems. As such, they are a source of potentially rich data sets that cover phenomena across a wide variety of time-scales, from slow thermal drifts to fast beam loss faults, and across many subsystems with correlations that may or may not be apparent. Data of this kind is described as ``big data'', that is, data sets so large or complex that it is not amenable to traditional data processing techniques. 

\subsubsection{SRF Fault Classification}

Recently, machine learning was applied in CEBAF for classification of SRF cavity faults \cite{DTenn}.
As a user-facility, the goal at CEBAF is to maximize beam time to the experimental halls.
Currently, a significant contributor to machine downtime are beam trips caused by SRF system faults.
During FY2018 there were an average of 6 RF trips per hour with a mean recovery time of 0.5 minute per trip.
Consequently, over an hour of beam time is lost every day. The amount of data lost in the experimental halls
is even greater because during analysis of the data 30 seconds of data before the trip and 30 seconds
after recovery are discarded.

The C100 modules, in particular, were responsible for 33\% of the downtime due to short trips across all accelerator subsystems. In order to better understand the nature and frequency of these faults, a waveform harvester was implemented in each of the eleven C100 cryomodules. For each C100 cavity fault, the system automatically writes 17 RF signals from each of the eight cavities in the cryomodule to file. The recorded time-series data allows subject matter experts to analyze the data and determine which of the eight cavities within the cryomodule went unstable first and classify the type of cavity fault. Due to the diligent work of system experts, more than 20,000 labeled examples exist; that is, time-series signals from cavities have corresponding labels indicating the first cavity to trip and the fault type. With the existence of this data there is a clear motivation to utilize supervised machine learning to automate the process. Real-time --- rather than post-mortem --- identification of the offending cavity and classification of the fault type would give control room operators valuable feedback for corrective action planning. Improving the stability of the RF system naturally translates into higher beam-on-target time. It also provides performance metrics that can be used to improve cavity designs \cite{tpas1}.

Initial efforts utilized ensemble machine learning, specifically random forests, to train a model on several hundred
labeled cavity faults. The models performed well on test data, achieving accuracy scores over 95\% and 96\% for
identifying the cavity which faulted first and for classifying the type of fault, respectively \cite{ETenn}. Encouraged
by these initial results, a prototype software system has been developed to deploy trained machine learning models
to run online \cite{FTenn}. Commissioning, testing and first results were completed in early 2020.

Future effort will be aimed at replacing current machine learning models with their deep learning counterparts \cite{GTennExtra}.
Deep learning is a sub-field of machine learning which is based on learning successive layers of increasingly
meaningful representations of the data. (The ``deep'' in deep learning refers to multiple hidden layers in
the network architecture). The primary advantage of methods based on learning data representations is that
it avoids the computationally costly feature engineering step. Efforts are also being made to understand
the relevant cavity fault time scales to see if preventative measures can be taken to avoid a fault if predicted
early enough. Longer term, plans are in place to upgrade all CEBAF cryomodules (not just C100s) with the same
digital LLRF system. This would allow data collection not only from the eleven C100 cryomodules, but the
remaining 39 cryomodules as well. With more data and information, there is increased potential for improving CEBAF availability.

\subsubsection{Other Applications}

Building on the initial success applying machine learning for SRF fault classification, one goal is to continue
to find ways to leverage machine learning to improve beam availability and machine reliability. Several such projects are being developed at CEBAF.

For example, one project is directed to uncovering latent knowledge in a large and complex data set, specifically
in CEBAF's archived data. The archiver represents a potentially rich source of information --- particularly given
the 25 years of operational data at CEBAF --- which is under-utilized.
The goals are two-fold; (1) mining useful information to improve the performance of the machine and (2) identifying
how the archiver and associated control systems need to evolve to keep pace with the rapid growth in machine learning.

Another promising application for machine learning is to guide machine tuning, a process which often relies on
brute force methods that can be slow to converge. Our efforts have been patterned on encouraging results  demonstrated from several fourth-generation light sources where the
machine is tuned via machine learning methods to optimize FEL power \cite{HTenn, ITenn}, but now optimizing based on other beam quality metrics important to the end users of CEBAF.

As a final example, remote monitoring using autonomous machines represents a novel intersection of robotics with AI \cite{KTenn,LTenn}. The use of robots in potentially hazardous environments, such as accelerator enclosures, would improve personnel safety at CEBAF and could automate time-consuming tasks.

As accelerators grow in complexity to meet the scientific requirements of users, machine learning will be a
necessary tool to help meet those demands.

\subsection{Enhancements for Physics}

Moving into the future, this section documents plans foreseen to enhance the physics reach of the lab. These plans focus  on a new generation of experiments at up to 12 GeV beam energy, but rely on
enhancements to the present beam parameters or maximize exploiting the present parameters by improving the detector
systems used.

\subsubsection{CLAS12, Generation 2}

%Reaching further into the Generalized Parton Distributions (GPDs), Transverse Momentum Dependent (TMD) distributions functions and spin-dependent nuclear distribution functions than what was envisioned with the present CLAS12 detector, requires upgrading it to handle higher particle luminosities, more complex event geometries and a much higher amount of data. Various task groups are looking into new technologies for these upgrades like detector streaming readout to handle the data volume while being able to efficiently extract events of interest with minimal dead time, Gas Electron Multiplier tracking detectors to be able to handle the larger particle flux of more complex events, transition radiation detectors for particle identification and others.

Reaching further into the Generalized Parton Distributions (GPDs), Transverse Momentum Dependent (TMDs) parton distributions,
and spin-dependent nuclear distribution functions than what was envisioned with the present
CLAS12 detector in Hall B, requires upgrading it to handle higher particle luminosities, more complex event geometries, and a much higher amount of data. Various task groups are looking into new technologies for these
upgrades like detector streaming readout to handle the data volume while being able to efficiently extract
events of interest with minimal dead time, Gas Electron Multiplier tracking detectors to be able to handle
the higher particle flux of more complex events, transition radiation detectors for particle identification and
others. The luminosity upgrade will greatly benefit  of the new technologies that the lab is investing on: 
AI-supported algorithms for particle tracking, electromagnetic calorimeter clustering, on-line data reconstruction 
and data preservation. A stronger integration with the IT group to exploit resources available on-site and 
off-site will provide the necessary computing power  for the next generation experiments.
The CLAS12 detector has the unique opportunity to test future EIC technologies. Replacing partially or in full 
the current components, it will be possible to deploy the proposed detectors and test them on-beam in 
conditions even more demanding than what expected at the Electron Ion Collider.
The  experience gained with the future Hall-B experimental program will be extremely useful to efficiently 
run any new projects (included SOLID and MOLLER), and optimize data collection, physics analysis and data preservation.

\subsubsection{MOLLER}

The MOLLER experiment ({\bf M}easurement {\bf O}f a {\bf L}epton {\bf L}epton {\bf E}lectroweak
{\bf R}eaction)\cite{MPRef3} aims to measure the parity-violating asymmetry $A_{PV}$ in polarized
electron - unpolarized electron (M{\o}ller) scattering. In the Standard Model of particle physics,
$A_{PV}$ is due to the interference between the electromagnetic amplitude, mediated by a photon $\gamma$,
and the weak neutral current mediated by a $Z_0$ boson. The experiment aims to measure the predicted value
of $A_{PV} \sim 33$ parts per billion (ppb) at the experiment kinematics with a precision of about 2\% of that
value. With such precision, the measurement would be sensitive to the interference of the photon with new neutral
current amplitudes which may exist from as yet undiscovered dynamics beyond the Standard Model. New MeV-scale
and multi-TeV-scale vector bosons, electron compositeness, supersymmetry and doubly charged scalars are some
examples of the new physics that could be reached by this experiment.

\begin{figure} [t]
\center{
\includegraphics[width=1.0\linewidth, angle=0]{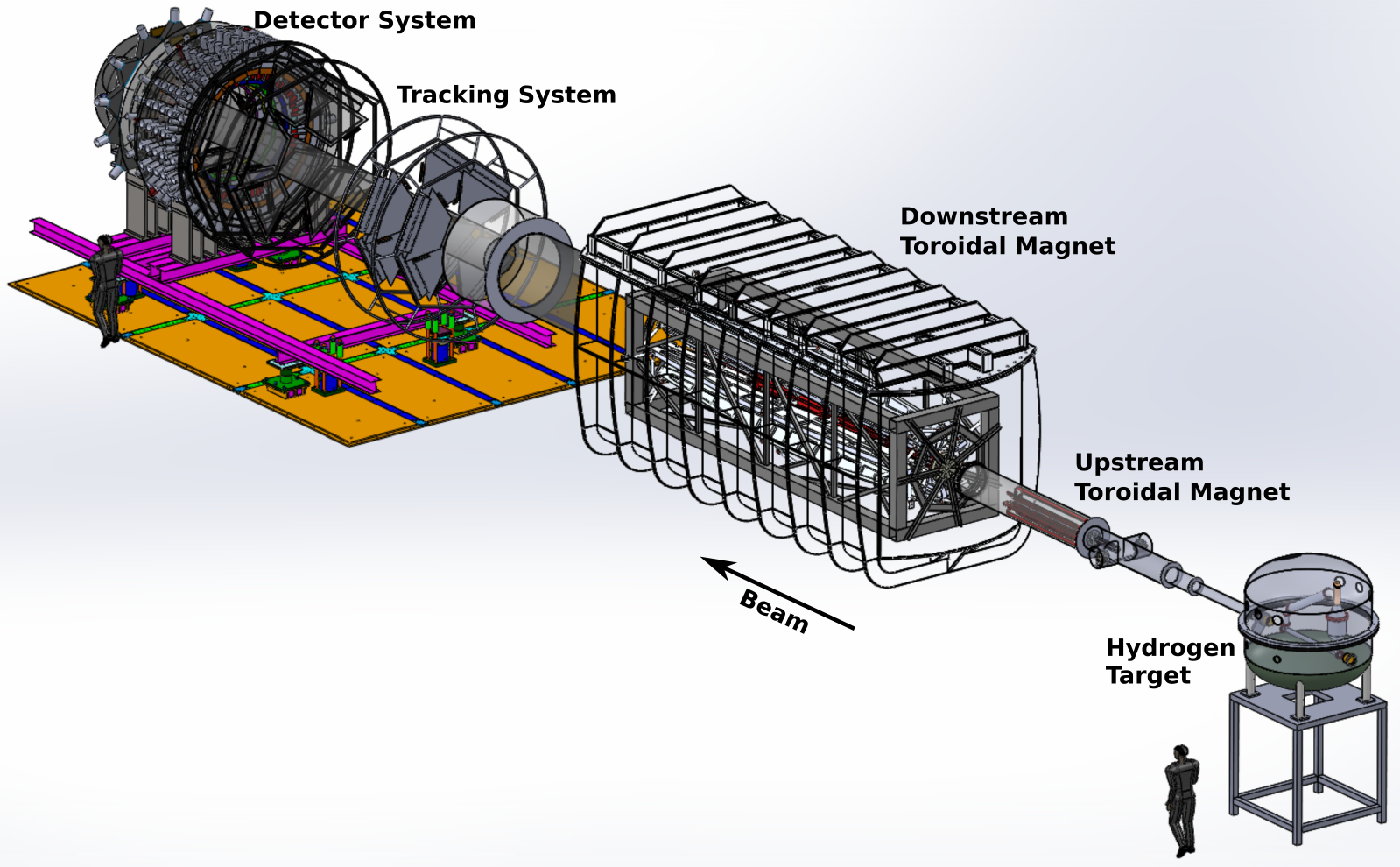}
\caption{\label{L20191021114558} Conceptual layout of the MOLLER experiment. The apparatus measures
approximately 30 meters from the center of the hydrogen target to the back of the detector system.}}
\end{figure}

Figure \ref{L20191021114558} shows the conceptual layout of the MOLLER experiment. Given its length, the experiment
is envisioned to take place in Hall A, the largest of the existing experimental halls with 53 meters inside diameter.
The experiment plans to use a polarized electron beam of 11 GeV, the highest beam energy that can be sent to
Halls A, B and C, with beam currents of 70 \textmu A. The experiment requires high beam polarizations ($>$84\%) and
high-frequency ($\sim$2 kHz) helicity flip to achieve its goals. While the beam energy, current and polarization
requested are standard parameters at JLab, the high-frequency helicity flip is not. The helicity flip frequency
affects the amount of noise the detectors see when comparing buckets of electrons with spins opposite to each other.
The noise originates from density changes in the 1.25 meters long liquid hydrogen target due to, basically, micron
sized bubbles generated by beam heating of the liquid, a total of about 4 kW. For comparison, it is usual for
parity-violating experiments to flip the beam helicity at 30 Hz. To achieve a 2 kHz helicity flip rate, new
electro-optical materials must be used to rotate the laser polarization and produce polarized electrons from a
photocathode. New polarization rotation cells of Rubidium Titanyle Phosphate (RTP) have been implemented to replace
the previously used cells of Potassium Dihydrogen Phosphate (KD*P). 

Besides the high frequency helicity flip, to be able to measure physics asymmetries of the order of 33 ppb
(about eight times smaller than any other previous JLab parity-violating experiment), many helicity correlated
errors like position, incident angle, beam size and charge differences between buckets of different helicity
must be reduced by a factor of about four or better compared with previous experiments, mostly performed during
the 6 GeV era.

Such improvements require changes in the injector and its coupling to the first linac as well as a more refined
understanding of the 12 GeV machine optics. The MOLLER experiment received Critical Decision-0 (CD0, Mission Need
Statement) from the Department of Energy on December 2016 but due to fiscal budget constraints work on it was
paused until recently. MOLLER is seeking to receive CD2/3 in 2023.

\subsubsection{SoLID}

\begin{figure} [ht]
\center{
\includegraphics[width=0.95\linewidth, angle=0]{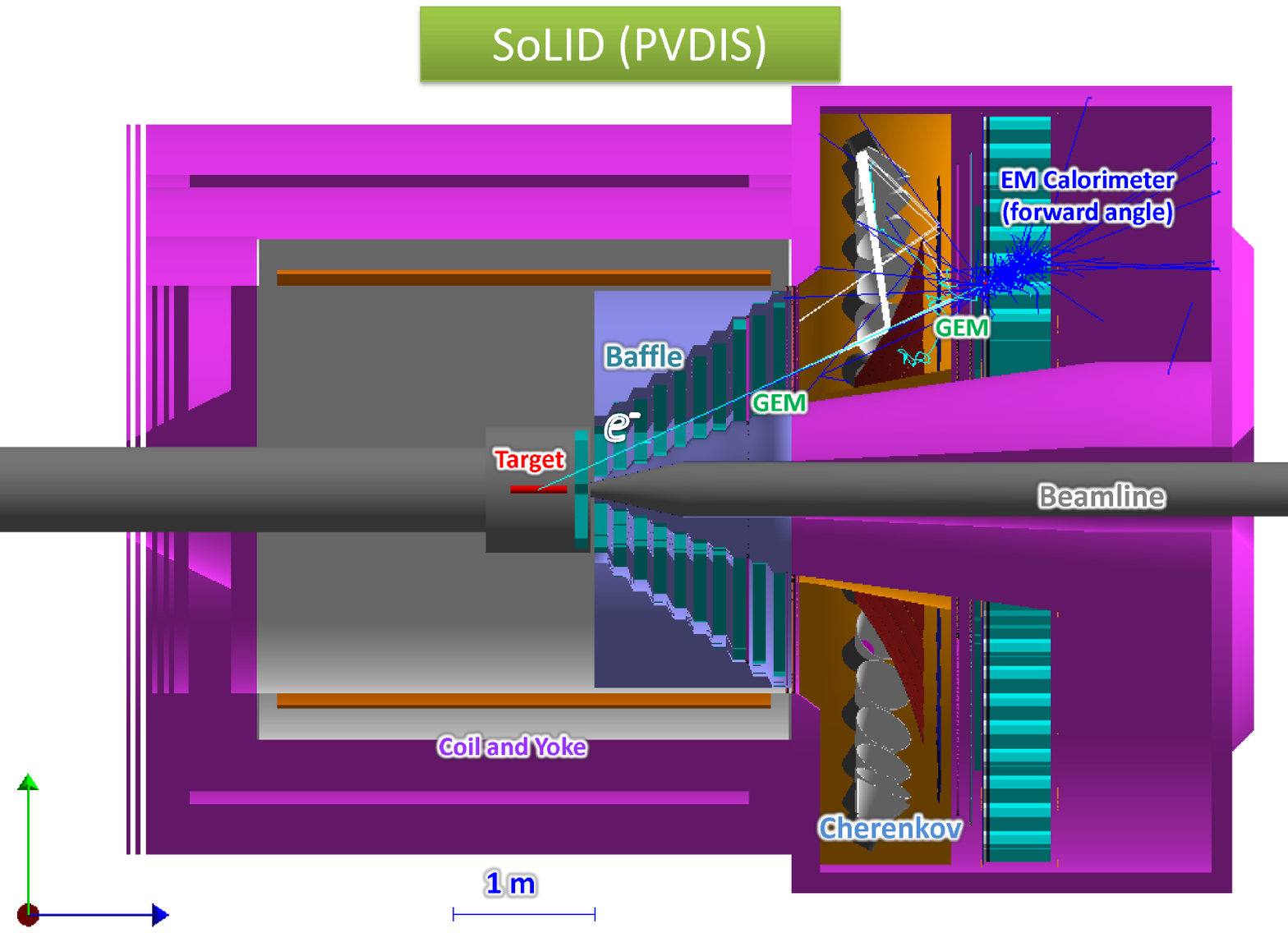}
\includegraphics[width=0.95\linewidth, angle=0]{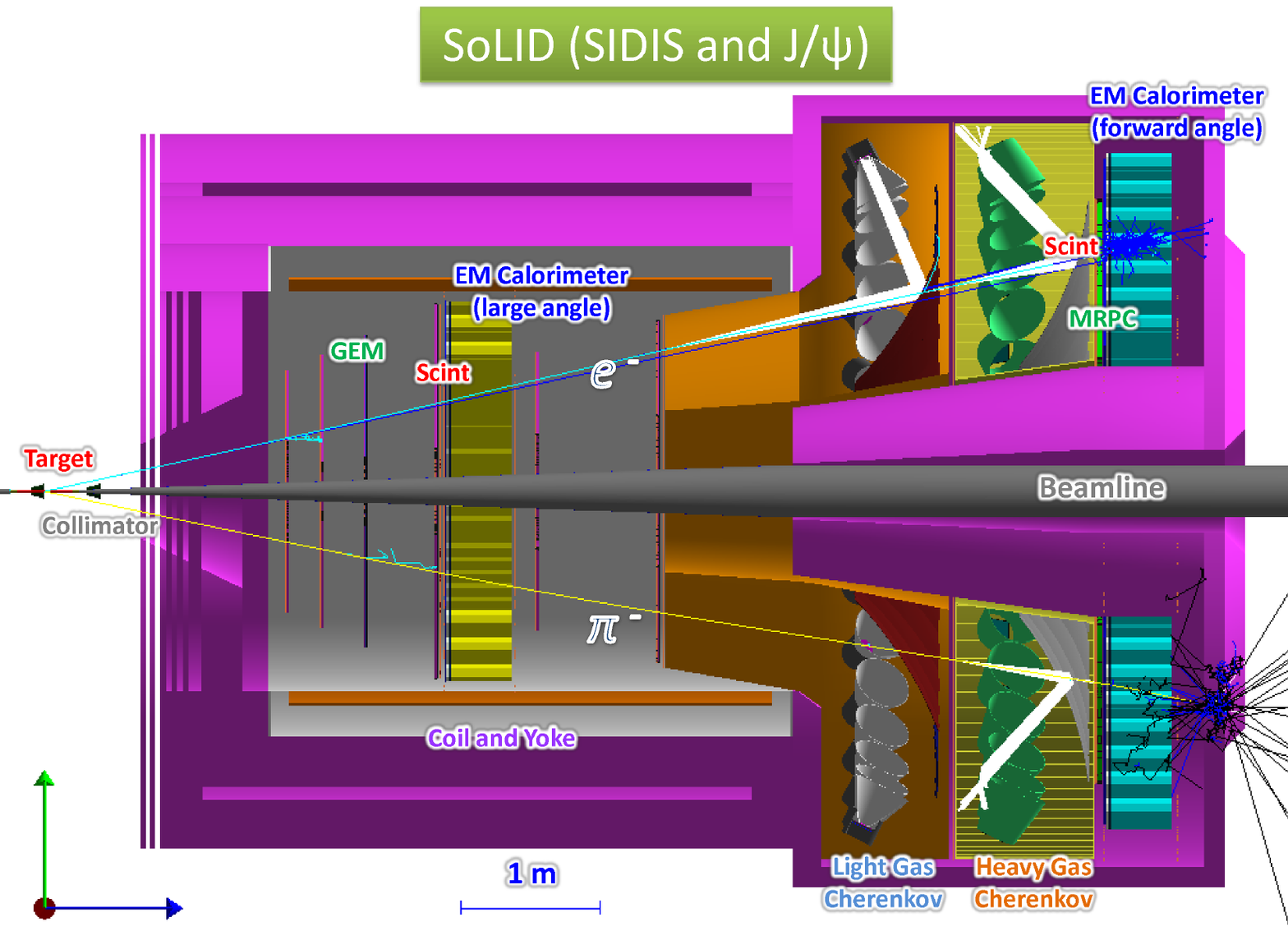}
\caption{\label{L20191021180931} Conceptual cross-section layout of the SoLID detector. The incident beam moves from left to right. The device has cylindrical shape with its major axis along the beam. It measures about eight meters along the beam direction and it has a diameter of about 5.6 m. At its core is the CLEO-II superconducting solenoid magnet. The magnet produces a field of up to 1.5 T, has an inner bore of 2.9 m and a length of about 3.5 m. }}
\end{figure}

A new, large acceptance, high luminosity detector,
SoLID ({\bf So}lenoidal {\bf L}arge {\bf I}ntensity {\bf D}evice)\cite{JGr2} has been proposed to fully exploit the
potential of JLab 12 GeV energy upgrade. As the name indicates, the key of this detector is to be able to operate
at luminosities much higher than possible in Halls B and D which also have large acceptance detectors. The core
research program approved by the PAC so far for such a device consists of one parity-violating, three semi-inclusive
deep-inelastic and a $J/\psi$ production proposal. Research programs with polarized targets are also being developed.
Figure \ref{L20191021180931} shows a conceptual layout of the detector. Two configurations are shown.

The top panel shows the proposed configuration to carry the parity-violating deep-inelastic scattering (PVDIS)
experiment while the bottom panel shows the general configuration used for the rest of the program. Note that to
be able to reach the design luminosity of $10^{39}$ cm$^{-2}$ sec$^{-1}$ required by the PVDIS experiments, a set of baffles is
required to block unwanted photons and hadrons originating in the target. The magnetic field must then be strong
enough to spiral the several GeV DIS electrons through the gaps in the baffles and also provide sufficient curvature
in the tracks so that their momentum can be reconstructed.  Both requirements can be met with a field integral along
the flight path on the order of 2.5 T-m.

\subsubsection{Compact Photon Source}

A compact, high intensity, multi-GeV photon source (CPS)\cite{JGr3} is being developed to gain access to new
lines of research in both Halls C and D. 
The research program in Hall C is focused on Deep-Virtual Compton Scattering (DVCS), Wide-Angle Compton
Scattering (WACS), Semi-Inclusive Deep Inelastic Scattering (SIDIS) meson production and neutral pion photoproduction
with regular and polarized targets. For this research, a Neutral Particle Spectrometer (NPS)\cite{JGr4} is being
developed to complement the existing High Momentum and Super-High Momentum Spectrometers (HMS and SHMS respectively)
of Hall C.

The research program in Hall D would use a CPS to produce a beam of neutral kaons \cite{JGr5} directed to the
existing GlueX detector system. Basically, this will be a new ``facility''. A flux of up to $10^4\ K_L/s$ is
expected, about three orders of magnitude larger than achieved in the past at other facilities. Such large fluxes
will allow one to perform measurements of both differential cross sections and self-analyzing polarizations of the
$\Lambda$, $\Sigma$, $\Xi$ and $\Omega$ hyperons produced. The data is expected to cement the orbitally excited
states in the $\Xi$ and $\Omega$ spectra as well as aid to constraint the partial wave analysis. It is also expected
to have a large impact in our understanding of the strange meson sector.

\subsection{Injector Improvements\label{MPrefsec}}

The present CEBAF injector \cite{AHRK123MeV} has a long history of reliability but there are improvements that can be made to support the 12 GeV physics program. 
In particular, the quarter cryomodule used to accelerate beam from 500 keV to 5 MeV introduces unwanted $x/y$
coupling as a result of the asymmetrical designs of the rf power couplers for early CEBAF cavities. 
This $x/y$ coupling makes it difficult to match the beam envelope across the quarter cryomodule \cite{MPRef2} which
in turn makes it difficult to obtain the maximum desired adiabatic damping required for parity-violation experiments
that have demanding helicity-correlated beam requirements \cite{MPRef3}.
 
\begin{figure}
\centering
\includegraphics[width=3.4in]{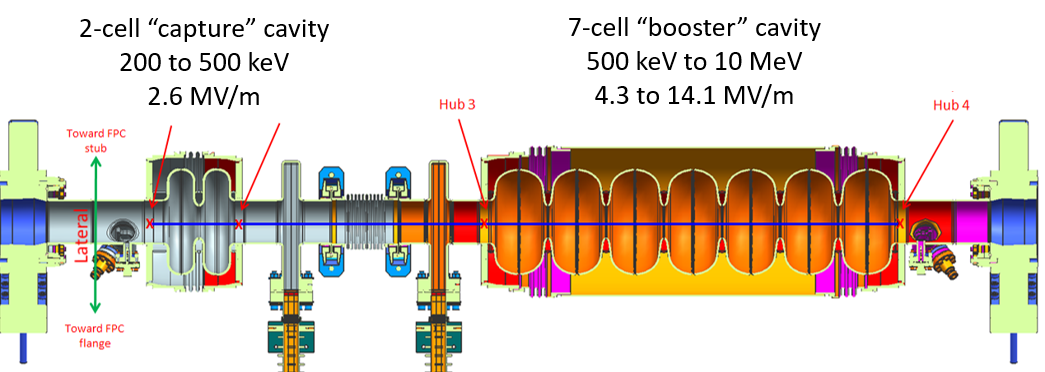}
\caption{Schematic diagram of new booster layout \label{fig:MPfig1}} 
\end{figure}

\begin{figure}
\centering
\includegraphics[width=3.4in]{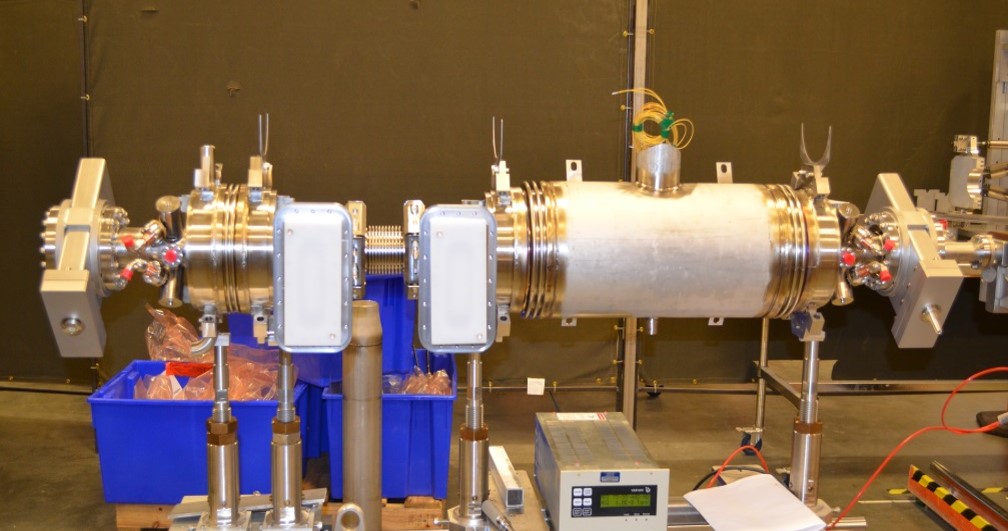}
\caption{Booster beam-line before enclosing it into a quarter cryomodule \label{fig:MPfig2}} 
\end{figure}

A new ``booster'' cryomodule \cite{MPRef4,MPRef5} was constructed and has been tested in a stand-alone
injector test facility for performance testing. The booster cryomodule is composed of two SRF cavities:
a 2-cell ``capture'' cavity to accelerate 200 keV beam from the photogun to 1.2~MeV total energy and a 7-cell cavity
to accelerate the 1.2~MeV beam to 10 MeV, as shown schematically in Fig.~\ref{fig:MPfig1}.
The booster cryomodule eliminates the only copper accelerating
structure used at CEBAF, and it is expected, the $x/y$ coupling problematic of the earlier quarter cryomodule.  We anticipate
the booster cryomodule will improve beam quality for parity-violation experiments and simplify injector
setup because there will be fewer RF-components required to accelerate beams to relativistic energy and
because the photogun will be operating at a 200 kV bias voltage, providing stiffer and more manageable beam.
A photograph of the two accelerating cavities on a beam-line before they were incorporated into
the quarter cryomodule is given in Figure~\ref{fig:MPfig2}.

The new booster cryomodule was installed at CEBAF in the summer of 2023. The plan is to increase the photogun bias voltage to 200 kV, instead of 130 kV used to present \cite{MPRef6}.  Modifying the photogun electrostatic
design \cite{MPRef7} is required. Similarly the electrostatic features of the Wein filters \cite{napac2022-wepa20}
that are part of the 4$\pi$ spin manipulator \cite{MPRef9} must be modified to maintain capability of
90~degree spin rotation for 200 keV beam.  

Another worthwhile improvement that supports the 12~GeV CEBAF physics program relates to the injector
RF ``chopper'' system \cite{MPRef10} originally used to create the required RF time structure on
DC beam produced from a thermionic gun that has since been removed from the accelerator.  Although the
photogun provides RF time structure directly, the chopper system is still used to remove nanoampere level
DC beam produced by low level DC light from the drive lasers \cite{MPRef11}.  

The present RF chopper system operates at the third subharmonic of the CEBAF accelerating frequency to support beam
delivery to three experiment halls.  The 12 GeV CEBAF provides beam to four experiment halls \cite{MPRef12}
and this means two halls must operate at 249.5 MHz with interleaved beams passing through the same chopper
slit.  An improved 12 GeV chopper system would provide independent chopper slits for each experiment hall.
For example, a chopper system operating at the sixth subharmonic of 249.5 MHz could provide independent beams to six experiment halls
\cite{MPRef13}.

\subsection{CEBAF Performance Plan}\label{CPP}

CEBAF has been run at the full 12 GeV project specification with pulsed electron beams not suitable for nuclear physics experiments. When CW beam is required, as for the experiments, CEBAF is unable to deliver the full 12 GeV beam to Hall D. For example, during the Spring 2022 physics running period the beam energy in Hall D was 11.6 GeV.
Effectively executing the 12 GeV experimental program is crucial in maintaining CEBAF as the world leader in experimental nuclear physics. 

The CEBAF Performance Plan (CPP) is an internal technical document \cite{CPPREF1} authored as a performance improvement strategy for CEBAF systems published soon after regular physics running began.  This document presents a plan for addressing the known performance gaps as soon as possible, and addressing obsolete systems. The plan places a priority on addressing the performance gaps up front so that the majority of the 12~GeV program can benefit from reliable CEBAF operations at design beam parameters.

Gap analysis was performed on several aspects of CEBAF operations \cite{CPPREF4}. Gaps are identified with respect to CEBAF operational goals outlined within the technical note. There are three subsections: CEBAF availability, energy reach, and operations performance. An outlined performance plan found in subsequent sections of the document map actions to close the performance gaps for realizing the stated goals. Frequent critical system failures, CEBAF energy degradation, and reduced weeks of operations driven by funding issues highlight a few of the topics sought to be addressed by the CEBAF Performance Plan. The gap in CEBAF energy reach is not insurmountable nor large enough to warrant a halt in 12 GeV operations, but it has been significant enough to place the effective execution of the 12 GeV experimental program at risk.  The plans presented in the CPP are meant to mitigate this risk \cite{CPPREF1,CPPREF4}.

In 2018 funding was allocated to support parts of the CPP.  A ``CEBAF Reliability Plan FY18-21” was submitted to the Department of Energy from Jefferson Lab leadership aimed at improving CEBAF reliability \cite{CPPREF2}. The subcategories funded include critical spare parts for accelerator and cryogenic systems, an RF klystron purchase agreement to address end-of-life components, immediate investments in higher risk obsolete systems, energy reach efforts to counter continued LINAC gradient degradation, and projects aimed at optimizing maintenance practices.  Significant progress has been made with implementing the CPP strategy within the aforementioned categories, though early in the strategy implementation to fully realize true reliability improvement.  An internal technical note, “CEBAF Performance Plan Implementation Summary'' JLAB-TN-20-012, highlights specific success and challenges in executing the CPP strategy through 2020 \cite{CPPREF3}. 
\begin{figure}
\centering
\includegraphics[width=3.4in]{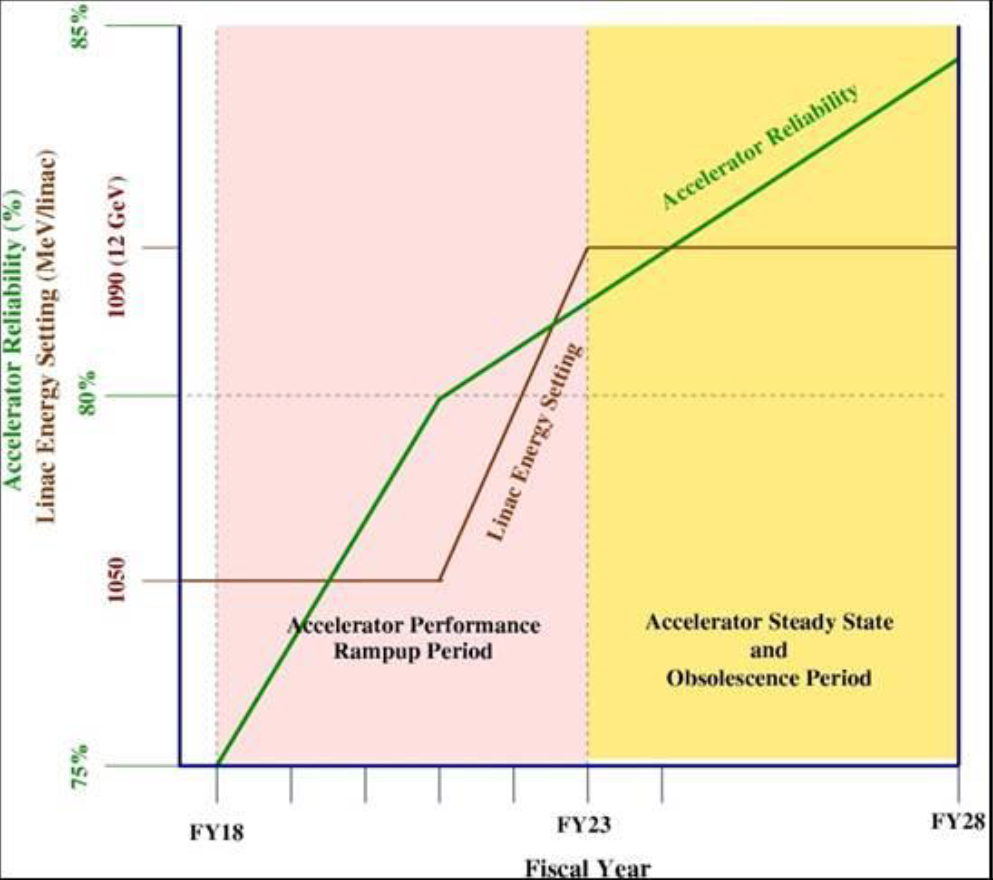}
\caption{Future reliability and energy reach predictions from executing the CEBAF Performance Plan. (From \cite{CPPREF4})  \label{fig:CPPFIGL}} 
\end{figure}

To address energy reach, and to build operating margin into the beam acceleration systems to the point that 12 GeV running is supported even with the loss of one full C100 cryomodule, a long-term plan has been developed involving several elements: (1) upgrading individual old-style C20 modules so that they can run at gradients approaching 75 MeV, (2) systematically refurbishing worst-performing C100 modules, and (3) understanding through performance analysis studies and measurements observed decay of superconducting cavity maximum field. Over the next five years, 7 ``C75'' cryomodule replacement/upgrades \cite{ciovati:srf2021} and 5 C100 cryomodule refurbishments will be completed. As these tasks are performed, the best recent understanding of cavity processing will be incorporated into any newly installed cryomodules \cite{CPPREF3}. Figure~\ref{fig:CPPFIGL} provides summary estimates of accelerator availability and energy reach over the next half decade.

% 
% Add a short parameter or two on Positrons
%& 22 GeV upgrade

%\subsection{Positrons and 22 GeV Upgrade}
%
%Looking to the further future, there are two upgrades %planned for the CEABF accelerator that would %dramatically increase the physics reach of Jefferson %Lab.    
%
%The first is a positron 
%
%

\section{Summary}
Jefferson Lab's CEBAF accelerator has been upgraded and operated at 12 GeV beam energy. This achievement was made possible by continuous improvement in the performance capabilities in niobium superconducting cavities that have arisen in the 25 years since the first CEBAF was completed. Individual ``C100" cryomodules of length 10.4 m capable of providing over 100 MeV CW beam energy gain were designed and built.
The {\it Renasence} cryomodule was the first SRF accelerator cryomodule to accelerate CW electron beam by 100 MeV. In addition to the upgrades of the linac, significant upgrades to other accelerator systems needed to be made: the recirculating arc magnets now operate at double the field previously, the cooling capacity of the main helium liquifier has been doubled, the upper beam energy of the injector has been enhanced, and the site electrification and cooling has been upgraded to allow beam operations at elevated energies. In addition, as part of the upgrade project a new experimental Hall D devoted to ``QCD spectroscopy'' was added to CEBAF and the beam preparation systems and extraction systems have been modified so that all experimental halls at CEBAF can operate with beam simultaneously. Experiments at 12 GeV have led to greater and deeper understanding of the atomic nucleus and its constituents, and the strong force that holds it together \cite{mckeown}.

\section{Acknowledgements}
This material is based upon work supported by the U.~S.~Department of Energy, Office of Science, Office of Nuclear Physics under contract DE-AC05-06OR23177. As is typical for review papers of this type, authors, who may have worked at one time at Jefferson Lab on some aspect of the 12 GeV project may no longer be working there. In the affiliation list for this paper the present or final institutional association of the author is listed.

\appendix*
\section{Experimental Hall Beam Performance Requirements}
A document was compiled summarizing the beam delivery performance requirements
for each of the CEBAF experimental halls and shared with individual
hall physics users to assist them in preparing proposals for beam time \cite{AFetal}.
For reference purposes, the individual tables are added to this paper.
Except for the caveat mentioned in Section~\ref{CPP}
regarding beam energy, these tables document the beam quality achieved and routinely delivered to
experiment users in the 12 GeV CEBAF era.

\begin{table*}[h]
\caption{\label{tab:table3}Delivery Beam Parameter Table for Hall A.}
\begin{ruledtabular}
\begin{tabular}{lcc}
 Beam property&Nominal value/range&Stability over 8 hours\\ \hline
 Spot size at target (rms) [\textmu m]&horizontal $<$
\footnote{Interpret '$<$' as 'not to exceed'.} 250, vertical $<$ 200&horizontal 20, vertical 20\\
 Angular divergence at target [\textmu rad]&$<$ 20
 &$<$ 2\\
 Beam current [\textmu A]&1-120&10\% of nominal\\
 Charge per beam bunch [fC]&4-480&10\% of nominal\\
 Bunch repetition rate [MHz] &249.5&\\
Beam position&\vbox{\hbox{all locations within}\hbox{2.5 mm of target center}}&
\vbox{\hbox{$<$ 40 \textmu m with slow lock}\hbox{ and 20 \textmu m at 60 Hz}}\\
Relative energy spread ({rms})&\vbox{\hbox{pass 1}\hbox{$< 10^{-4}$}}, \vbox{\hbox{pass 2}\hbox{$< 10^{-4}$}},
\vbox{\hbox{pass 3}\hbox{$< 10^{-4}$}},
\vbox{\hbox{pass 4}\hbox{$< 3\times 10^{-4}$}}, \vbox{\hbox{pass 5}\hbox{$< 5\times 10^{-4}$}}&$~$ 10\% of nominal\\
 Beam direction [\textmu rad]&$\pm$300&$<$ 2\\
 Energy range [GeV]&1-11&\\
 Energy accuracy (rms)&$3\times10^{-3}$&\\
 Beam polarization&up to 85\%&\\
 Charge asymmetry&$<$ 0.1\% &\\ 
 Background beam halo&$<$ 0.1\% &\\ 
 Beam availability (including RF trips)&60\% &\\
\end{tabular}
\end{ruledtabular}
\end{table*}

\begin{table*}
\caption{\label{tab:table4}Delivery Beam Parameter Table for Hall B.}
\begin{ruledtabular}
\begin{tabular}{lcc}
 Beam property&Nominal value/range&Stability over 8 hours\\ \hline
 Spot size at wire scanner (rms) [\textmu m]&$<$
\footnote{Interpret '$<$' as 'not to exceed'.} 100 for 1-6 GeV, $<$ 200 for 7-11 GeV&User defined measurement frequency\\
 Angular divergence at target [\textmu rad]&$<$ 100
 &$<$ 2\\
 Beam current [nA]&1-160&$<$ 5\% when $>$ 5 nA\\
 Charge per beam bunch [fC]&$4\times10^{-3}-0.64$&$<$ 5\% when $>$ 5 nA\\
 Bunch repetition rate [MHz] &249.5&\\
Beam position&\vbox{\hbox{all locations within}\hbox{2 mm of beam axis}}&
\vbox{\hbox{$<$ 40 \textmu m with slow lock}\hbox{ and 20 \textmu m at 60 Hz}}\\
Relative energy spread (rms)&\vbox{\hbox{pass 1}\hbox{$< 10^{-4}$}}, \vbox{\hbox{pass 2}\hbox{$< 10^{-4}$}},
\vbox{\hbox{pass 3}\hbox{$< 10^{-4}$}},
\vbox{\hbox{pass 4}\hbox{$< 3\times 10^{-4}$}}, \vbox{\hbox{pass 5}\hbox{$< 5\times 10^{-4}$}}&$~$ 10\% of nominal\\
 Beam direction [\textmu rad]&$\pm$300&$<$ 2\\
 Energy range [GeV]&1-11&\\
 Energy accuracy (rms)&$3\times10^{-3}$&\\
 Beam polarization&up to 85\%&\\
 Charge asymmetry&$<$ 0.1\% &\\ 
 Background beam halo&$<$ 0.1\% &\\ 
 Beam availability (including RF trips)&60\% &\\
\end{tabular}
\end{ruledtabular}
\end{table*}

\begin{table*}
\caption{\label{tab:table5}Delivery Beam Parameter Table for Hall C.}
\begin{ruledtabular}
\begin{tabular}{lcc}
 Beam property&Nominal value/range&Stability over 8 hours\\ \hline
 Spot size at target (rms) [\textmu m]&horizontal $<$
\footnote{Interpret '$<$' as 'not to exceed'.} 250, vertical $<$ 200&horizontal 20, vertical 20\\
 Angular divergence at target [\textmu rad]&$<$ 20
 &$<$ 2\\
 Beam current [\textmu A]&1-120&10\% of nominal\\
 Charge per beam bunch [fC]&4-480&10\% of nominal\\
 Bunch repetition rate [MHz] &249.5&\\
Beam position&\vbox{\hbox{all locations within}\hbox{2.5 mm of target center}}&
\vbox{\hbox{$<$ 40 \textmu m with slow lock}\hbox{ and 20 \textmu m at 60 Hz}}\\
Relative energy spread (rms)&\vbox{\hbox{pass 1}\hbox{$< 10^{-4}$}}, \vbox{\hbox{pass 2}\hbox{$< 10^{-4}$}},
\vbox{\hbox{pass 3}\hbox{$< 10^{-4}$}},
\vbox{\hbox{pass 4}\hbox{$< 3\times 10^{-4}$}}, \vbox{\hbox{pass 5}\hbox{$< 5\times 10^{-4}$}}&$~$ 10\% of nominal\\
 Beam direction [\textmu rad]&$\pm$300&$<$ 2\\
 Energy range [GeV]&1-11&\\
 Energy accuracy (rms)&$3\times10^{-3}$&\\
 Beam polarization&up to 85\%&\\
 Charge asymmetry&$<$ 0.1\% &\\ 
 Background beam halo&$<$ 0.1\% &\\ 
 Beam availability (including RF trips)&60\% &\\
\end{tabular}
\end{ruledtabular}
\end{table*}

\begin{table*}
\caption{\label{tab:table6}Delivery Beam Parameter Table for Hall D.}
\begin{ruledtabular}
\begin{tabular}{lcc}
 Beam property&Nominal value/range&Stability over 8 hours\\ \hline
 Spot size at target (rms) [\textmu m]&horizontal $<$
\footnote{Interpret '$<$' as 'not to exceed'.} 1000, vertical $<$ 500&horizontal 100, vertical 100\\
 Angular divergence at target [\textmu rad]&$<$ 15
 &$<$ 1\\
 Beam current [nA]&1-2000&10\% of nominal\\
 Charge per beam bunch [fC]&$4\times10^{-3}-8$&10\% of nominal\\
 Bunch repetition rate [MHz] &249.5&\\
Beam position&$\pm$ 1 mm&$<$ 40 \textmu m\\
Relative energy spread (rms)&$2-3\times10^{-3}$&$~$ 10\% of nominal\\
 Beam direction [\textmu rad]&$\pm$30&$<$ 2\\
 Energy range [GeV]&8.8-12.1&\\
 Energy accuracy (rms)&$3\times10^{-3}$&\\
 Background beam halo&$<$ 0.1\% &\\ 
 Beam availability (including RF trips)&60\% &\\
\end{tabular}
\end{ruledtabular}
\end{table*}

\bibliography{12GeVPap}
\end{document}